\pdfoutput=1

\documentclass[11pt,twoside,a4paper,cmspaper,final,collab]{cms-tdr}
\def\svnVersion{482967P}\def\cmsCernNoTag{CERN-EP-2018-149}\def\cmsCernDate{\today}\def\cmsMessage{Published in the Journal of High Energy Physics as \href{http://dx.doi.org/10.1007/JHEP11(2018)151}{\doi{10.1007/JHEP11(2018)151.}}}

\begin{document}\cmsNoteHeader{SUS-17-003}

\hyphenation{had-ron-i-za-tion}
\hyphenation{cal-or-i-me-ter}
\hyphenation{de-vices}
\RCS$HeadURL: svn+ssh://svn.cern.ch/reps/tdr2/papers/SUS-17-003/trunk/SUS-17-003.tex $
\RCS$Id: SUS-17-003.tex 474543 2018-09-10 20:09:32Z vdutta $

\newlength\cmsFigWidth
\ifthenelse{\boolean{cms@external}}{\setlength\cmsFigWidth{0.85\columnwidth}}{\setlength\cmsFigWidth{0.4\textwidth}}
\ifthenelse{\boolean{cms@external}}{\providecommand{\cmsLeft}{top\xspace}}{\providecommand{\cmsLeft}{left\xspace}}
\ifthenelse{\boolean{cms@external}}{\providecommand{\cmsRight}{bottom\xspace}}{\providecommand{\cmsRight}{right\xspace}}
\ifthenelse{\boolean{cms@external}}{\providecommand{\cmsTable}{\relax}}{\providecommand{\cmsTable}[1]{\resizebox{\textwidth}{!}{#1}}}
\ifthenelse{\boolean{cms@external}}{\providecommand{\CL}{C.L.\xspace}}{\providecommand{\CL}{CL\xspace}}
\providecommand{\NA}{\ensuremath{\text{---}}}
\newlength\cmsTabSkip\setlength{\cmsTabSkip}{1ex}
\newcommand{\distau}{\ensuremath{\PSGt\PSGt}\xspace}
\newcommand{\tausneu}{\ensuremath{\widetilde{\nu}_{\tau}}\xspace}
\newcommand{\neutralinos}{\ensuremath{\chiz_{\mathrm{i}}}\xspace}
\newcommand{\ntwo}{\PSGczDt}
\newcommand{\charginos}{\ensuremath{\chipm_{\mathrm{i}}}\xspace}
\newcommand{\cone}{\PSGcpmDo}
\newcommand{\conemp}{\ensuremath{\PSGc^\mp_1}\xspace}
\newcommand{\etau}{\ensuremath{\Pe\tauh}\xspace}
\newcommand{\mutau}{\ensuremath{\mu\tauh}\xspace}
\newcommand{\emu}{\ensuremath{\Pe\mu}\xspace}
\newcommand{\tautau}{\ensuremath{\tauh\tauh}\xspace}
\newcommand{\dxy}{\ensuremath{d_{xy}}\xspace}
\newcommand{\dz}{\ensuremath{d_{z}}\xspace}
\newcommand{\irel}{\ensuremath{I_{\text{rel}}}\xspace}
\newcommand{\dR}{\ensuremath{\Delta R}\xspace}
\newcommand{\MT}{\ensuremath{m_{\mathrm{T}}}\xspace}
\newcommand{\mll}{\ensuremath{m(\ell_1,\ell_2)}\xspace}
\newcommand{\dphill}{\ensuremath{\Delta\phi(\ell_1,\ell_2)}\xspace}
\newcommand{\wjets}{\ensuremath{\PW}+jets\xspace}
\newcommand{\zjets}{\ensuremath{\PZ}+jets\xspace}
\newcommand{\ztautau}{\ensuremath{\PZ\to\PGt\PGt}\xspace}
\newcommand{\zmumu}{\ensuremath{\PZ\to\mu\mu}\xspace}
\newcommand{\mttwo}{\ensuremath{m_{\mathrm{T2}}}\xspace}
\newcommand{\DZ}{\ensuremath{D_{\zeta}}\xspace}
\newcommand{\sumMT}{\ensuremath{\Sigma\MT}\xspace}
\newcommand{\njet}{\ensuremath{N_{\text{jet}}}\xspace}

\cmsNoteHeader{SUS-17-003}
\title{Search for supersymmetry in events with a \PGt lepton pair and missing transverse momentum in proton-proton collisions at $\sqrt{s}=13\TeV$}

\date{\today}

\abstract{A search for the electroweak production of supersymmetric particles in proton-proton collisions at a center-of-mass energy of $13\TeV$ is presented in final states with a \PGt lepton pair. Both hadronic and leptonic decay modes are considered for the \PGt leptons. Scenarios involving the direct pair production of \PGt sleptons, or their indirect production via the decays of charginos and neutralinos, are investigated. The data correspond to an integrated luminosity of $35.9\fbinv$ collected with the CMS detector in 2016. The observed number of events is consistent with the standard model background expectation. The results are interpreted as upper limits on the cross section for \PGt slepton pair production in different scenarios. The strongest limits are observed in the scenario of a purely left-handed low mass \PGt slepton decaying to a nearly massless neutralino. Exclusion limits are also set in the context of simplified models of chargino-neutralino and chargino pair production with decays to \PGt leptons, and range up to 710 and 630\GeV, respectively.}

\hypersetup{
pdfauthor={CMS Collaboration},
pdftitle={Search for supersymmetry in events with a tau lepton pair and missing transverse momentum in proton-proton collisions at sqrt(s)=13 TeV},
pdfsubject={CMS},
pdfkeywords={CMS, physics, SUSY}}

\maketitle

\section{Introduction}
\label{sec:intro}
Supersymmetry (SUSY)~\cite{Ramond:1971gb,Golfand:1971iw,Neveu:1971rx,Volkov:1972jx,Wess:1973kz,Wess:1974tw,Fayet:1974pd,Nilles:1983ge} is an attractive extension of the standard model (SM) of particle physics. It potentially provides solutions to some of the shortcomings affecting the SM, such as the need for fine tuning~\cite{'tHooft:1979bh,Witten:1981nf,Dine:1981za,Dimopoulos:1981au,Dimopoulos:1981zb,Kaul:1981hi} to explain the observed value of the Higgs boson mass~\cite{Aad:2012tfa,Chatrchyan:2012ufa,Chatrchyan:2013lba,Aad:2014aba,Khachatryan:2014jba,Aad:2015zhl}, and the absence of a dark matter (DM) candidate. Supersymmetric models are characterized by the presence of a superpartner for every SM particle with the same quantum numbers except that its spin differs from that of its SM counterpart by half a unit. The cancellation of quadratic divergences in quantum corrections to the Higgs boson mass from SM particles and their superpartners could resolve the fine-tuning problem. In SUSY models with $R$-parity conservation~\cite{Farrar:1978xj}, the lightest supersymmetric particle (LSP) is stable~\cite{djoua,carena} and could be a DM candidate~\cite{darkmatter}. The superpartners of the electroweak gauge and Higgs bosons, namely the bino, winos, and Higgsinos, mix to form neutral and charged mass eigenstates, referred to as the neutralinos (\neutralinos) and charginos (\charginos), respectively. In this paper we assume \PSGczDo, the lightest neutralino, to be the LSP.

The analysis reported in this paper investigates the production of the hypothetical \PGt slepton (\PSGt), the superpartner of the \PGt lepton. Supersymmetric scenarios in which the \PSGt is light lead to the possibility of \PGt lepton rich final states~\cite{Belanger:2012jn,Arganda:2018hdn}. Coannihilation scenarios involving a light \PSGt that has a small mass splitting with an LSP that is almost purely bino lead to a DM relic density consistent with cosmological observations~\cite{WMAP,Griest:1990kh,Vasquez,King,Battaglia:2001zp,Arnowitt:2008bz}, making the search for new physics in these final states particularly interesting. In this analysis, we examine simplified SUSY models~\cite{Simp,Alwall:2008ag,Alwall:2008va,Alves:2011wf} in which the \PSGt can be produced either directly, through pair production, or indirectly, in the decay chains of charginos and neutralinos. In all cases, we assume that the \PSGt decays to a \PGt lepton and \PSGczDo. The most sensitive searches for direct \PSGt pair production to date were performed at the CERN LEP collider~\cite{Heister:2001nk,Abdallah:2003xe,Achard:2003ge,Abbiendi:2003ji,lepsusy}. At the CERN LHC, the ATLAS~\cite{Aad:2014yka,Aad:2015eda} and CMS~\cite{SUS14022,Khachatryan:2015kxa} Collaborations have both performed searches for direct and indirect \PSGt production with 8\TeV LHC data. The ATLAS Collaboration has also recently reported the results of a search for SUSY in final states with \PGt leptons, probing indirect \PSGt production in models of chargino-neutralino and chargino pair production, using data collected at $\sqrt{s} = 13\TeV$~\cite{Aaboud:2017nhr}.

\begin{figure}[htb]
\centering
\includegraphics[width=0.32\textwidth]{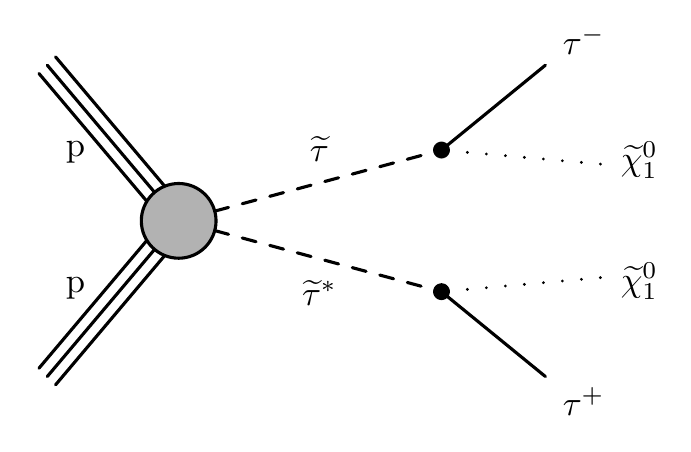}
\includegraphics[width=0.32\textwidth]{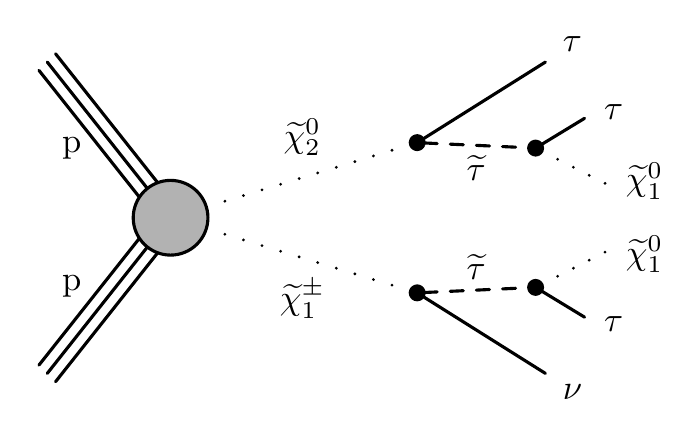}
\includegraphics[width=0.32\textwidth]{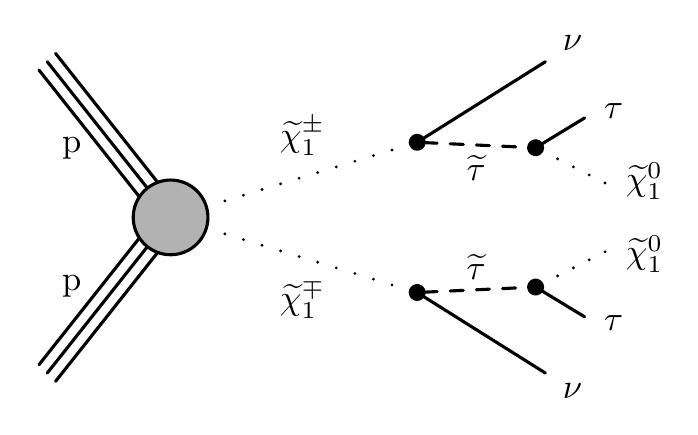}
\caption{\label{fig:diagram} Diagrams for the simplified models studied in this paper: direct \PSGt pair production followed by each \PSGt decaying to a \PGt lepton and \PSGczDo (left), and chargino-neutralino (middle) and chargino pair (right) production with subsequent decays leading to \PGt leptons in the final state.}
\end{figure}

The cross section for direct \PSGt pair production depends strongly on the chirality of the SM partner~\cite{Fuks:2013lya}, while the experimental acceptance also changes considerably due to differences in the polarization of the \PGt leptons. We use the terms left- or right-handed \PSGt to refer to a \PSGt that is the superpartner of a left- or right-handed chiral state, respectively. In the case of a purely right-handed \PSGt, the decay products of hadronically decaying \PGt leptons originating from \PSGt decays have larger visible transverse momentum (\pt) than in the purely left-handed scenario, while the reverse is true for leptonically decaying \PGt leptons. Three different scenarios of direct \PSGt pair production are considered in this paper: (i) a purely left-handed \PSGt ($\PSGt_{\mathrm{L}}$), (ii) a purely right-handed \PSGt ($\PSGt_{\mathrm{R}}$), and (iii) maximal mixing between the right- and left-handed eigenstates. We also consider simplified models of mass-degenerate chargino-neutralino (\cone \ntwo) and chargino pair (\cone \conemp) production. We assume that \ntwo (the second-lightest neutralino mass eigenstate) decays through the chain $\ntwo \to \PGt \PSGt \to \PGt \PGt \PSGczDo$, and that \cone(the lightest chargino) decays as $\cone \to \PSGt \PGnGt / \tausneu \PGt \to \PGt \PGnGt \PSGczDo$, with equal branching fractions assumed for each of the two possible \cone decay chains. For these indirect \PSGt production mechanisms, we assume the \PSGt to be in the maximally mixed state, and the degenerate \PSGt and \tausneu masses to be halfway between the mass of the produced particles ($\cone/\ntwo$) and the \PSGczDo mass. Diagrams illustrating these simplified models of direct and indirect \PSGt production are shown in Fig.~\ref{fig:diagram}.

The results reported in this paper are based on data collected with the CMS detector at the LHC during 2016 in proton-proton (\Pp\Pp) collisions at a center-of-mass energy of 13\TeV, corresponding to an integrated luminosity of 35.9\fbinv. We study events with two \PGt leptons in the final state, taking into account both hadronic and leptonic decay modes of the \PGt lepton. The following reconstructed visible final states are considered: \emu, \etau, \mutau, and \tautau, where \tauh denotes a hadronically decaying \PGt lepton. For the purposes of this paper, we will occasionally refer to the \tautau final state as the all-hadronic final state, and the \emu, \etau, and \mutau final states collectively as the leptonic final states. In most cases, we require the presence of significant missing transverse momentum, which can arise from the presence of stable neutralinos produced at the end of the SUSY particle decay cascades, as well as from the neutrinos produced in \PGt lepton decays.

The structure of this paper is as follows. A brief description of the CMS detector is presented in Section~\ref{sec:detector}, followed by a discussion of the event reconstruction and simulation in Section~\ref{sec:evtreco}. We describe the event selection for the search in Section~\ref{sec:evtsel}, the background estimation strategy in Section~\ref{sec:bkgest}, and the systematic uncertainties affecting the analysis in Section~\ref{sec:sysunc}. Finally, the results of the search and their statistical interpretation are presented in Section~\ref{sec:results}, followed by a summary in Section~\ref{sec:summary}.

\section{The CMS detector}
\label{sec:detector}
The central feature of the CMS apparatus is a superconducting solenoid of 6\unit{m} internal diameter, providing a magnetic field of 3.8\unit{T}. Within the solenoid volume are a silicon pixel and strip tracker, a lead tungstate crystal electromagnetic calorimeter (ECAL), and a brass and scintillator hadron calorimeter, each composed of a barrel and two endcap sections. Forward calorimeters extend the pseudorapidity ($\eta$) coverage provided by the barrel and endcap detectors. Muons are detected in gas-ionization chambers embedded in the steel flux-return yoke outside the solenoid. Events of interest are selected using a two-tiered trigger system~\cite{Khachatryan:2016bia}. The first level, composed of custom hardware processors, uses information from the calorimeters and muon detectors to select events at a rate of around 100\unit{kHz} within a time interval of less than 4\mus. The second level, known as the high-level trigger, consists of a farm of processors running a version of the full event reconstruction software optimized for fast processing, and reduces the event rate to around 1\unit{kHz} before data storage. A more detailed description of the CMS detector, together with a definition of the coordinate system used and the relevant kinematic variables, can be found in Ref.~\cite{Chatrchyan:2008zzk}.

\section{Event reconstruction and simulated samples}
\label{sec:evtreco}
Event reconstruction uses a particle-flow (PF) algorithm~\cite{Sirunyan:2017ulk}, combining information from the tracker, calorimeter, and muon systems to identify charged and neutral hadrons, photons, electrons, and muons in an event. The missing transverse momentum, \ptvecmiss, is computed as the negative vector sum of the \pt of all PF candidates reconstructed in an event, and its magnitude \ptmiss is an important discriminator between signal and SM background. Events selected for the search are required to pass filters~\cite{CMS-PAS-JME-16-004} designed to remove detector- and beam-related noise and must have at least one reconstructed vertex. Usually more than one such vertex is reconstructed, due to pileup, \ie, multiple $\Pp\Pp$ collisions within the same or neighboring bunch crossings. The reconstructed vertex with the largest value of summed physics-object $\pt^2$ is selected to be the primary $\Pp\Pp$ interaction vertex. The physics objects are the jets, clustered using a jet finding algorithm~\cite{Cacciari:2008gp,Cacciari:2011ma} with the tracks assigned to the vertex as inputs, and the associated \ptvecmiss.

Charged particles that originate from the primary vertex, photons, and neutral hadrons are clustered into jets using the anti-\kt algorithm~\cite{Cacciari:2008gp} with a distance parameter of 0.4, as implemented in the {\sc FastJet} package~\cite{Cacciari:2011ma}. The jet energy is corrected to account for the contribution of additional pileup interactions in an event and to compensate for variations in detector response~\cite{Cacciari:2011ma,pileup}. Jets considered in the searches are required to have their axes within the tracker volume, within the range $\abs{\eta} < 2.4$.  We also require them to have $\pt>20$\GeV. Jets are required to be separated from electron, muon, or \tauh candidates that are selected for the analysis by $\dR \equiv \sqrt{\smash[b]{(\Delta\eta)^2 + (\Delta\phi)^2} } > 0.4$ in order to avoid double counting of objects.

Jets originating from the hadronization of \PQb~quarks are identified, or ``tagged", with the combined secondary vertex (CSV) algorithm~\cite{Chatrchyan:2012jua,Sirunyan:2017ezt} using two different working points, referred to as ``loose" and ``medium". The \PQb~tagging efficiency for jets originating from \PQb~quarks is measured in simulation to be about 81 (63)\% for the loose (medium) working point, while the misidentification rates for jets from charm quarks, and from light quarks or gluons, are about 37 and 9\% (12 and 1\%), respectively.

Electron candidates are reconstructed by first matching clusters of energy deposited in the ECAL to reconstructed tracks. Selection criteria based on the distribution of the shower shape, track--cluster matching, and consistency between the cluster energy and track momentum are then used in the identification of electron candidates~\cite{Khachatryan:2015hwa}. Muon candidates are reconstructed by requiring consistent measurement patterns in the tracker and muon systems~\cite{Chatrchyan:2012xi}. Electron and muon candidates are required to be consistent with originating from the primary vertex by imposing restrictions on the magnitude of the impact parameters of their tracks with respect to the primary vertex in the transverse plane (\dxy), and on the longitudinal displacement (\dz) of those impact points. To ensure that the electron or muon candidate is isolated from any jet activity, the relative isolation quantity (\irel), defined as the ratio of the scalar \pt sum of the particles in an $\eta$--$\phi$ cone around the candidate to the candidate \pt, is required to be below a threshold appropriate for the selection under consideration. An area-based estimate~\cite{pileup} of the pileup energy deposition in the cone is used to correct \irel for contributions from particles originating from pileup interactions.

The \tauh candidates are reconstructed using the CMS hadron-plus-strips algorithm~\cite{Chatrchyan:2012zz,Khachatryan:2015dfa}. The constituents of the reconstructed jets are used to identify individual \PGt lepton decay modes with one charged hadron and up to two neutral pions, or three charged hadrons. The presence of extra particles within the jet, not compatible with the reconstructed decay mode, is used as a criterion to discriminate \tauh decays from other jets.  A multivariate discriminant \cite{CMS-PAS-TAU-16-002}, which contains isolation as well as lifetime information, is used to suppress the rate for quark and gluon jets to be misidentified as \tauh candidates. The working point used for the analysis in the \etau and \mutau final states, referred to as the ``tight" working point, typically has an efficiency of around 50\% for genuine \tauh, with a misidentification rate of approximately 0.03\% for light-quark or gluon jets. A more stringent (``very tight") working point is used for the analysis in the \tautau final state in order to suppress the background from SM events comprised uniquely of jets produced through the strong interaction, referred to as quantum chromodynamics (QCD) multijet events. The very tight working point corresponds to typical efficiencies of around 40\% for genuine \tauh, and a misidentification rate of approximately 0.01\% for light-quark or gluon jets. We also employ a relaxed (``loose") working point in the extrapolation procedures used to estimate the contributions of events to the background in which light-quark or gluon jets are misidentified as \tauh. The loose working point corresponds to an efficiency of $\approx$65\% for genuine \tauh, and a misidentification rate of $\approx$0.07\%.  Electrons and muons misidentified as \tauh are suppressed using dedicated criteria based on the consistency between the measurements in the tracker, calorimeters, and muon detectors~\cite{Khachatryan:2015dfa,CMS-PAS-TAU-16-002}.

Significant contributions to the SM background for this search originate from Drell-Yan+jets (DY+jets), \wjets, \ttbar, and diboson processes, as well as from QCD multijet events. Smaller contributions arise from rare SM processes such as triboson and Higgs boson production, single top quark production, and top quark pair production in association with vector bosons. We rely on a combination of data control samples and Monte Carlo (MC) simulations to estimate the contributions of each background source. MC simulations are also used to model the signal processes.

{\tolerance=800 The \MGvATNLO 2.3.3~\cite{Alwall:2014hca} event generator is used at leading order (LO) precision to produce simulated samples of the \wjets and DY+jets processes, based on the NNPDF3.0LO~\cite{Ball:2014uwa} set of parton distribution functions (PDFs). Top quark pair production, diboson and triboson production, and rare SM processes like single top production or top quark pair production with associated bosons, are generated at next-to-leading order (NLO) precision with \MGvATNLO and {\POWHEG}v2.0~\cite{Nason:2004rx,Frixione:2007vw,Alioli:2010xd,Re:2010bp}, using the NNPDF3.0NLO~\cite{Ball:2014uwa} set of PDFs. Showering and hadronization are carried out by the \PYTHIA~8.205 package~\cite{Sjostrand:2014zea}, while a detailed simulation  of the CMS detector is based on the \GEANTfour~\cite{geant4} package. Finally, renormalization and factorization scale and PDF uncertainties have been derived with the use of the \textsc{SysCalc} package~\cite{Kalogeropoulos:2018cke}.\par}

Signal models of direct \PSGt pair production are generated with \MGvATNLO at LO precision up to the production of \PGt leptons, which are then decayed with \PYTHIA~8.212. For the models of chargino-neutralino pair production that are also studied, \PYTHIA~8.212 is used to describe the decays of the parent charginos and neutralinos produced by \MGvATNLO at LO precision. The NNPDF3.0LO set of PDFs is used in the generation of all signal models. The CMS fast simulation package~\cite{fastsim} is used to simulate the CMS detector for the signal samples.

Event reconstruction in simulated samples is performed in a similar manner as for data. A nominal distribution of pileup interactions is used when producing the simulated samples. The samples are then reweighted to match the pileup profile observed in the collected data. The signal production cross sections are calculated at NLO with next-to-leading logarithmic (NLL) soft-gluon resummation calculations~\cite{Fuks:2013lya,Fuks:2012qx,Fuks:2013vua}. The most precise cross section calculations that are available are used to normalize the SM simulated samples, corresponding most often to next-to-next-to-leading order (NNLO) accuracy.

\section{Event selection}
\label{sec:evtsel}
The data used for this search are selected with various triggers that require the presence of isolated electrons, muons, or \tauh candidates. In the case of the \etau final state, the trigger used relies on the presence of an isolated electron with $\pt>25\GeV$ satisfying stringent identification criteria, while for the \mutau final state, the trigger is based on the presence of an isolated muon with $\pt>24\GeV$. A combination of triggers is used for the events selected in the \emu final state, requiring the presence of an electron and a muon. These triggers require the leading lepton to have \pt greater than 23\GeV and the subleading lepton to have \pt greater than 8 or 12\GeV for an electron or muon, respectively. Data in the \tautau final state are selected with a trigger requiring the presence of two \tauh candidates, each with $\pt>35\GeV$. Trigger efficiencies are measured in data and simulation. We apply scale factors accounting for any discrepancies, parameterized in the \pt and $\eta$ of the reconstructed electrons, muons, and \tauh candidates, to the simulation. The efficiencies measured in data are applied directly as correction factors to simulated signal samples, which are produced using the fast simulation package and for which the trigger simulation is not available. The trigger efficiencies range from 60 to 95\%, depending on the final state and the \pt and $\eta$ range under consideration.

Subsequent to the trigger criteria, the event selection for each final state requires the presence of exactly two reconstructed leptons with opposite charges, corresponding to the \emu, \etau, \mutau, or \tautau final states. The various lepton selection requirements implemented in the analysis are summarized in Table~\ref{tab:lepselection}. The \pt and $\abs{\eta}$ thresholds implemented when selecting these objects are dictated by the corresponding trigger thresholds described above. We require all selected leptons to be isolated. In the case of electron and muon candidates, the isolation requirement is enforced by placing an upper bound on the relative isolation quantity, \irel. For \tauh candidates, we use a multivariate discriminant. In order to ensure consistency with the primary vertex, upper bounds are placed on the absolute values of the electron and muon \dxy and \dz. We avoid overlaps between the two reconstructed leptons in the mixed final states (\emu, \etau, and \mutau) by requiring them to have a minimum separation in \dR of at least 0.3. In order to ensure orthogonality between the different final states and suppress background, we reject events with additional electrons or muons beyond the two selected leptons that satisfy slightly less stringent selection criteria. These criteria are summarized in Table~\ref{tab:lepveto}.

\begin{table}[!!hb]
\centering
\topcaption{Summary of lepton selection requirements for the analysis. Entries with a second value in parentheses refer to the lepton with the higher (lower) \pt.}
\label{tab:lepselection}
\begin{tabular}{lcccc}
\hline
Selection requirement & \emu & \etau & \mutau & \tautau \\
\hline
Electron \pt [\GeVns{}] & ${>}24\,(13)$ & ${>}26$ & \NA & \NA \\
Electron $\abs{\eta}$     & ${<}2.5$ & ${<}2.1$ & \NA & \NA \\
Electron $\abs{\dxy}$ [cm] & ${<}0.045$ & ${<}0.045$ & \NA & \NA \\
Electron $\abs{\dz}$ [cm]  & ${<}0.2$ & ${<}0.2$ & \NA & \NA \\
Electron \irel        & ${<}0.1$ & ${<}0.1$ & \NA & \NA \\
Muon \pt [\GeVns{}]     & ${>}24\,(10)$ & \NA & ${>}25$ & \NA \\
Muon $\abs{\eta}$         & ${<}2.4$ & \NA & ${<}2.4$ & \NA \\
Muon $\abs{\dxy}$ [cm] & ${<}0.045$ & \NA & ${<}0.045$ & \NA \\
Muon $\abs{\dz}$ [cm]  & ${<}0.2$ & \NA & ${<}0.2$ & \NA \\
Muon \irel            & ${<}0.15$ & \NA & ${<}0.15$ & \NA \\
\tauh \pt [\GeVns{}]     & \NA & ${>}20$ & ${>}20$ & ${>}40$ \\
$\tauh \abs{\eta}$        & \NA & ${<}2.3$ & ${<}2.3$ & ${<}2.1$ \\
\tauh isolation working point & \NA & Tight & Tight & Very tight \\
\hline
\end{tabular}
\end{table}

\begin{table}[!h]
\centering
\topcaption{Summary of requirements for identifying additional electrons and muons.}
\label{tab:lepveto}
\begin{tabular}{lcccc}
\hline
Selection requirement & \emu & \etau & \mutau & \tautau \\
\hline
Electron \pt [\GeVns{}]          & ${>}15$ & ${>}15$ & ${>}10$ & ${>}20$ \\
Electron $\abs{\eta}$     & ${<}2.5$ & ${<}2.5$ & ${<}2.5$ & ${<}2.5$ \\
Electron $\abs{\dxy}$ [cm]     & ${<}0.045$ & ${<}0.045$ & ${<}0.045$ & ${<}0.1$ \\
Electron $\abs{\dz}$ [cm]      & ${<}0.2$ & ${<}0.2$ & ${<}0.2$ & ${<}0.2$ \\
Electron \irel        & ${<}0.3$ & ${<}0.3$ & ${<}0.3$ & ${<}0.175$ \\
Muon \pt [\GeVns{}]              & ${>}15$ & ${>}10$ & ${>}15$ & ${>}20$ \\
Muon $\abs{\eta}$         & ${<}2.4$ & ${<}2.4$ & ${<}2.4$ & ${<}2.4$ \\
Muon $\abs{\dxy}$ [cm]         & ${<}0.045$ & ${<}0.045$ & ${<}0.045$ & ${<}0.045$ \\
Muon $\abs{\dz}$ [cm]          & ${<}0.2$ & ${<}0.2$ & ${<}0.2$ & ${<}0.2$ \\
Muon \irel            & ${<}0.3$ & ${<}0.3$ & ${<}0.3$ & ${<}0.25$ \\
\hline
\end{tabular}
\end{table}

A subsequent set of selection criteria is imposed for each final state to further suppress background and enhance the search sensitivity. Differences in the background compositions between the different final states play a role in the determination of the corresponding selection criteria which, together with the selection requirements described above, define the ``baseline selection".

In all final states, we require $\abs{\dphill} > 1.5$, with additional requirements of $\dR(\ell_1,\ell_2) < 3.5$ and $\abs{\Delta\eta(\ell_1,\ell_2)} < 2$ being applied for the leptonic final states to suppress the QCD multijet background. Here $\ell_1$ and $\ell_2$ represent the leading and trailing reconstructed electrons, muons, or \tauh candidates, respectively. In order to suppress backgrounds with top quarks, we veto events containing any \PQb-tagged jet with $\pt > 30\GeV$ identified with the loose CSV working point in the \tautau final state. In the leptonic final states, these backgrounds are reduced by vetoing any event that contains either more than one jet with $\pt > 20\GeV$, or any such jet that is \PQb~tagged using the medium CSV working point. One-jet events in these final states are required to have a separation in $\abs{\Delta\eta}$ of less than 3 between the jet and the reconstructed leptons and, in the case of the \etau and \mutau final states, a separation in \dR of less than 4 between the jet and the \tauh. Background events from low-mass resonances are removed in these final states by requiring the invariant mass of the two leptons, \mll, to exceed 50\GeV. In the \emu final state, \mll is required to lie in the window 90--250\GeV in order to suppress \zjets events with \ztautau, while the electron and muon \pt are required to be less than 200\GeV in order to suppress \ttbar and {\PW\PW} events, since the signal processes targeted are not expected to produce leptons with higher \pt.

In order to further improve discrimination against the SM background, we take advantage of the expected presence of two \PSGczDo in the final state for signal events, which would lead to additional \ptmiss. While background processes such as \wjets with $\PW \to \ell\nu$ can also produce genuine \ptmiss, the correlations between \ptvecmiss and the reconstructed leptons are expected to be different between signal and background processes, and these differences can be exploited. In particular, mass observables that can be calculated from the reconstructed leptons and the \ptvecmiss provide strong discriminants between signal and background. For a mother particle decaying to a visible and an invisible particle, the transverse mass (\MT), calculated using only the \ptvec of the decay products, should have a kinematic endpoint at the mass of the mother particle. Assuming that the \ptmiss corresponds to the \pt of the invisible particle, we calculate the \MT observable for the visible particle q and the invisible particle as follows:
\begin{equation}
\MT(\mathrm{q}, \ptvecmiss) \equiv \sqrt{2 p_{\text{T,q}} \ptmiss [1 - \cos \Delta\phi(\vec{p}_{\text{T,q}}, \ptvecmiss)]}.\label{eq:MT}
\end{equation}
By requiring $20 < \MT(\ell, \ptvecmiss) < 60\GeV$ or $\MT(\ell, \ptvecmiss) > 120\GeV$ where $\ell$ here represents the electron (muon) in the \etau (\mutau) final state, the \wjets background is significantly reduced. To further suppress the SM background in the leptonic final states, we require the sum of the transverse masses, \sumMT, to be at least 50\GeV. The \sumMT is defined as the scalar sum of $\MT(\ell_1,\ptvecmiss)$ and $\MT(\ell_2,\ptvecmiss)$.

\begin{table}[!!bh]
\centering
\topcaption{Summary of baseline selection requirements in each final state.}
\label{tab:basesel}
\cmsTable{
\begin{tabular}{lcccc}
\hline
Selection requirement & \emu & \etau & \mutau & \tautau \\
\hline
$\abs{\dphill}$ & ${>}1.5$ & ${>}1.5$ & ${>}1.5$ & ${>}1.5$ \\
$\abs{\Delta\eta(\ell_1,\ell_2)}$ & ${<}2$ & ${<}2$ & ${<}2$ & \NA \\
$\dR(\ell_1,\ell_2)$ & ${<}3.5$ & ${<}3.5$ & ${<}3.5$ & \NA \\
\multirow{2}{*}{\PQb-tagged jet veto} & $\pt > 20\GeV$, & $\pt > 20\GeV$, & $\pt > 20\GeV$, & $\pt > 30\GeV$, \\
 & medium CSV & medium CSV & medium CSV & loose CSV \\
Additional jet veto & ${>}1$ jet, $\pt > 20\GeV$ & ${>}1$ jet, $\pt > 20\GeV$ & ${>}1$ jet, $\pt > 20\GeV$ & \NA \\
$\abs{\Delta\eta(\text{jet},\ell_{\mathrm{i}})}$ (1--jet events) & ${<}3$ & ${<}3$ & ${<}3$ & \NA \\
$\dR(\text{jet},\tauh)$ (1--jet events) & \NA & ${<}4$ & ${<}4$ & \NA \\
\mll [\GeVns{}] & 90--250 & ${>}50$ & ${>}50$ & \NA \\
$\Pe/\mu$ \pt upper bound [\GeVns{}] & ${<}200$ & \NA & \NA & \NA \\
\multirow{2}{*}{$\MT(\Pe/\mu, \ptvecmiss)$ [\GeVns{}]} & \multirow{2}{*}{\NA} & 20--60 & 20--60 & \multirow{2}{*}{\NA} \\
 & & or ${>}120$ & or ${>}120$ & \\
\sumMT [\GeVns{}] & \NA & ${>}50$ & ${>}50$ & \NA \\
\hline
\end{tabular}
}
\end{table}

The baseline selection criteria described above are summarized in Table~\ref{tab:basesel}. We apply these criteria to obtain an optimized sample of events in each final state. These events are then further subdivided using discriminating kinematic variables into exclusive search regions (SRs) to improve the sensitivity of the search to a range of sparticle masses. One of these discriminating variables is the ``stransverse mass" \mttwo~\cite{MT2variable,MT2variable2}. This kinematic mass variable is a generalization of the variable \MT for situations with multiple invisible particles.  It serves as an estimator of the mass of pair-produced particles in situations in which both particles decay to a final state containing the same invisible particle. For direct \PSGt pair production, with both \PSGt decaying to a \PGt lepton and a \PSGczDo, \mttwo should be correlated with the \PSGt mass. Large values of \mttwo can therefore be used to discriminate between models with large \PSGt masses and the SM background. This variable is again calculated using the \ptvec of the different particles:
\begin{equation}
\mttwo = \min_{\ptvec^{\,\mathrm{X}(1)} + \ptvec^{\,\mathrm{X}(2)} = \ptvecmiss}
  \left[ \max \left( \MT^{(1)} , \MT^{(2)} \right) \right],
\label{eq:MT2}
\end{equation}
where $\ptvec^{\,\mathrm{X}(i)}$  (with $i$=1,2) are the unknown transverse momenta of the two undetected particles and $\MT^{(i)}$ are the transverse masses obtained by either pairing of the two hypothetical invisible particles with the two leptons.  The minimization is done over the possible momenta of the invisible particles, which should add up to the \ptvecmiss in the event.

Another variable that is used to distinguish signal from background, \DZ, is defined as:
\begin{equation}
\DZ = P_{\zeta, \text{miss}} -0.85 P_{\zeta, \text{vis}} \, ,
\end{equation}
where $P_{\zeta, \text{miss}} =  \ptvecmiss \cdot \vec{\zeta}$ and $P_{\zeta, \text{vis}}  =  (\ptvec^{\,\ell_1} + \ptvec^{\,\ell_2}) \cdot \vec{\zeta}$, with $\vec{\zeta}$ being the bisector between the directions of the two leptons. The \DZ variable helps to discriminate events in which \ptmiss originates from the decay of two \PGt leptons from other processes~\cite{CuencaAlmenar:2008zza,Khachatryan:2014wca}. Different background processes are characterized by different ranges of \DZ. For instance, the DY+jets background is largely expected to have positive \DZ values, while \wjets and \ttbar events may have negative values.

The more restrictive trigger requirements in the \tautau final state significantly reduce the signal acceptance, and the very low cross sections of the targeted \distau signal models result in very small expected signal event yields after the baseline selection. Events surviving the baseline selection in this final state are therefore categorized into only three SRs. These three SRs are exclusive and are optimized for sensitivity to different \PSGt mass ranges. For higher values of the \PSGt mass, a requirement of large \mttwo significantly improves the discrimination of signal from background. We therefore define a search region, designated SR1, by selecting events with $\mttwo > 90\GeV$. For lower \PSGt masses, \sumMT is found to be a more powerful discriminant than \mttwo. Two additional SRs, designated SR2 and SR3, are therefore defined by selecting events with moderate \mttwo ($40 < \mttwo < 90\GeV$), and further subdividing them into high and moderate \sumMT ranges: $>$350\GeV and 300--350\GeV, respectively. For these two SRs, we place a further requirement of $\ptmiss > 50\GeV$ to sufficiently suppress the QCD multijet background.

In the leptonic final states, events satisfying the baseline selection criteria are categorized into SRs based on a series of thresholds applied to the values of the discriminating observables \ptmiss, \mttwo, and \DZ. The SR binning is defined to be slightly different for events in the 0- and 1-jet categories and is chosen such that there are small variations in the relative background contributions in the different bins. This allows us to obtain stronger constraints on the background predictions in the final result, obtained from a simultaneous maximum likelihood fit to the data in all SRs. Tables~\ref{tab:srdefinition_0jet} to \ref{tab:srdefinition_1jet_emu} list the criteria used to define the SRs in the 0- and 1-jet categories, respectively. While the same binning is chosen for the \etau and \mutau final states, the SR bins chosen in the \emu final state are slightly different because of the different background composition.

\begin{table}[htbp]
\centering
\topcaption{Definition of SRs in the 0-jet category for the \etau and \mutau final states. }
\label{tab:srdefinition_0jet}
\begin{tabular}{cccc}
\hline
Bin name & \ptmiss [\GeVns{}] & \mttwo [\GeVns{}] & \DZ [\GeVns{}] \\
\hline
$\mathrm{0j}-{1}$  & ${<}40$ & ${<}40$ & $<-100$\\
$\mathrm{0j}-{2}$  & & ${>}40$  & $>-500$\\[\cmsTabSkip]
$\mathrm{0j}-{3}$  &  [40,80] & ${<}40$ & $<-100$\\
$\mathrm{0j}-{4}$  & &  & ${>}50$ \\
$\mathrm{0j}-{5}$  & & [40,80] & $<-100$\\
$\mathrm{0j}-{6}$  & &  & $>-100$\\
$\mathrm{0j}-{7}$  & & ${>}80$  & $>-500$\\[\cmsTabSkip]
$\mathrm{0j}-{8}$  &  [80,120] & ${<}40$ & $<-100$\\
$\mathrm{0j}-{9}$  & &  & $>-100$\\
$\mathrm{0j}-{10}$ & & [40,80] & $<-150$\\
$\mathrm{0j}-{11}$ & &  & $>-150$\\
$\mathrm{0j}-{12}$ & & ${>}80$  & $>-500$\\[\cmsTabSkip]
$\mathrm{0j}-{13}$ &[120,250] & ${<}40$ & $<-100$\\
$\mathrm{0j}-{14}$ & &  & $>-100$\\
$\mathrm{0j}-{15}$ & & [40,80] & $<-150$\\
$\mathrm{0j}-{16}$ & &  & [$-150$,$-100$] \\
$\mathrm{0j}-{17}$ & &  & $>-100$\\
$\mathrm{0j}-{18}$ & &[80,100]& $>-500$\\
$\mathrm{0j}-{19}$ & &   [100,120] & $>-500$\\
$\mathrm{0j}-{20}$ & & ${>}120$ & $>-500$\\[\cmsTabSkip]
$\mathrm{0j}-{21}$ & ${>}250$  & ${>}0$ & $>-500$\\
\hline
\end{tabular}
\end{table}

\begin{table}[htbp]
\centering
\topcaption{Definition of SRs in the 1-jet category for the \etau and \mutau final states. }
\label{tab:srdefinition_1jet}
\begin{tabular}{cccc}
\hline
Bin name & \ptmiss [\GeVns{}] & \mttwo [\GeVns{}] & \DZ [\GeVns{}] \\
\hline
$\mathrm{1j}-{1}$ & ${<}40$ & ${<}40$ & $<-150$\\
$\mathrm{1j}-{2}$ & &  & [$-150$,$100$] \\
$\mathrm{1j}-{3}$ & & ${>}40$  & $>-500$\\[\cmsTabSkip]
$\mathrm{1j}-{4}$ & [40,80] & ${<}40$ & $<-100$\\
$\mathrm{1j}-{5}$ & &  & ${>}50$ \\
$\mathrm{1j}-{6}$ & & [40,80] & $<-100$\\
$\mathrm{1j}-{7}$ & &  & $>-100$\\
$\mathrm{1j}-{8}$ & & ${>}80$  & $>-500$\\[\cmsTabSkip]
$\mathrm{1j}-{9}$ & [80,120] & ${<}40$ & $<-100$\\
$\mathrm{1j}-{10}$& & [40,80] & $<-150$\\
$\mathrm{1j}-{11}$& &  & $>-150$\\
$\mathrm{1j}-{12}$& & [80,120] & $>-500$\\
$\mathrm{1j}-{13}$& & ${>}120$ & $>-500$\\[\cmsTabSkip]
$\mathrm{1j}-{14}$& [120,250]& ${<}40$ & $<-150$\\
$\mathrm{1j}-{15}$& &  & [$-150$,$-100$] \\
$\mathrm{1j}-{16}$& &  & $>-100$\\
$\mathrm{1j}-{17}$& & [40,80] & $<-150$\\
$\mathrm{1j}-{18}$& &  & [$-150$,$-100$] \\
$\mathrm{1j}-{19}$& &  & $>-100$\\
$\mathrm{1j}-{20}$& &   [80,100] & $>-500$\\
$\mathrm{1j}-{21}$& &  [100,120] & $>-500$\\
$\mathrm{1j}-{22}$& & ${>}120$ & $>-500$\\[\cmsTabSkip]
$\mathrm{1j}-{23}$&  ${>}250$  & ${>}80$ & $>-500$\\
\hline
\end{tabular}
\end{table}

\begin{table}[htbp]
\centering
\topcaption{Definition of SRs in the 0-jet category for the \emu final state. }
\label{tab:srdefinition_0jet_emu}
\begin{tabular}{cccc}
\hline
Bin name & \ptmiss [\GeVns{}] & \mttwo [\GeVns{}] & \DZ [\GeVns{}] \\
\hline
$\mathrm{0j}-{1}$  & ${<}40$ & ${<}40$ & $<-100$\\
$\mathrm{0j}-{2}$  & &   & ${>}0$ \\
$\mathrm{0j}-{3}$  & & ${>}40$  & $>-500$\\[\cmsTabSkip]
$\mathrm{0j}-{4}$ &  [40,80] & ${<}40$ & $<-100$\\
$\mathrm{0j}-{5}$ & &  & ${>}50$ \\
$\mathrm{0j}-{6}$ & & [40,80] & $<-100$\\
$\mathrm{0j}-{7}$ & &  & $>-100$\\
$\mathrm{0j}-{8}$ & & ${>}80$  & $>-500$\\[\cmsTabSkip]
$\mathrm{0j}-{9}$ &  [80,120] & ${<}40$ & $<-100$\\
$\mathrm{0j}-{10}$& &  & $>-100$\\
$\mathrm{0j}-{11}$& & [40,80] & $<-150$\\
$\mathrm{0j}-{12}$& &  & $>-150$\\
$\mathrm{0j}-{13}$& & ${>}80$  & $>-500$\\[\cmsTabSkip]
$\mathrm{0j}-{14}$&[120,250] & ${<}40$ & $<-100$\\
$\mathrm{0j}-{15}$& &  & $>-100$\\
$\mathrm{0j}-{16}$& & [40,80] & $<-150$\\
$\mathrm{0j}-{17}$& &  & [$-150$,$-100$] \\
$\mathrm{0j}-{18}$& &  & $>-100$\\
$\mathrm{0j}-{19}$& &[80,100]& $>-500$\\
$\mathrm{0j}-{20}$& &[100,120] & $>-500$\\
$\mathrm{0j}-{21}$& & ${>}120$ & $>-500$\\[\cmsTabSkip]
$\mathrm{0j}-{22}$ & ${>}250$  & ${>}0$ & $>-500$\\  \hline
\end{tabular}
\end{table}

\begin{table}[htbp]
\centering
\topcaption{Definition of SRs in the 1-jet category for the \emu final state. }
\label{tab:srdefinition_1jet_emu}
\begin{tabular}{cccc}
\hline
Bin name & \ptmiss [\GeVns{}] & \mttwo [\GeVns{}] & \DZ [\GeVns{}] \\
\hline
$\mathrm{1j}-{1}$ & ${<}40$ & ${<}40$ & $<-150$\\
$\mathrm{1j}-{2}$ & &  & [$-150$,$100$] \\
$\mathrm{1j}-{3}$ & &  & ${>}0$ \\
$\mathrm{1j}-{4}$ & & ${>}40$  & $>-500$\\[\cmsTabSkip]
$\mathrm{1j}-{5}$ & [40,80] & ${<}40$ & $<-100$\\
$\mathrm{1j}-{6}$ & & & ${>}50$ \\
$\mathrm{1j}-{7}$ & & [40,80] & $>-100$\\
$\mathrm{1j}-{8}$ & & ${>}40$  & $>-500$\\[\cmsTabSkip]
$\mathrm{1j}-{9}$ & [80,120] & ${<}40$ & $<-100$\\
$\mathrm{1j}-{10}$& & [40,80] & $<-100$\\
$\mathrm{1j}-{11}$& & [80,120] & $>-500$\\
$\mathrm{1j}-{12}$& & ${>}120$ & $>-500$\\[\cmsTabSkip]
$\mathrm{1j}-{13}$& [120,250] & ${<}40$ & $<-150$\\
$\mathrm{1j}-{14}$& &  & [$-150$,$-100$] \\
$\mathrm{1j}-{15}$ & &  & $>-100$\\
$\mathrm{1j}-{16}$ & & [40,80] & $<-150$ \\
$\mathrm{1j}-{17}$ & &  & [$-150$,$-100$] \\
$\mathrm{1j}-{18}$ & &  &  $>-100$\\
$\mathrm{1j}-{19}$ & & [80,100] & $>-500$\\
$\mathrm{1j}-{20}$ & &  [100,120] & $>-500$\\
$\mathrm{1j}-{21}$ &  & ${>}120$  & $>-500$\\[\cmsTabSkip]
$\mathrm{1j}-{22}$ &  ${>}250$  & ${>}80$ & $>-500$\\
\hline
\end{tabular}
\end{table}

\section {Background estimation}
\label{sec:bkgest}
The dominant background sources for this search are DY+jets, \wjets, QCD multijet, \ttbar, and diboson processes. These background sources have different relative contributions in the different final states. For the \tautau final state, the dominant background consists of QCD multijet and \wjets processes, where one or more of the \tauh candidates originates from a parton and is misidentified as a prompt \tauh. This background is predicted using a data-driven method relying on a control region with a loose isolation requirement. For the \etau and \mutau final states, the main backgrounds after the baseline selection are DY+jets ($\approx$50\%), \wjets ($\approx$30\%), and QCD multijet ($\approx$10\%) events. The DY+jets background contribution, which usually consists of events with two prompt leptons, is determined from simulation after applying shape and normalization corrections that are determined from data. The \wjets and QCD multijet backgrounds usually contain a jet that is misidentified as \tauh, and are determined from a sideband sample using a data-driven method similar to the one used in the \tautau case. The main backgrounds in the \emu final state originate from \ttbar ($\approx$45\%) and {\PW\PW} ($\approx$35\%) events, and are estimated from simulation after applying corrections derived from data.  A detailed description of the procedures used to estimate the background contributions from the different sources follows.

\subsection{Estimation of the Drell-Yan+jets background}
\label{dyestimation}
\begin{figure}[htb]
\centering
\includegraphics[width=0.48\textwidth,height=0.48\textwidth]{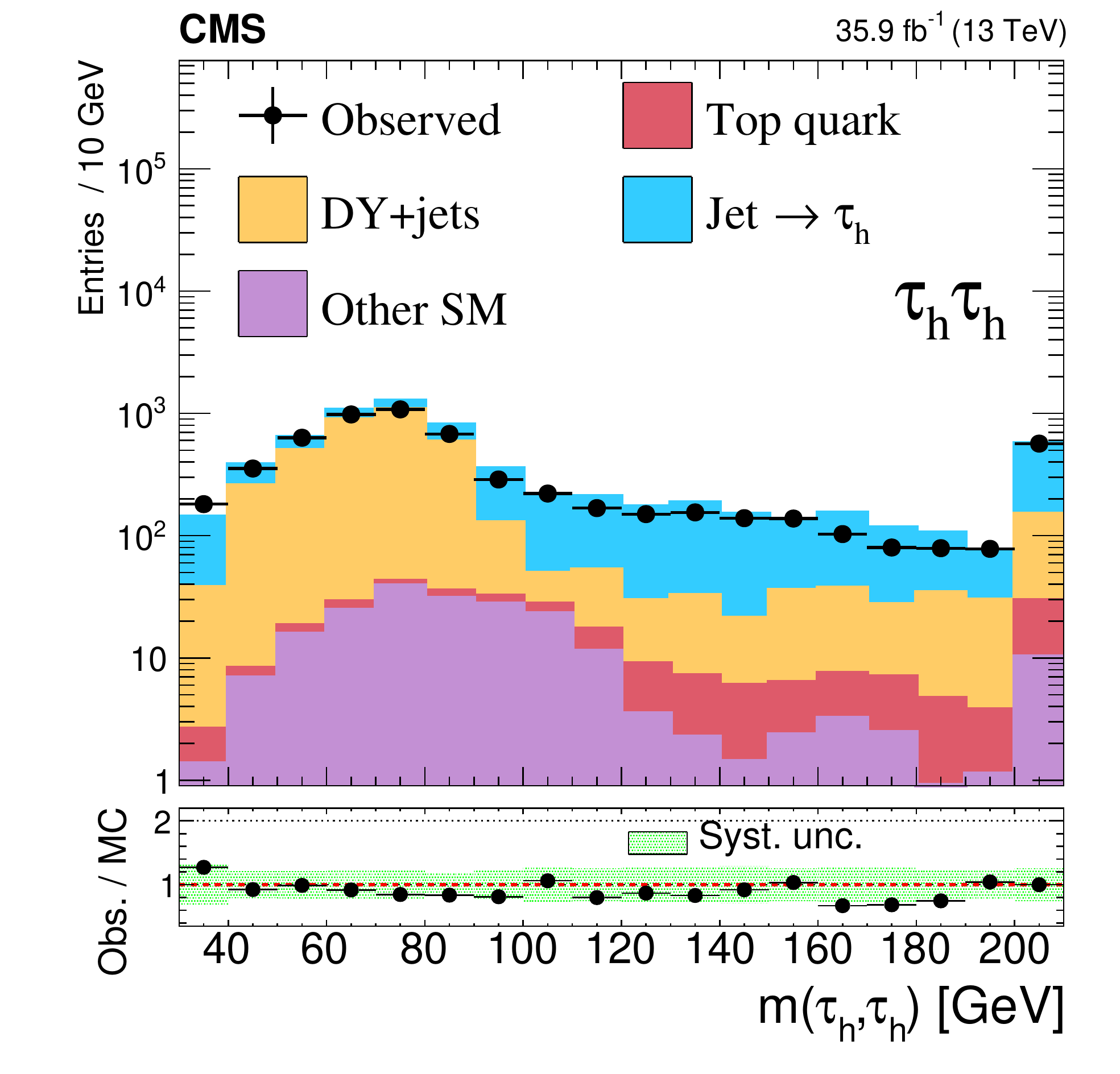}
\includegraphics[width=0.48\textwidth,height=0.48\textwidth]{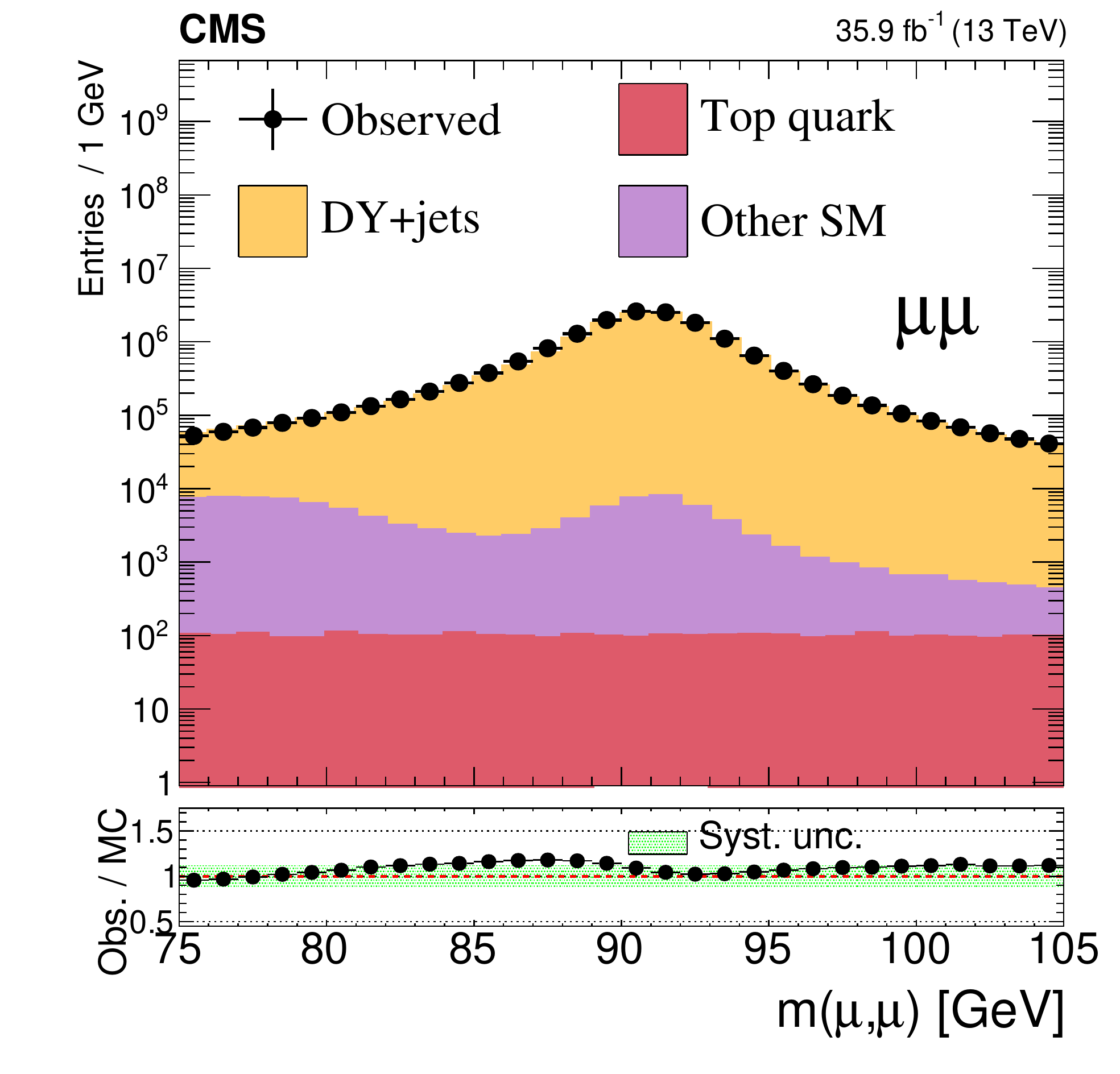}
\caption{\label{fig:dyvalidation} Left: visible mass spectrum used to validate the modeling of the DY+jets background in the \tautau final state in a \ztautau control sample selected with low \mttwo or \sumMT and a minimum \tautau system \pt of 50\GeV. The last bin includes overflows. Right: dimuon mass distribution in the high-purity \zmumu control sample after all estimated correction factors have been applied to the simulation. In the legend, ``Top quark" refers to the background originating from \ttbar and single top quark production.}
\end{figure}

The DY+jets background mainly originates from \ztautau decays. We estimate the contribution of this background from simulation after corrections based on control samples in data. If the \PZ boson mass shape or \pt spectrum are poorly modeled in the simulation, then distributions of the discriminating kinematic variables can differ significantly between data and simulation, especially at the high-end tails that are relevant for the SRs. We therefore use a high-purity \zmumu control sample to compare the dimuon mass and \pt spectra between data and simulation and apply the observed differences as corrections to the simulation in the search sample in the form of two-dimensional weights parameterized in the generator-level \PZ boson mass and \pt. The correction factors range up to 30\% for high mass and \pt values.  The full size of this correction is propagated as a systematic uncertainty.  The known differences in the electron, muon, and \tauh identification and isolation efficiencies, jet, electron, muon, and \tauh energy scales, and \PQb~tagging efficiency between data and simulation are taken into account.  The uncertainties corresponding to these corrections are also propagated to the final background estimate. The corrected simulation is validated in the \tautau final state using a \ztautau control sample selected by inverting either the \mttwo or \sumMT requirements used to define the SRs.  Additionally requiring a \pt of at least 50\GeV for the \tautau system reduces the QCD multijet background and improves the purity of this control sample.  Figure~\ref{fig:dyvalidation} (left) shows that the corrected simulation agrees with the data within the experimental uncertainties in this sample.

Finally, for the analysis in the leptonic final states, a normalization scale factor as well as corrections to the \PZ \pt distribution in the simulation are derived from a very pure \zmumu control sample in data. Events in this sample are selected by requiring two isolated muons and no additional leptons, fewer than two jets, no \PQb-tagged jets, and a dimuon mass window of 75--105\GeV to increase the probability that they originate from \zmumu decays to $>$99\%. After subtracting all other contributions estimated from simulation, a normalization scale factor of $0.96 \pm 0.05$ is extracted from the ratio of data to simulated events. The uncertainty in the scale factor is dominated by the systematic uncertainty. Figure~\ref{fig:dyvalidation} (right) shows a comparison of the dimuon mass distribution in data and simulation after all the corrections, including the normalization scale factor, have been applied.

\subsection{Estimation of the background from misidentified jets}
\label{sec:fakeestimation}

\subsubsection{Estimation in the \texorpdfstring{\tautau}{tau-tau} final state}
After requiring two high-\pt \tauh candidates, the dominant background for the search in the \tautau final state consists of QCD multijet and \wjets events, in which one or both of the \tauh candidates originate from a jet and are misidentified as prompt \tauh. This background is predicted using a method relying on extrapolation from a data sample selected with a loose isolation requirement. We estimate how frequently nonprompt or misidentified \tauh candidates that are selected with the loose isolation working point also pass the very tight isolation requirement applied in the SRs by studying a multijet-enriched control sample where we require both  \tauh candidates to have the same charge. The same-charge \tautau event sample is collected with the same trigger as the search sample, in order to take into account any biases from the isolation requirement present at the trigger level, which is not identical to the isolation requirement that corresponds to the final analysis selection criteria.  We also require \mttwo to be small ($<$40\GeV) to reduce any potential contributions from signal and \wjets events.

The final rate measured in this sample for misidentified \tauh selected with the loose isolation working point to pass the very tight isolation requirement is around 25\%, but it depends considerably on the \pt and the decay mode (one- or three-prong) of the \tauh candidate, and the parent jet flavor.  The extrapolation is measured in bins of \tauh  \pt and separately for the different decay modes to reduce any dependence on these factors. A systematic uncertainty of around 30\% is evaluated that accounts for the dependence of the misidentification rate on the jet flavor, based on studies performed in simulation.  We also noticed that the extrapolation is affected by whether or not the \tauh candidate other than the one for which the extrapolation is being applied is isolated. A correction and a corresponding systematic uncertainty are derived for this effect.

Since the isolation efficiency for prompt \tauh candidates is only around 65\%, processes with genuine \tauh may leak into the data sideband regions and need to be taken into account when calculating the final estimate for the background processes with misidentified \tauh.  To take this correctly into account, we define three categories for events that have at least two loosely isolated \tauh candidates: events with both \tauh candidates passing the very tight isolation requirement, events with one passing and one failing the very tight isolation requirement, and finally events with both \tauh candidates failing the very tight isolation requirement. We then equate these observable quantities with the expected sum totals of contributions from events with two prompt \tauh candidates, two misidentified \tauh candidates, or one prompt and one misidentified \tauh candidate to each of these populations. The contributions of background events with one or two misidentified \tauh candidates in the SRs can then be determined analytically by inverting this set of equations.  A closure test is performed in events with two oppositely charged \tauh candidates. where the \mttwo or \sumMT requirements used to define the SRs are explicitly inverted to avoid any overlap with the SRs.  Figure~\ref{fig:fakevalidation} (left), which shows the \mttwo distribution in this sample, confirms that the background estimation method is able to predict the background with misidentified \tauh candidates within the systematic uncertainties.

\subsubsection{Estimation in the \texorpdfstring{\etau and \mutau}{e-tau and mu-tau} final states}
The misidentification of jets as \tauh candidates also gives rise to a major source of background for the search in the \etau and \mutau final states, mainly from \wjets events with leptonic {\PW} boson decays. We estimate this background from a sideband sample in data selected by applying the SR selections, with the exception that the \tauh candidates are required to satisfy the loose but not the tight isolation working point. A transfer factor for the extrapolation in \tauh isolation is determined from a \wjets control sample selected from events with one muon and at least one \tauh candidate that passes the loose isolation requirement. In events with more than one \tauh candidate, the most isolated candidate is used in the determination of the transfer factor. Events with additional electrons or muons satisfying the criteria listed in Table~\ref{tab:lepveto} are rejected. In order to increase the purity of \wjets events in this sample by reducing the contribution of \ttbar and QCD multijet events, we require $60 < \MT < 120\GeV$, $\ptmiss > 40\GeV$, no more than two jets, and an azimuthal separation of at least 2.5 radians between any jet and the {\PW} boson reconstructed from the muon and the \ptvecmiss. The remaining sample has an expected purity of $82\%$ for \wjets events. The transfer factor, $R$, is then determined from this control sample, after subtracting the remaining non-\wjets background contributions estimated from simulation, as follows:
\begin{equation}
\label{eq:tranferFactor}
R = \frac{N^{\mathrm{CS}}_{\text{data}}({\mathrm{T}})-N^{\mathrm{CS}}_{\text{MC  no \PW}}({\mathrm{T}})}{N^{\mathrm{CS}}_{\text{data}} ({\mathrm{L\&!T}})-N^{\mathrm{CS}}_{\text{MC  no \PW}} ({\mathrm{L\&!T}})}.
\end{equation}
Here, $N^{\mathrm{CS}}_{\text{data}}$ corresponds to the number of events in the control sample in data. The parenthetical arguments ${\mathrm{T}}$ and ${\mathrm{L\&!T}}$ denote events in which the \tauh candidate satisfies the tight isolation working point, and the loose but not the tight working point, respectively. The transfer factor is determined in bins of \pt and $\eta$ of the \tauh candidate, as tabulated in Table~\ref{tab:tfr}.

\begin{table}[!tbh]
\centering
\topcaption{Transfer factor $R$ determined from the \wjets control sample according to Eq.~(\ref{eq:tranferFactor}), as a function of \pt and $\eta$ of the \tauh candidate. The uncertainties are statistical only. }
\label{tab:tfr}
\begin{tabular}{lccc}
\hline
$(\abs{\eta},\pt) $                            & 20--30\GeV      & 30--40\GeV     & ${>}40\GeV$ \\ \hline
$\abs{\eta}<0.80$                & $0.74 \pm 0.07$      & $0.66 \pm 0.01$       & $0.56 \pm 0.02$          \\
$0.80<\abs{\eta}< 1.44$   & $0.68  \pm 0.01$     & $0.61 \pm 0.01$      & $0.39 \pm0.03$           \\
$1.44<\abs{\eta}<1.57$ & \multicolumn{2}{c}{$0.68 \pm 0.03$} & $0.64  \pm 0.08$          \\
$1.57<\abs{\eta}<2.30$ & \multicolumn{2}{c}{$0.59 \pm 0.02$} & $0.61  \pm 0.01$          \\
\hline
\end{tabular}
\end{table}

The contribution of the background originating from a jet misidentified as a \tauh candidate in each SR is then determined from the corresponding data sideband region selected by requiring the \tauh candidate to satisfy the loose but not the tight isolation working point as follows:
\begin{equation}
N^{\mathrm{SR}}(\text{jet} \to \PGt) = R \, (N^{\text{sideband}}_{\text{data}} - N^{\text{sideband}}_{\mathrm{MC}}(\text{genuine}\,\PGt)),
\end{equation}
where $N^{\text{sideband}}_{\text{data}}$ represents the number of data events in the sideband region, from which $N^{\text{sideband}}_{\mathrm{MC}}(\text{genuine}~\PGt)$, the expected contribution of events with genuine \PGt leptons determined from simulation with generator-level matching, is subtracted. Figure~\ref{fig:fakevalidation} (middle) shows a comparison of the data with the background prediction in the \etau final state for the \sumMT distribution for the baseline selection, where the ratio of signal to background is expected to be small.

\subsection{Estimation in the \texorpdfstring{\emu}{e-mu} final state}
Jets may also be misidentified as electrons or muons, although the misidentification probabilities for these objects are smaller than for \tauh. The contribution of the background from misidentified jets in the \emu final state is determined from data using a matrix method. For each SR selection we define four regions $A$, $B$, $C$, and $D$, which contain events with two leptons of either the same or opposite charge. We designate two categories for the leptons: well-isolated (electrons with $\irel < 0.1$, muons with $\irel < 0.15$), or loosely-isolated ($0.1 < \irel < 0.2$ for electrons, $0.15 < \irel < 0.30$ for muons). In order to enrich the QCD multijet contribution in events in the loosely-isolated category, we also invert the baseline selection requirements affecting the separation between the two leptons, \ie, we now require $\Delta R(\ell_1, \ell_2) >3.5$ and $\abs{ \Delta \eta (\ell_1, \ell_2)} > 2$. We use the designations $A$ ($B$) for the regions with two well-isolated leptons of the same (opposite) charge, and $C$ ($D$) for the corresponding regions with a loosely-isolated lepton. Region $B$ constitutes the search region. The purity of the $C$ and $D$ regions in QCD multijet events is $>$90\%, while that of the $A$ regions is $\approx$55\% after the SR selections.

\begin{figure}[htbp]
\centering
\includegraphics[width=0.48\textwidth, height=0.48\textwidth]{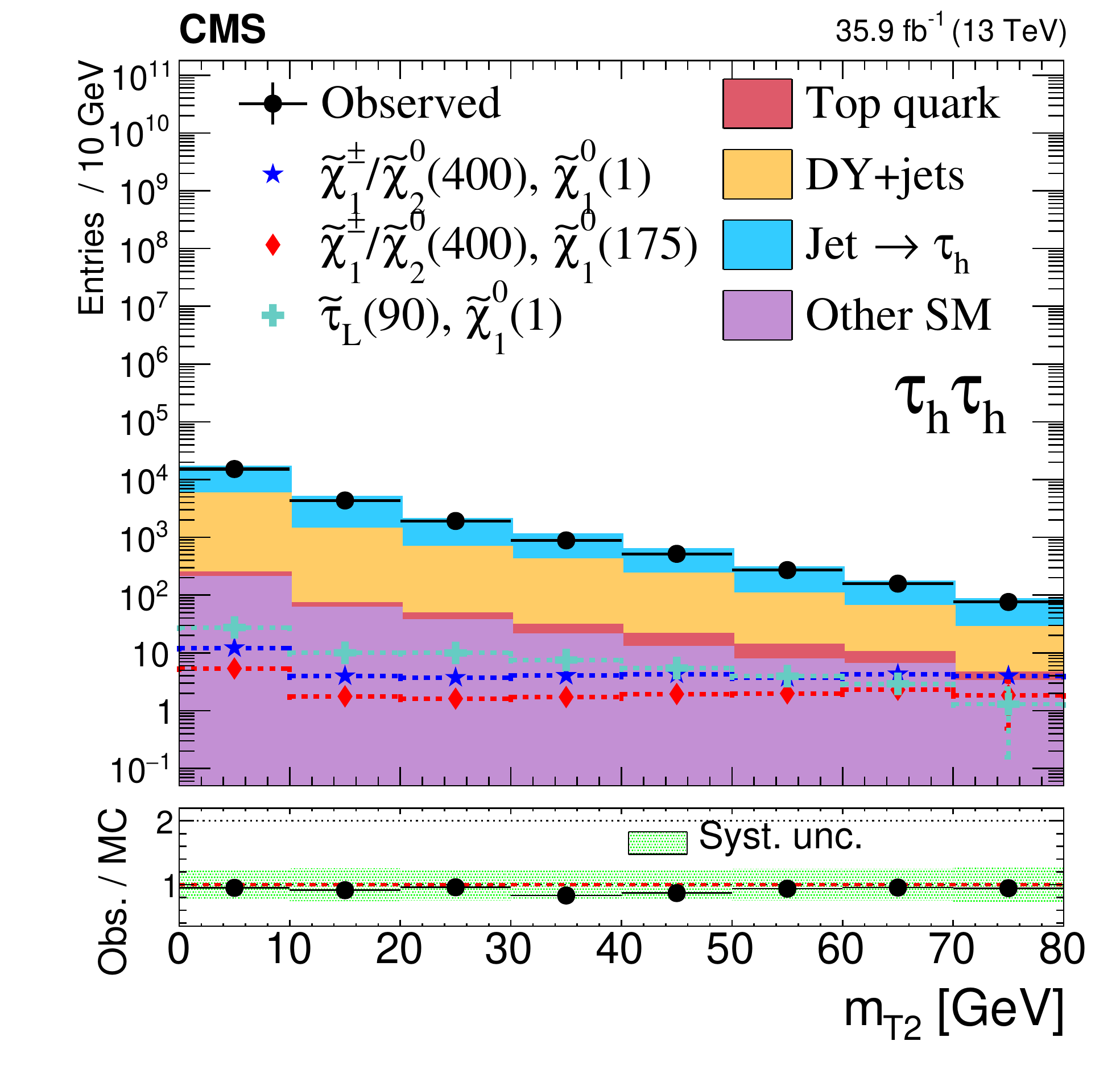}
\includegraphics[width=0.48\textwidth, height=0.48\textwidth]{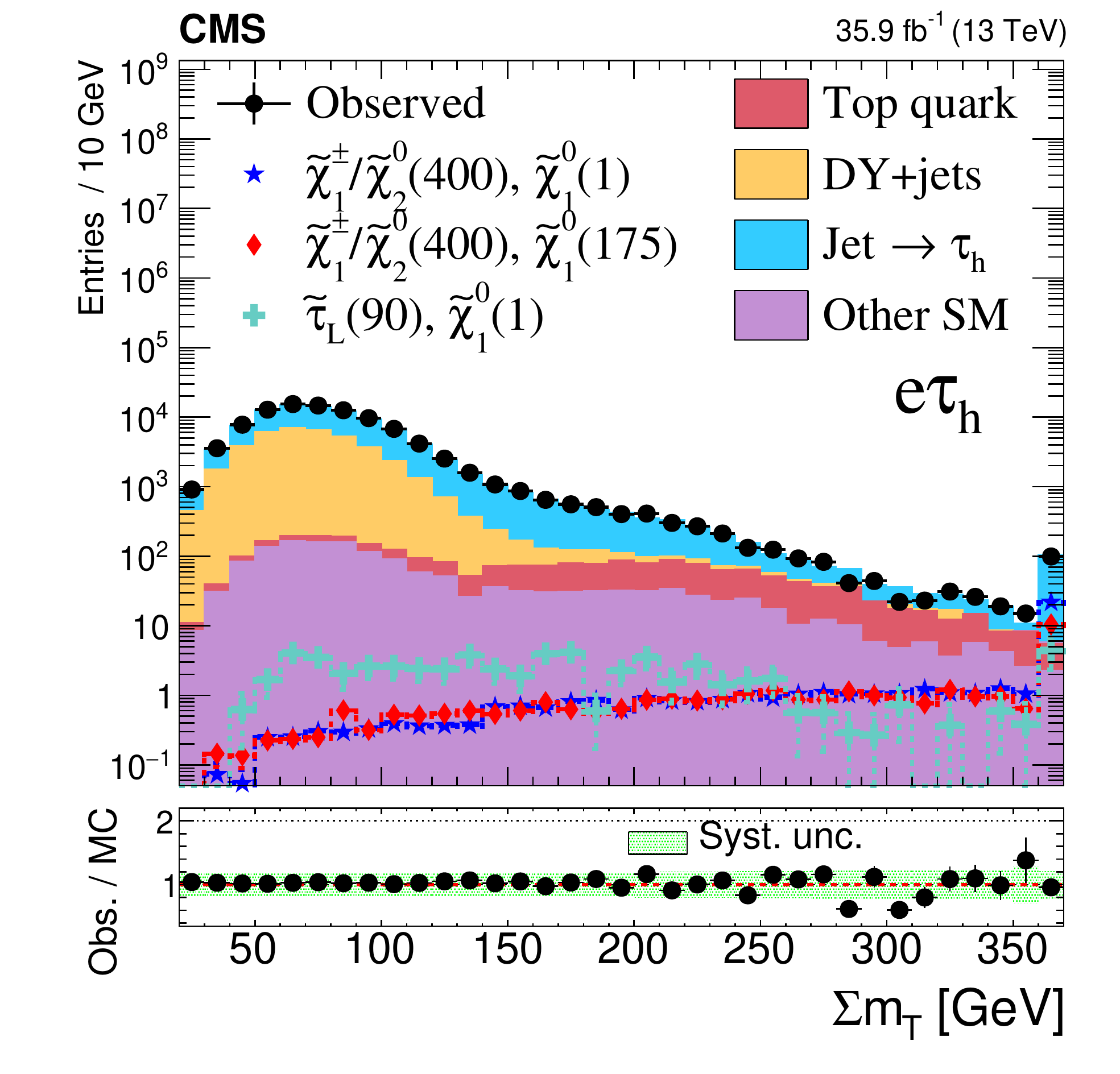} \\
\includegraphics[width=0.48\textwidth, height=0.48\textwidth]{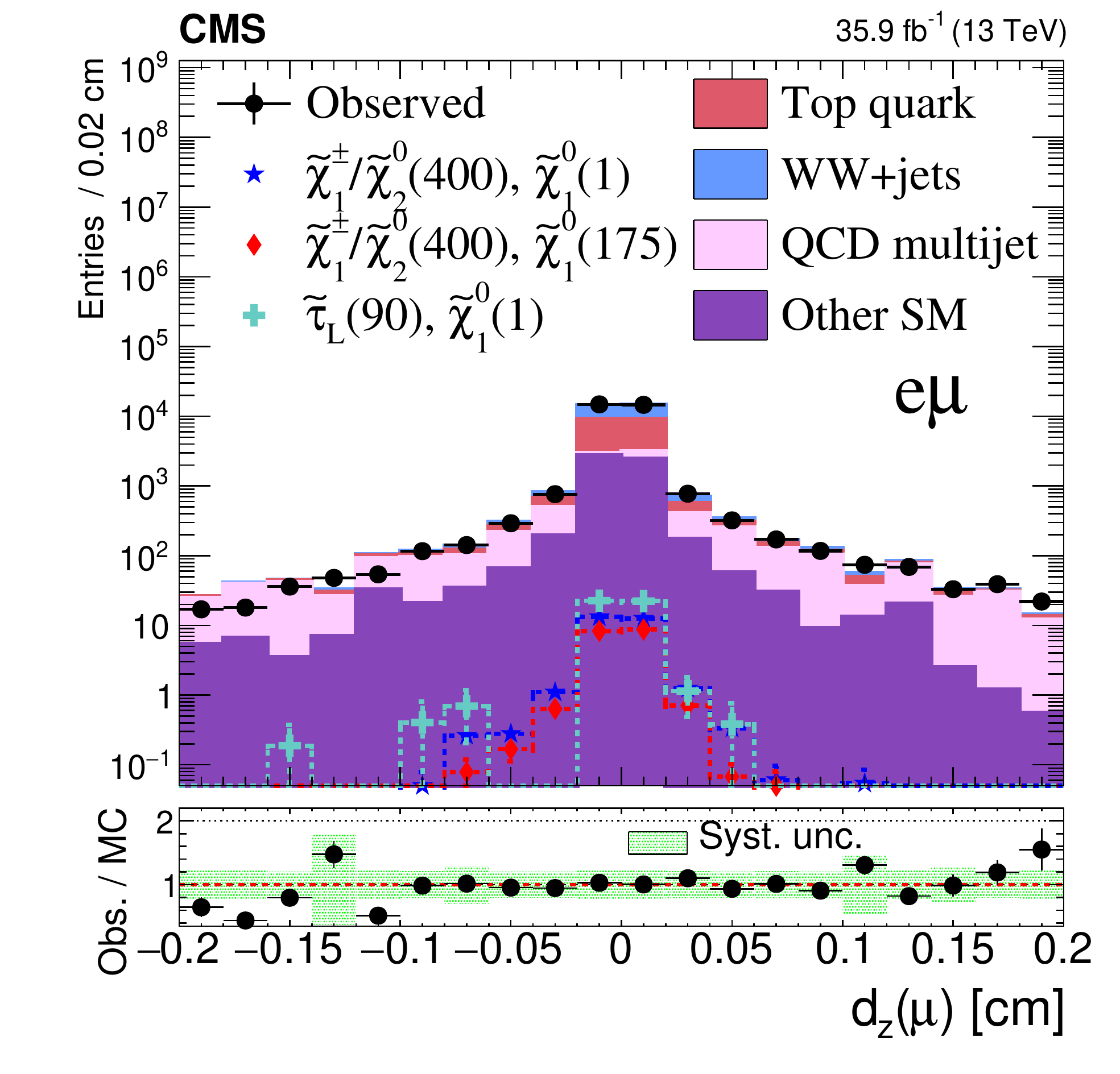}
\caption{\label{fig:fakevalidation} Top left: closure test for the method used to estimate the \tauh misidentification rate for the \tautau final state in a data control sample where the \mttwo or \sumMT requirements used in the SRs are inverted. Top right: \sumMT distribution for events in the \etau final state after the baseline selection, showing the estimation of the background with a jet misidentified as a \tauh, which is determined in a signal depleted control region. The last bin includes overflows. Bottom: distribution of the muon \dz in the \emu final state after the baseline selection, showing the estimation of the QCD multijet background using the matrix method. In the legend,``Top quark" refers to the background originating from \ttbar and single top quark production. In all cases, the predicted and observed yields show good agreement. Distributions for two benchmark models of chargino-neutralino production, and one of direct left-handed \PSGt pair production, are overlaid. The ratio of signal to background is expected to be small for these selections. The numbers within parentheses in the legend correspond to the masses of the parent SUSY particle and the \PSGczDo in GeV for these benchmark models.}
\end{figure}

The charge and the isolation of misidentified leptons are expected to be uncorrelated. However, we expect a correlation to be present for the other backgrounds in these regions, \eg, prompt leptons from \ttbar events are expected to have opposite charge. In order to account for this effect, we subtract the contributions expected from simulation for all other backgrounds from the observed numbers of events in the $A, C$, and $D$ regions to obtain the estimate of the background originating from misidentified leptons in the SRs, $N_{B}$, as follows:
\begin{equation}
N_B = (N^{\text{data}}_A - N^{\text{MC}}_A) \, \frac{N^{\text{data}}_D - N^{\text{MC}}_D}{N^{\text{data}}_C-N^{\text{MC}}_C}.
\end{equation}
The distribution of the muon \dz is shown in Fig.~\ref{fig:fakevalidation} (right) for events in the \emu final state and illustrates the estimation of the QCD multijet background using the matrix method. The data agree well with the predicted background.

\subsection{Estimation of other backgrounds}
Smaller contributions exist from other SM backgrounds, including other diboson processes, such as \PW\PZ+jets, triboson, and Higgs boson processes. There are also contributions from top quark processes: \ttbar and single top quark production, or top quark pair production in association with vector bosons.  These are estimated from simulation, using the known efficiency and energy scale corrections and evaluating both experimental and theoretical uncertainties as described in Section~\ref{sec:sysunc}. The shape of the top quark \pt spectrum is known to be different between simulation and data from studies of the differential \ttbar cross section~\cite{CMS-TOP-12-028,CMS-TOP-16-007}. The simulation is therefore reweighted by a correction factor parameterized in the top quark \pt to improve the modeling of the \ttbar background, and the full size of the correction is propagated as a systematic uncertainty. The normalization of this background is checked in an \emu control sample enriched in \ttbar events, selected by requiring the presence of at least two jets, at least one of which should be \PQb~tagged. The ratio of data to simulation for \ttbar events is found to be $1.00\pm0.05\syst\pm0.01\stat$, \ie, consistent with unity.

\section{Systematic uncertainties}
\label{sec:sysunc}
We rely on control samples in data in various ways for the estimation of the major backgrounds in the analysis. The dominant uncertainties affecting these estimates are therefore often statistical in nature, driven by the limited event yields in the corresponding control samples. For the estimates that rely on simulation, we also propagate systematic uncertainties corresponding to the different corrections that are applied, as well as statistical uncertainties related to the limited size of simulated samples. A more detailed discussion of the assessment of systematic uncertainties affecting the individual background sources follows.

In the \tautau final state, we rely on an extrapolation in the \tauh isolation to obtain an estimate of the background with misidentified \tauh candidates. The uncertainty in this extrapolation is driven by the uncertainty introduced by the dependence of the isolation on the jet flavor.  It also includes the statistical uncertainty in the control regions from which this extrapolation is measured.  The uncertainty in the identification and isolation efficiency for prompt \tauh candidates is also propagated to the final estimate.  Finally an additional uncertainty is assessed for the fact that the extrapolations for both \tauh candidates are correlated, leading to an overall systematic uncertainty of 30--37\% for this background estimate, depending on the SR. In the estimation of the background from jets misidentified as \tauh in the \etau and \mutau final states, for which the transfer factor is estimated in a \wjets control sample, the purity of this control sample is $\approx$85\%, and the remaining $\approx$15\% is propagated as a systematic uncertainty. A systematic uncertainty of up to 5\% is considered for the rate of leptons misidentified as \tauh candidates in the leptonic final states.

The effects of different sources of uncertainty, such as uncertainties related to the jet energy scale; unclustered energy contributing to \ptmiss; and muon, electron, and \tauh energy scales that affect the simulated event samples used in the evaluation of the transfer factor are also propagated to the final background estimate. In the \emu final state, the largest source of uncertainty in the estimation of the background with misidentified leptons is the contamination from other background processes in the control regions $A, C$, and $D$ used for the background estimation. While the $C$ and $D$ regions are quite pure in QCD multijet events ($>$90\%), the level of contamination can be as high as $\approx$45\% in the $A$ region. A $50\%$ uncertainty is assigned to the QCD multijet background prediction in this final state to cover the potential effects of this contamination.

We rely mostly on simulation to obtain estimates of the other background contributions and the signal yields.  We propagate uncertainties related to the \PQb~tagging, trigger, and selection efficiencies, renormalization and factorization scale uncertainties, PDF uncertainties, and uncertainties in the jet energy scale, jet energy resolution, unclustered energy contributing to \ptmiss, and the energy scales of electrons, muons, and \tauh. For the DY+jets background, we have an additional uncertainty related to the corrections applied to the mass shape and \pt distribution, while for the \ttbar background, we propagate an uncertainty arising from the corrections to the top quark \pt spectrum. In the leptonic final states, we derive normalization scale factors for the DY+jets and \ttbar backgrounds in high-purity control samples. We assess uncertainties in these scale factors arising from the various systematic effects mentioned above and propagate them to the corresponding background estimates. We also monitor the trends of these scale factors by applying a series of selection requirements on the discriminating kinematic variables that are as close as possible to the selections applied in the SRs. In the \tautau final state, where the SRs are selected with stringent criteria applied to kinematic variables, we assign a 20\% normalization uncertainty for the production cross sections of these backgrounds, as well as for other SM processes. In the leptonic final states, an uncertainty of $10\%$ is assigned to the normalization of rare SM backgrounds to cover potential variations between the different SRs. As the {\PW\PW} background contribution can be sizeable in the leptonic final states and in particular for the \emu final state, a normalization uncertainty of $25\%$ is considered for this contribution. These uncertainties have been determined from sideband regions that are defined by the same baseline cuts as those that define the search bins, except considering only those bins of the search variables that are not used in the fit for the signal extraction.

The uncertainty of 2.5\%~\cite{CMS-PAS-LUM-17-001} in the integrated luminosity measurement is taken into account in all background estimates for which we do not derive normalization scale factors in dedicated data control samples, as well as for signal processes. In the case of the signal models we assign additional uncertainties due to differences between the fast simulation used for the signal models and the full simulation used for the background estimates that affect the \ptmiss resolution and lepton efficiencies.  We also checked the effects of possible mismodeling of the initial-state radiation (ISR), which affects the total transverse momentum ($\pt^{\mathrm{ISR}}$) of the system of SUSY particles, for the signal processes by reweighting the $\pt^{\mathrm{ISR}}$ distribution of simulated signal events. This reweighting procedure is based on studies of the transverse momentum of \PZ boson events~\cite{Chatrchyan:2013xna}. However these effects were found to be negligible for our SR definitions. The main systematic uncertainties for the signal models and background estimates are summarized in Table~\ref{tab:systematics}.

\begin{table}[tbh]
\centering
\topcaption{\label{tab:systematics} Systematic uncertainties in the analysis for the signal models and the different SM background predictions. The uncertainty values are evaluated separately for each signal model and mass hypothesis studied and are listed as percentages.}
\cmsTable{
\begin{tabular}{lccccc}
\hline
Uncertainty (\%)          & Signal & Misidentified $\Pe/\mu/\tauh$ & DY+jets & Top quark backgrounds & Rare SM    \\
\hline
\tauh efficiency & 5--11 &  0.1--5 & 5--10 & 4--10 & 0.1--10 \\
Electron efficiency (\emu, \etau) & 3 & \NA & 3 & 3 & 3 \\
Muon efficiency (\emu, \mutau) & 2 & \NA & 2 & 2 & 2 \\
Isolation extrapolation (\etau, \mutau, \tautau) & \NA &  15--35 & \NA & \NA & \NA \\
Misidentified \tauh correlations (\tautau)  & \NA &  8--13 & \NA & \NA & \NA \\
QCD multijet normalization (\emu) & \NA &  50 & \NA & \NA & \NA \\
\tauh energy scale (\etau, \mutau, \tautau)  & 0.1--23 &  \NA & 1--34 & 0.1--24 & 0.1--33 \\
Jet energy scale  & 0.1--45 &  \NA & 0.5--24 & 0.5--39 & 0.1--67 \\
Jet energy resolution  & 1--4 &  \NA & 29--61 & 3--10 & 11--31 \\
Unclustered energy  & 0.1--41 &  \NA & 2--42 & 0.1--41 & 0.1--100 \\
Electron energy scale (\emu, \etau)  & 0.1--22 &  \NA & 0.5--5 & 0.1--13 & 0.1--100 \\
Muon energy scale (\emu, \mutau)  & 0.1--11 &  \NA & 0.1--18 & 0.1--11 & 0.1--100 \\
\PQb~tagging  & 0.5--3 &  1--4 & 0.1--3 & 4--20 & 0.1--2 \\
Drell-Yan mass and \pt   & \NA &  \NA & 0.5--29 & \NA & \NA \\
Background cross sections & \NA &  \NA & 2--20 & 5--20 & 10--20 \\
Fast vs. full simulation  & 1--30 &  \NA & \NA & \NA & \NA \\
Integrated luminosity & 2.5 & \NA & \NA & \NA & 2.5 \\
\hline
\end{tabular}
}
\end{table}

\section{Results and interpretation}
\label{sec:results}
The results of the analysis in the \tautau final state are summarized in Table~\ref{tab:results_tautau}. The background estimates for the different SM processes are shown with the full uncertainty, the quadratic sum of the statistical and systematic uncertainties. As discussed in Section~\ref{sec:sysunc}, the uncertainties in the \tautau final state are dominated by the statistical uncertainties in the data control regions and the number of simulated events produced. These uncertainties are modeled in the likelihood function used for the statistical interpretation of the results with gamma distributions~\cite{CMS-NOTE-2011-005}. If there is no event in the control region used to obtain a given background estimate for any SR or no event in the simulated sample surviving the SR selection criteria, then the one standard deviation (s.d.) upper bound evaluated for that background contribution is presented in the table.  No significant excess is observed in any of the SRs.

\begin{table}[tbh]
\centering
\topcaption{Final predicted and observed event yields in the three SRs defined for the \tautau final state with all statistical and systematic uncertainties combined.  For the background estimates with no event in the corresponding data control region or in the simulated sample after the SR selection, the predicted yield is indicated as being less than the one standard deviation upper bound evaluated for that estimate. The central value and the uncertainties for the total background estimate are then extracted from the full pre-fit likelihood. Expected yields are also given for signal models of direct \PSGt pair production in the purely left- and right-handed scenarios and in the maximally mixed scenario, with the \PSGt and \PSGczDo masses in GeV indicated in parentheses.}
\label{tab:results_tautau}
\begin{tabular}{lccc}
\hline
        & SR1 & SR2  & SR3      \\
\hline
Nonprompt and misidentified \tauh & 0.68 $^{+0.90}_{-0.68}$ & $2.49 \pm 1.83$  & ${<}1.24$
\rule{0pt}{2.6ex}\\
Drell-Yan+jets background & 0.80$^{+0.97}_{-0.80}$ & ${<}0.71$ & ${<}0.71$
\rule{0pt}{2.6ex} \\
Top quark backgrounds  & 0.02$^{+0.03}_{-0.02}$ & $0.73 \pm 0.31$ & $1.76 \pm 0.68$
\rule{0pt}{2.6ex} \\
Rare SM processes  & $0.72 \pm 0.38$ & $0.20 \pm 0.15$ & 0.20 $^{+0.25}_{-0.20}$
\rule{0pt}{2.6ex}\rule[-1.2ex]{0pt}{0pt} \\[\cmsTabSkip]
Total background  & 2.22$^{+1.37}_{-1.12}$ & 4.35$^{+1.75}_{-1.53}$  & 3.70$^{+1.52}_{-1.08}$
\rule{0pt}{2.6ex}\rule[-1.2ex]{0pt}{0pt}\\[\cmsTabSkip]
Left (150,1)  & $1.25 \pm 0.40$ & $2.91 \pm 0.59$ & $1.53 \pm 0.33$ \\
Right (150,1)  & $1.09 \pm 0.26$  & $1.27 \pm 0.20$ & $0.74 \pm 0.17$ \\
Mixed (150,1)  & $1.04 \pm 0.22$  & $1.39 \pm 0.27$ & $0.92 \pm 0.15$ \\[\cmsTabSkip]
Observed & 0 & 5  & 2
\\
\hline
\end{tabular}
\end{table}

\begin{figure}[htbp]
\centering
\includegraphics[width=.48\textwidth]{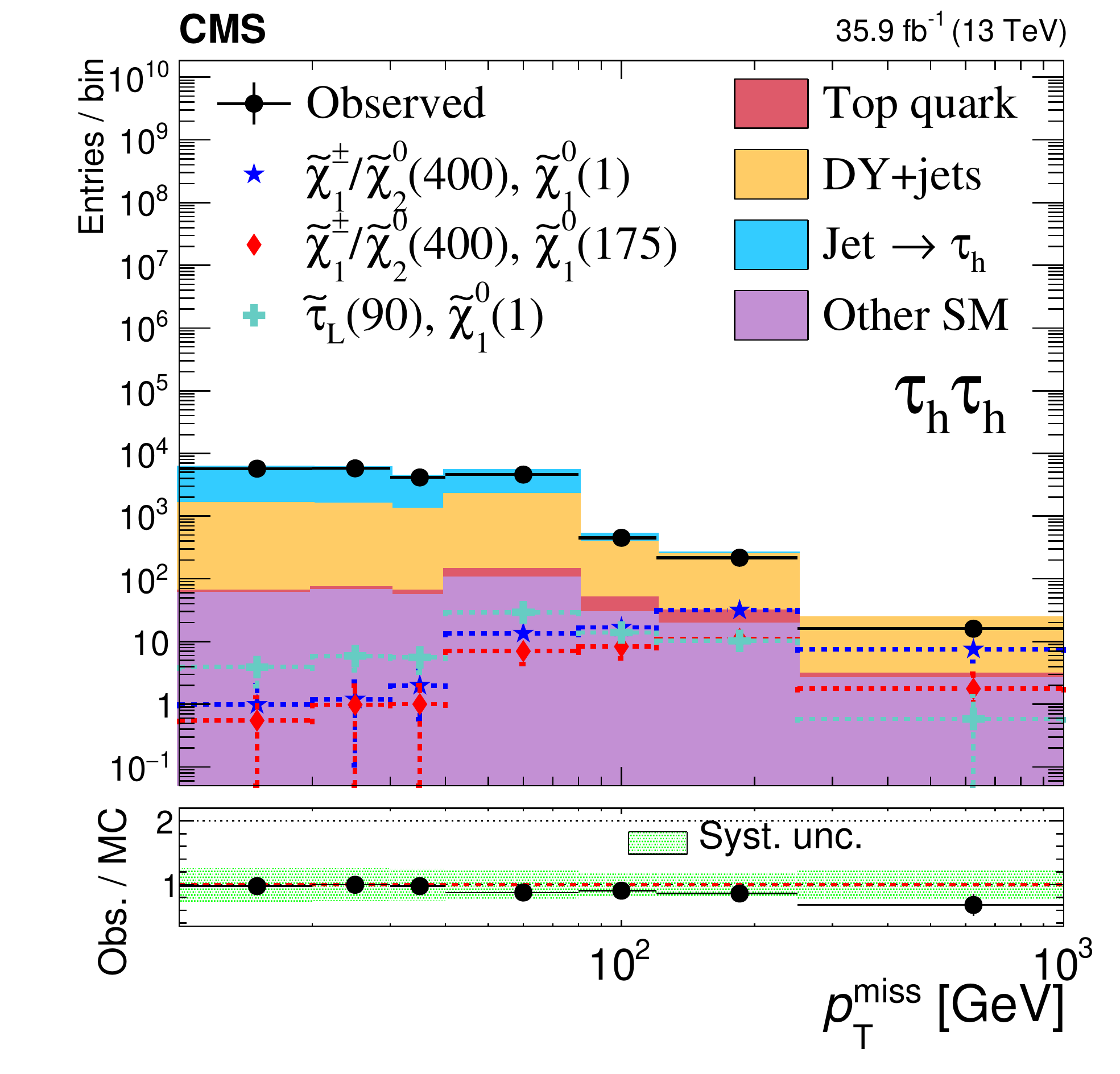}
\includegraphics[width=.48\textwidth]{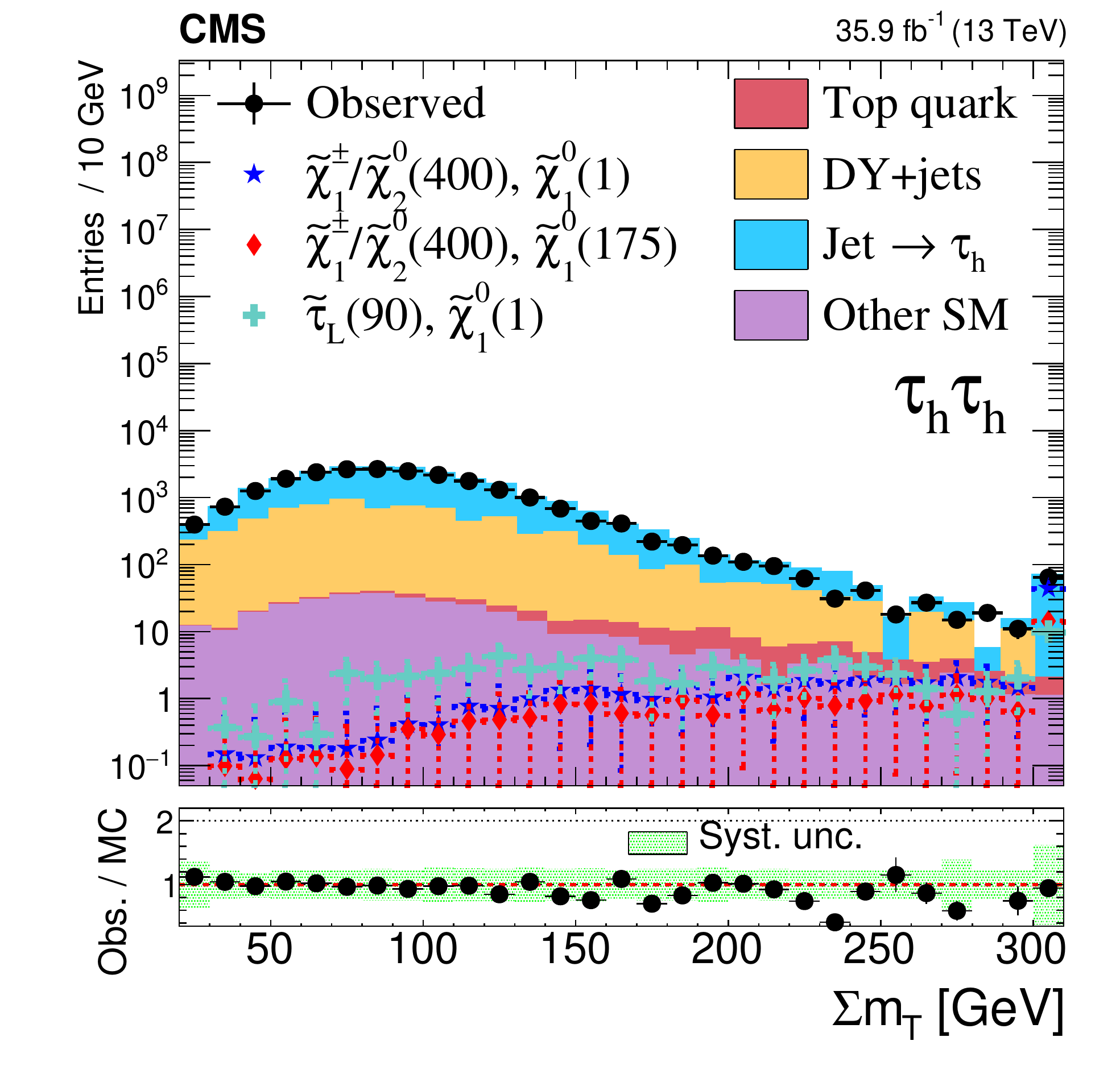}
\caption{Distributions of the search variables \ptmiss (left) and \sumMT (right) for the \tautau final state for events after the baseline selection. The black points show the data. The background estimates are represented with stacked histograms. Distributions for two benchmark models of chargino-neutralino production, and one of direct left-handed \PSGt pair production, are overlaid. The numbers within parentheses in the legend correspond to the masses of the parent SUSY particle and the \PSGczDo in GeV for these benchmark models. In both cases, the last bin includes overflows.
}
\label{fig:search_variables_tautau}
\end{figure}

\begin{figure}[htbp]
\centering
\includegraphics[width=.48\textwidth]{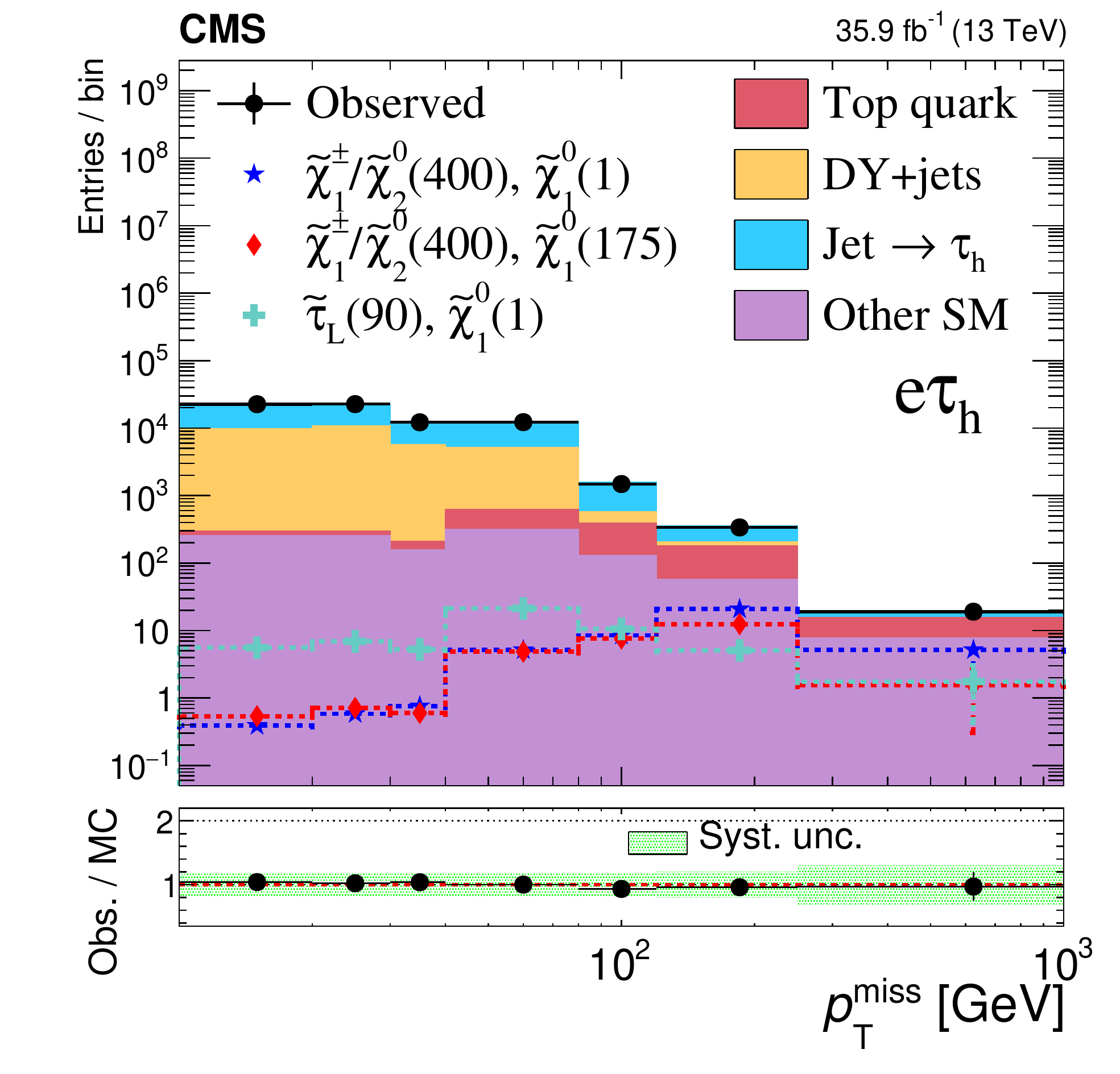}
\includegraphics[width=.48\textwidth]{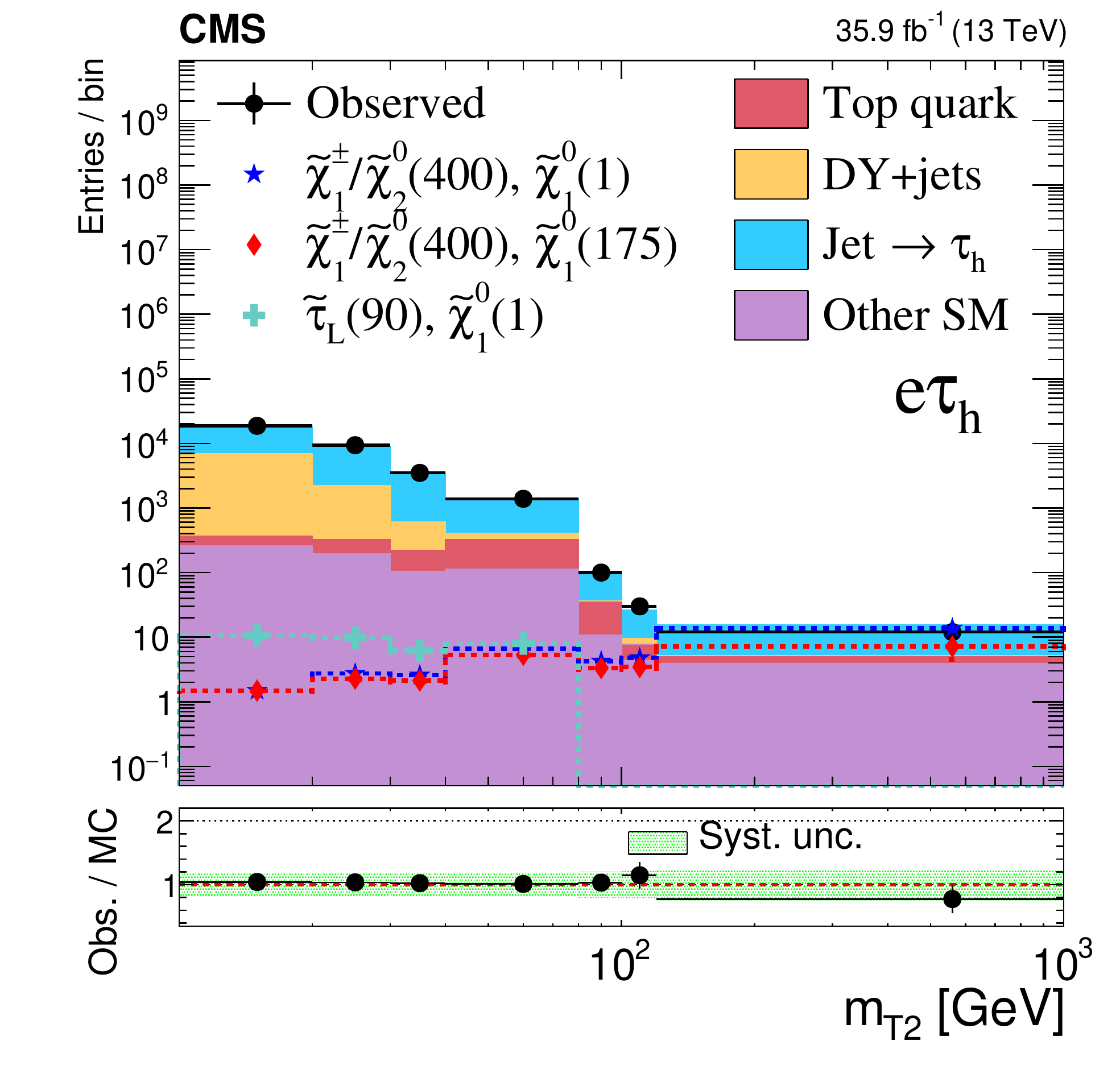} \\
\includegraphics[width=.48\textwidth]{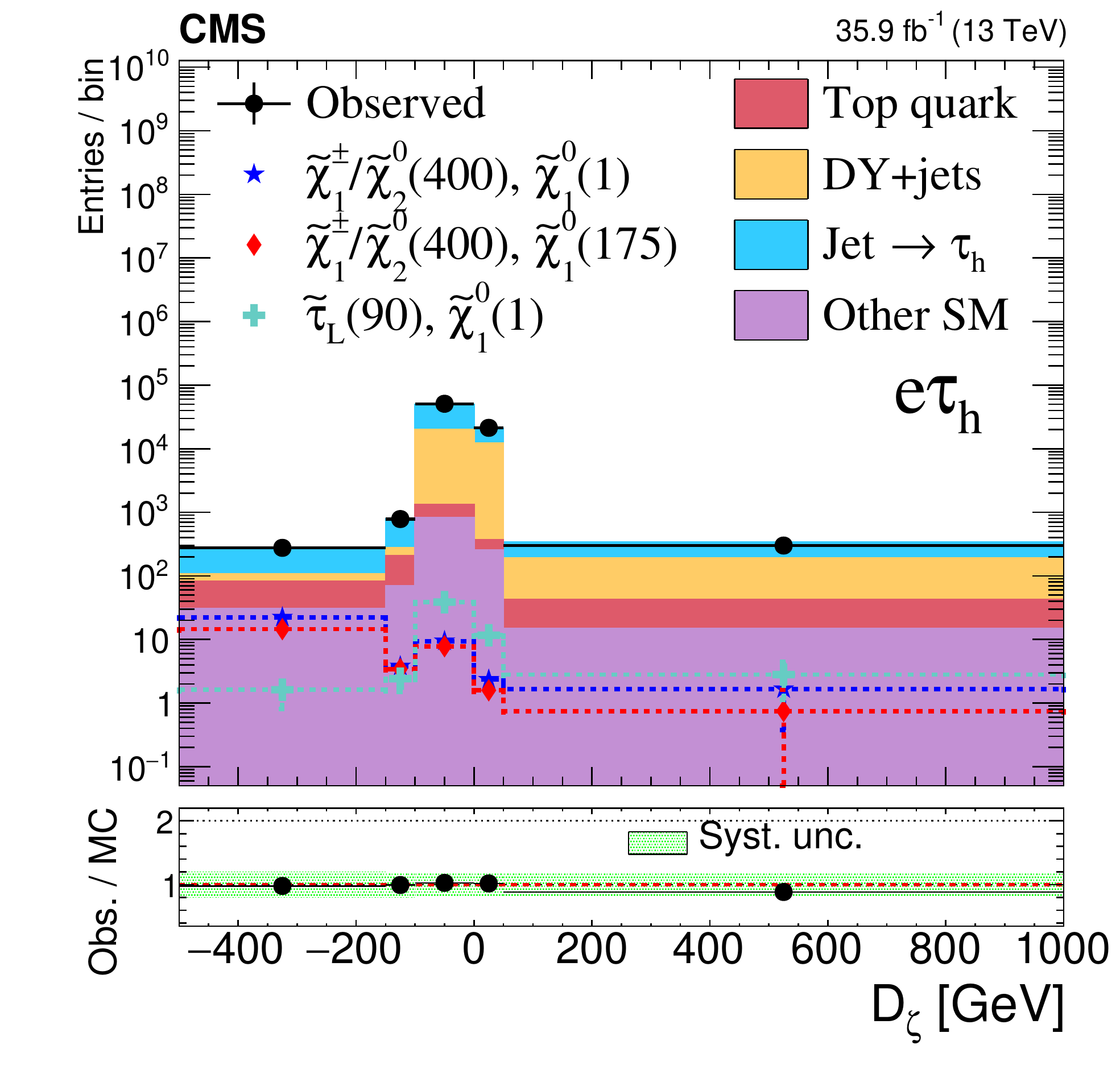}
\caption{Distributions of the search variables \ptmiss (top left), \mttwo (top right), and \DZ (bottom) for the \etau final state for events after the baseline selection. The black points show the data. The background estimates are represented with stacked histograms. Distributions for two benchmark models of chargino-neutralino production, and one of direct left-handed \PSGt pair production, are overlaid. The numbers within parentheses in the legend correspond to the masses of the parent SUSY particle and the \PSGczDo in GeV for these benchmark models. In all cases, the last bin includes overflows.
}
\label{fig:search_variables_etau}
\end{figure}

\begin{figure}[htbp]
\centering
\includegraphics[width=.48\textwidth]{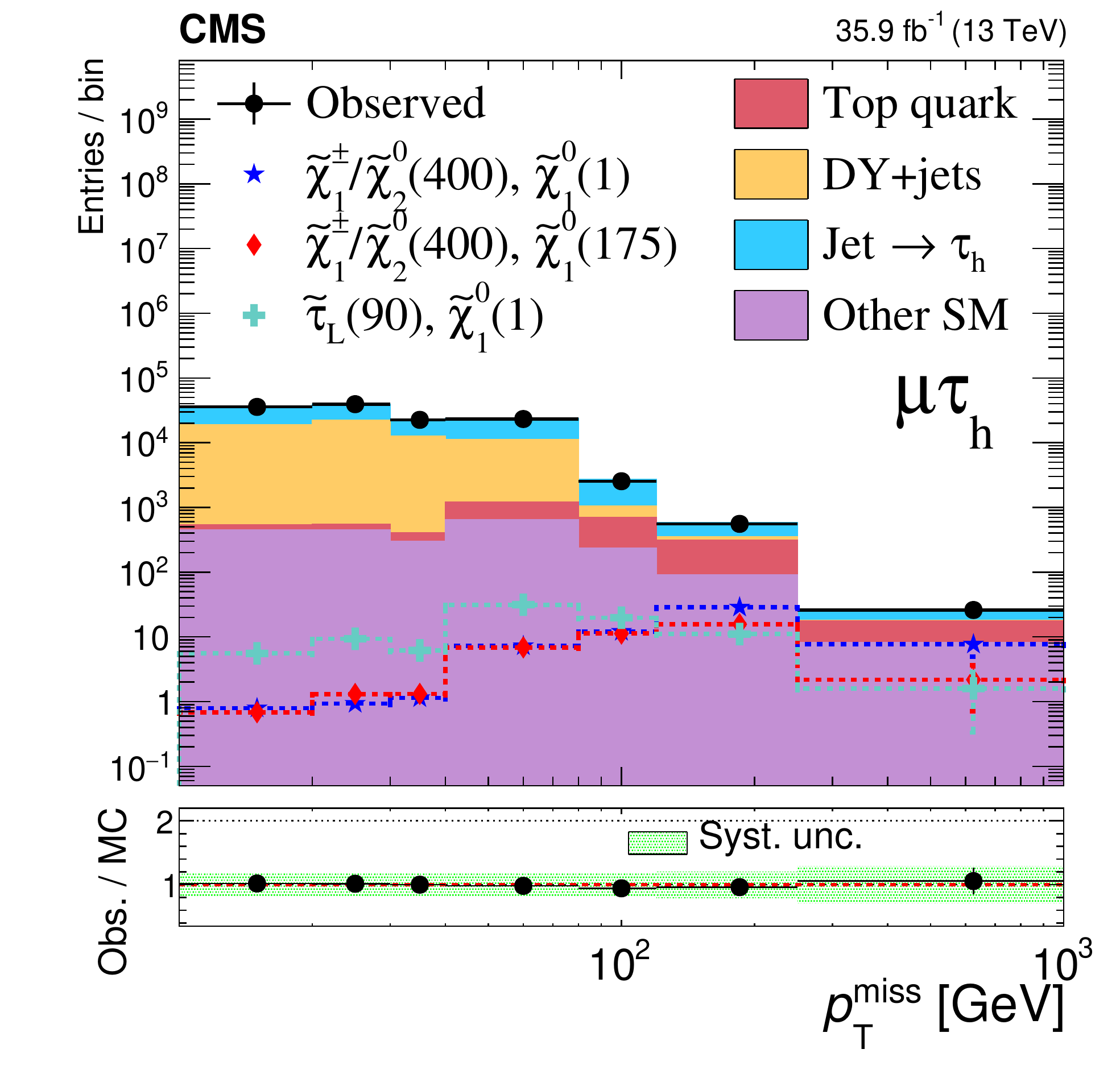}
\includegraphics[width=.48\textwidth]{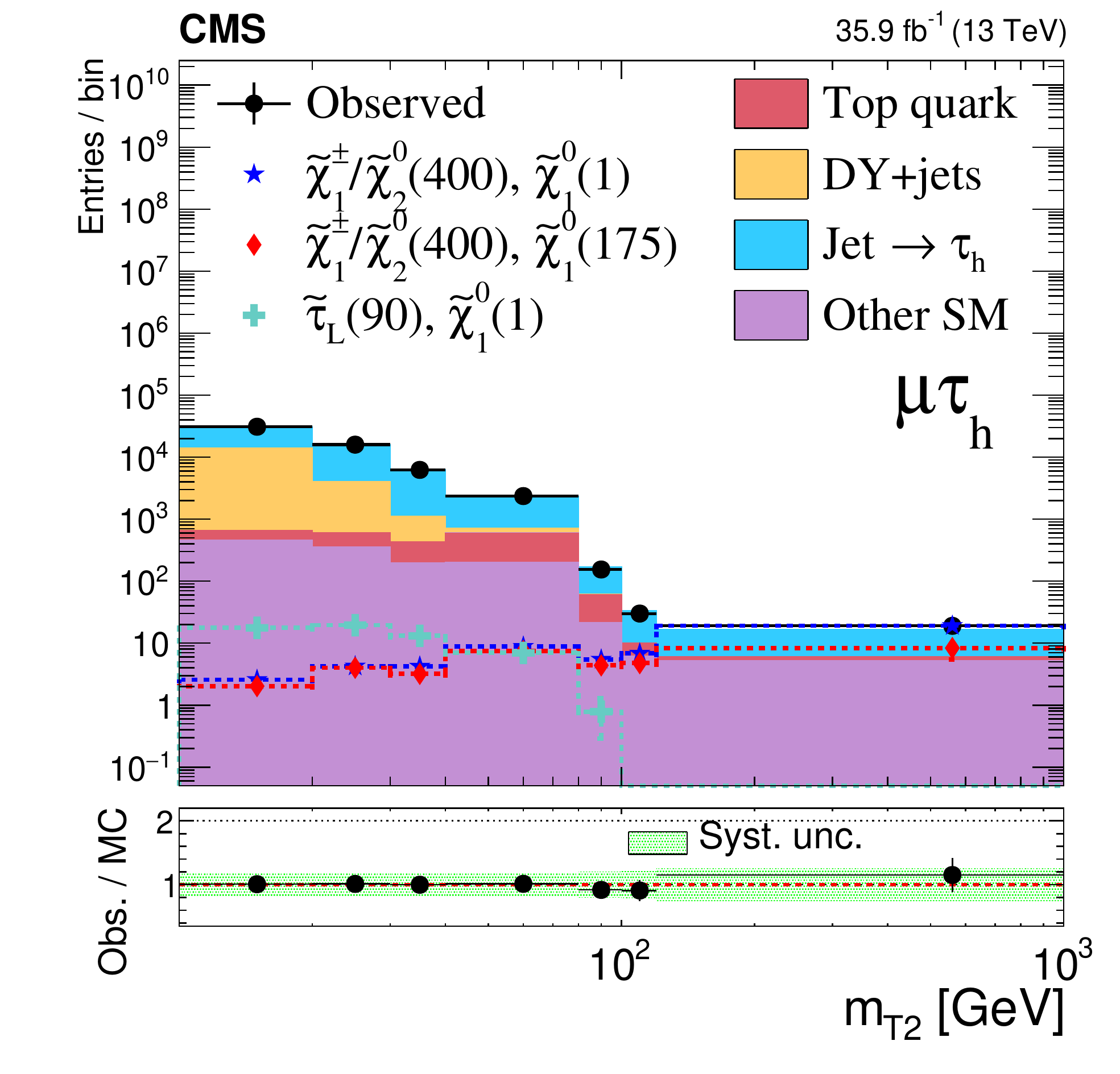} \\
\includegraphics[width=.48\textwidth]{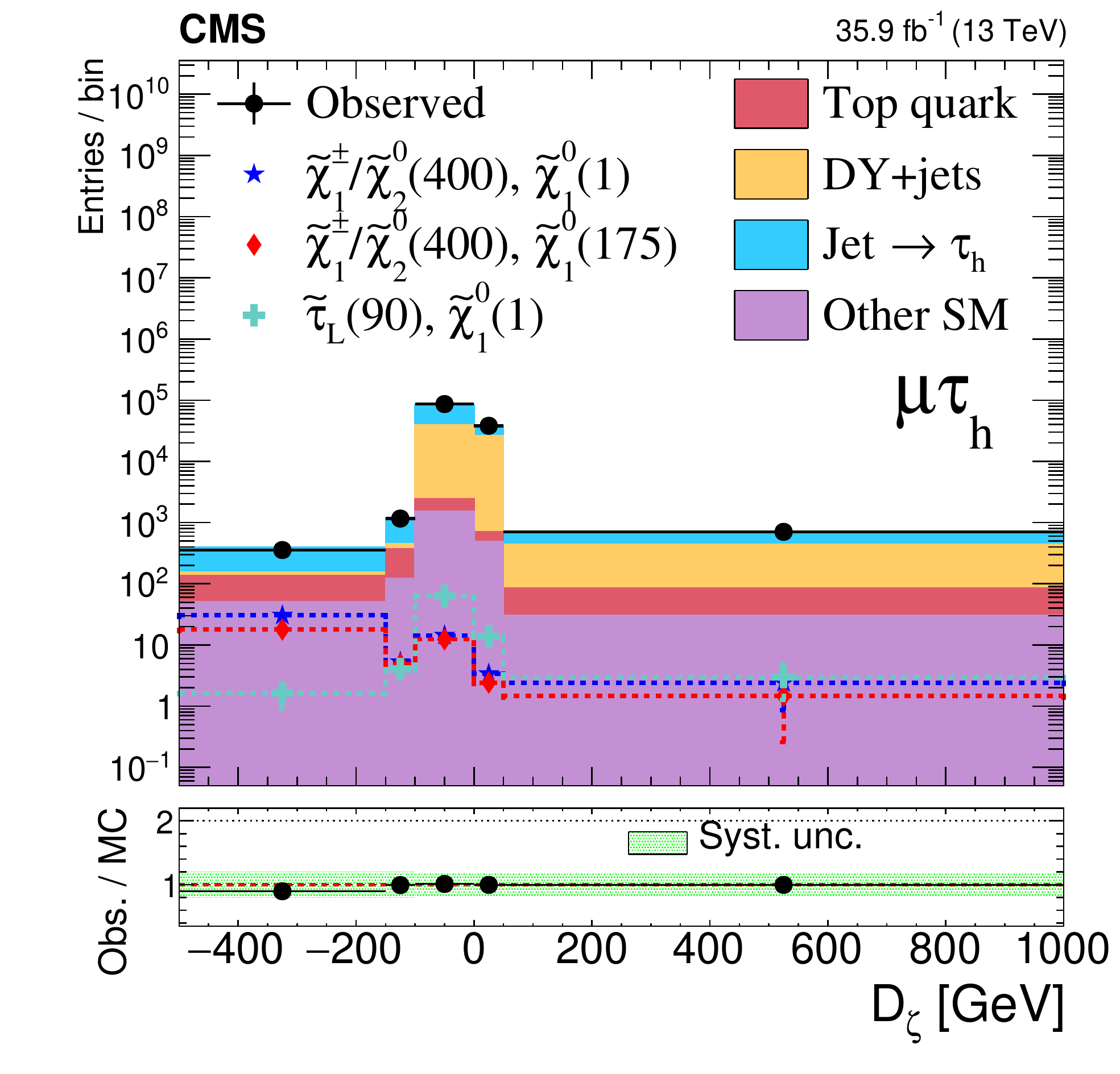}
\caption{Distributions of the search variables \ptmiss (top left), \mttwo (top right), and \DZ (bottom) for the \mutau final state for events after the baseline selection. The black points show the data. The background estimates are represented with stacked histograms. Distributions for two benchmark models of chargino-neutralino production, and one of direct left-handed \PSGt pair production, are overlaid. The numbers within parentheses in the legend correspond to the masses of the parent SUSY particle and the \PSGczDo in GeV for these benchmark models. In all cases, the last bin includes overflows.
}
\label{fig:search_variables_mutau}
\end{figure}

\begin{figure}[htbp]
\centering
\includegraphics[width=.48\textwidth]{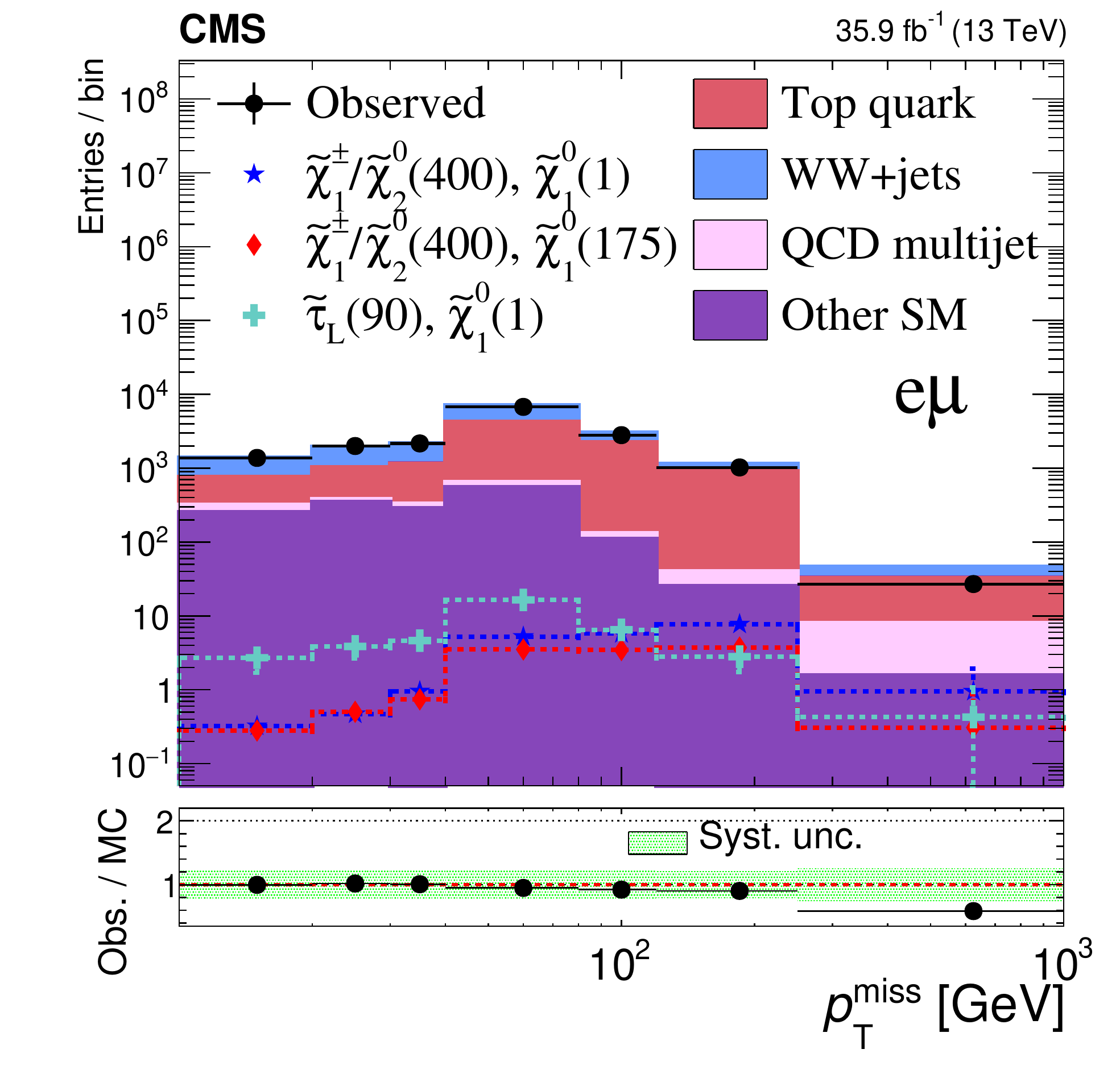}
\includegraphics[width=.48\textwidth]{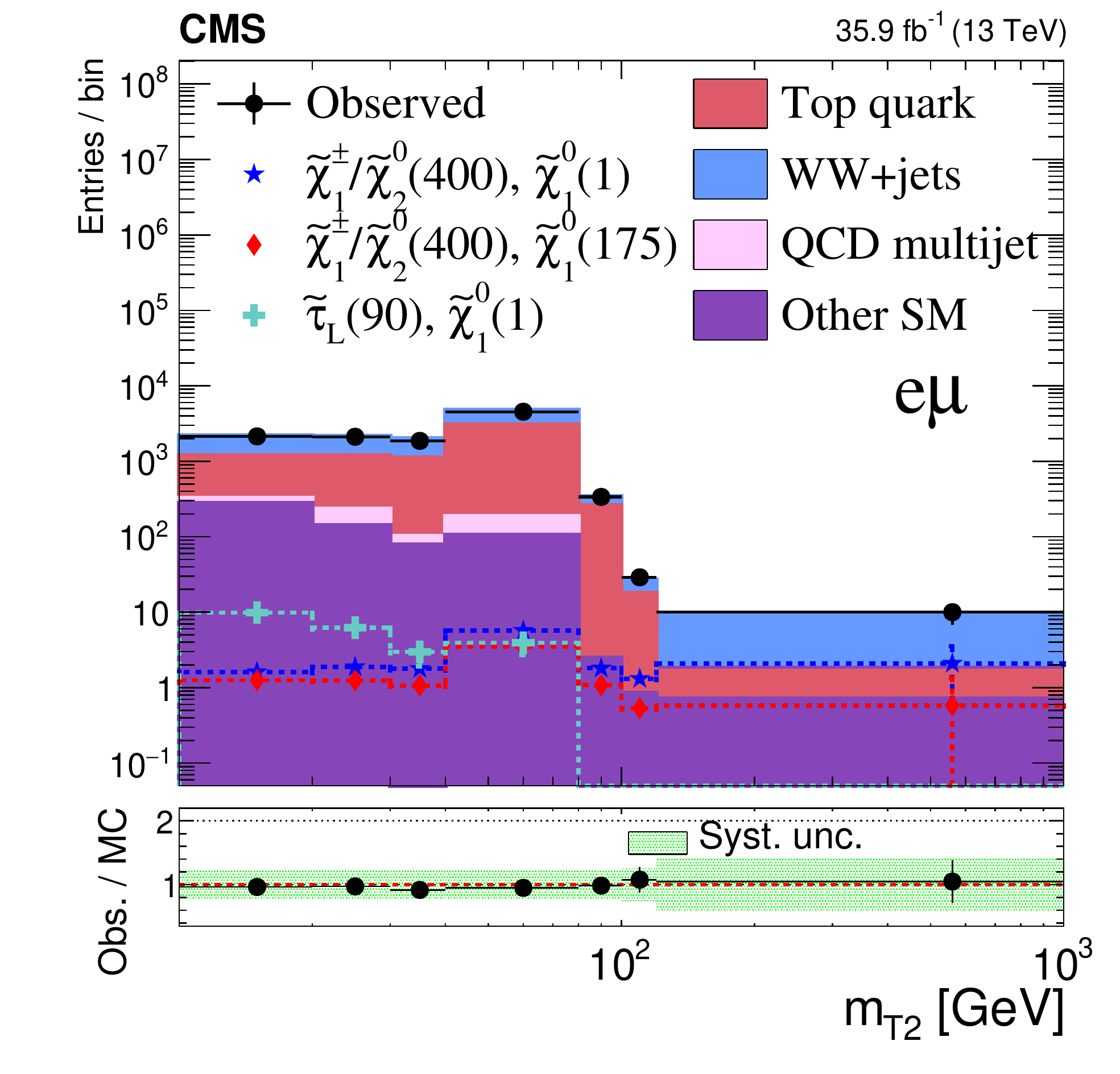} \\
\includegraphics[width=.48\textwidth]{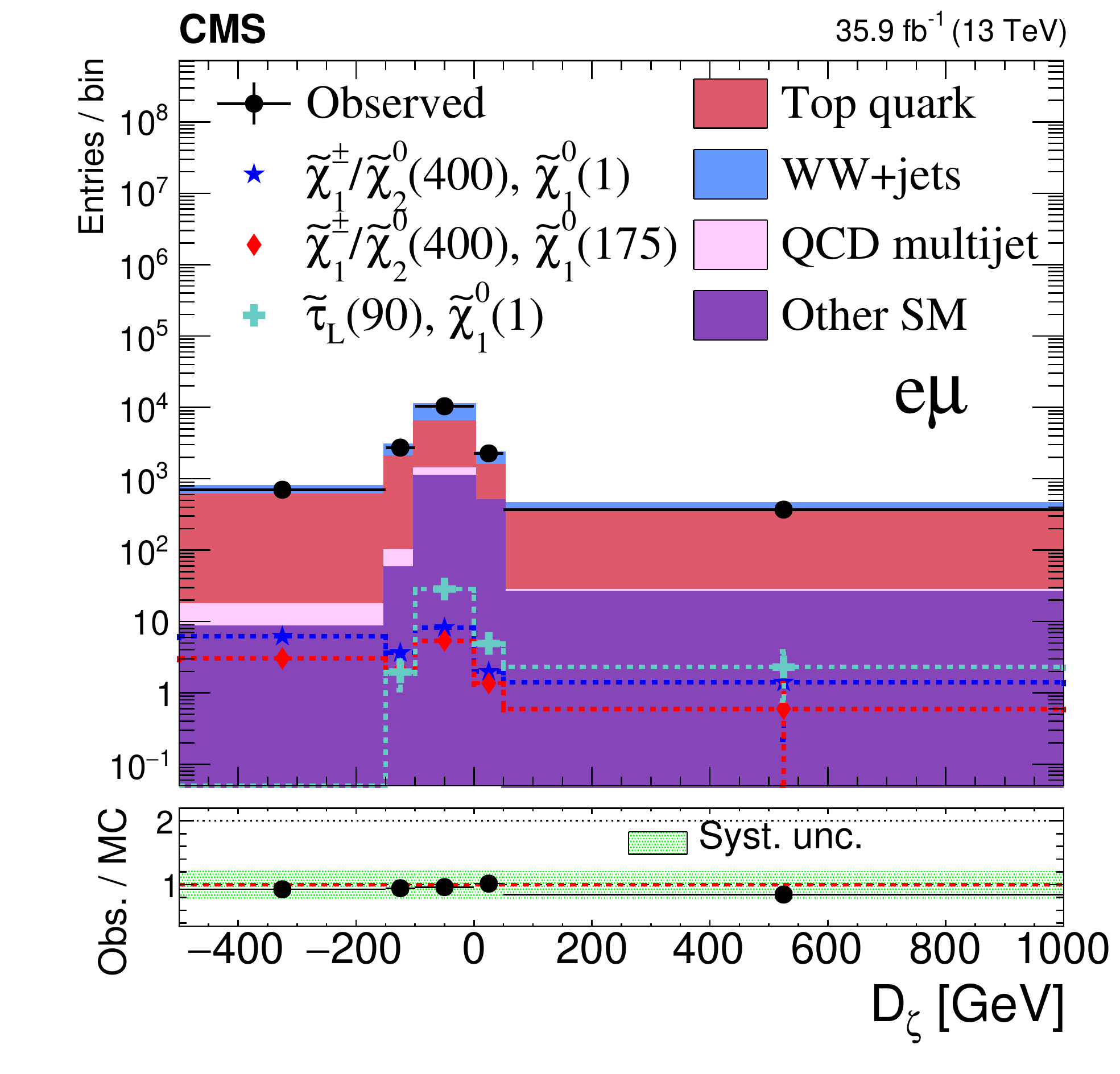}
\caption{Distributions of the search variables \ptmiss (top left), \mttwo (top right), and \DZ (bottom) for the \emu final state for events after the baseline selection. The black points show the data. The background estimates are represented with stacked histograms. Distributions for two benchmark models of chargino-neutralino production, and one of direct left-handed \PSGt pair production, are overlaid. The numbers within parentheses in the legend correspond to the masses of the parent SUSY particle and the \PSGczDo in GeV for these benchmark models. In all cases, the last bin includes overflows.
}
\label{fig:search_variables_emu}
\end{figure}

\begin{figure}[htbp]
\centering
\includegraphics[width=0.95\textwidth]{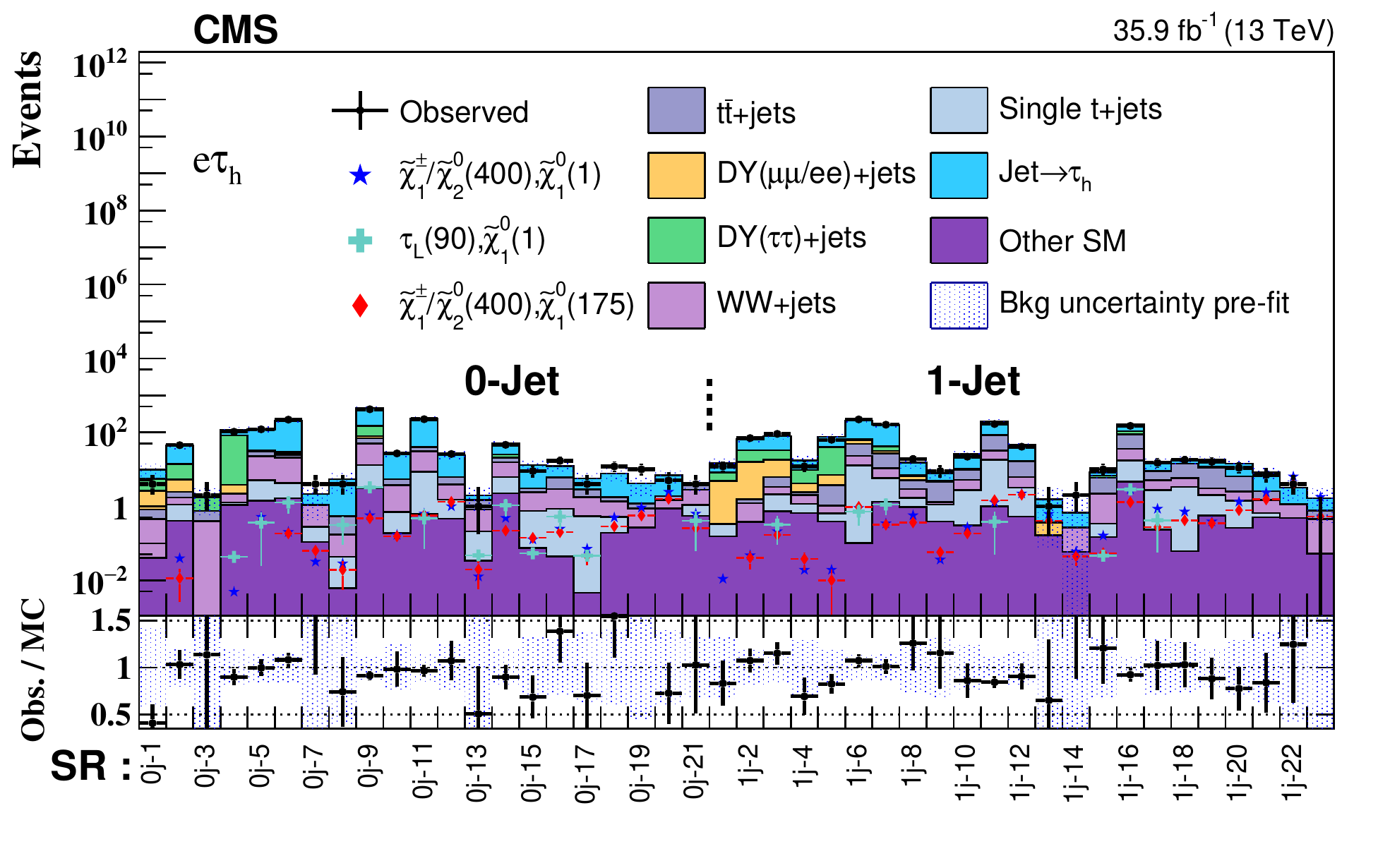} \\
\includegraphics[width=0.95\textwidth]{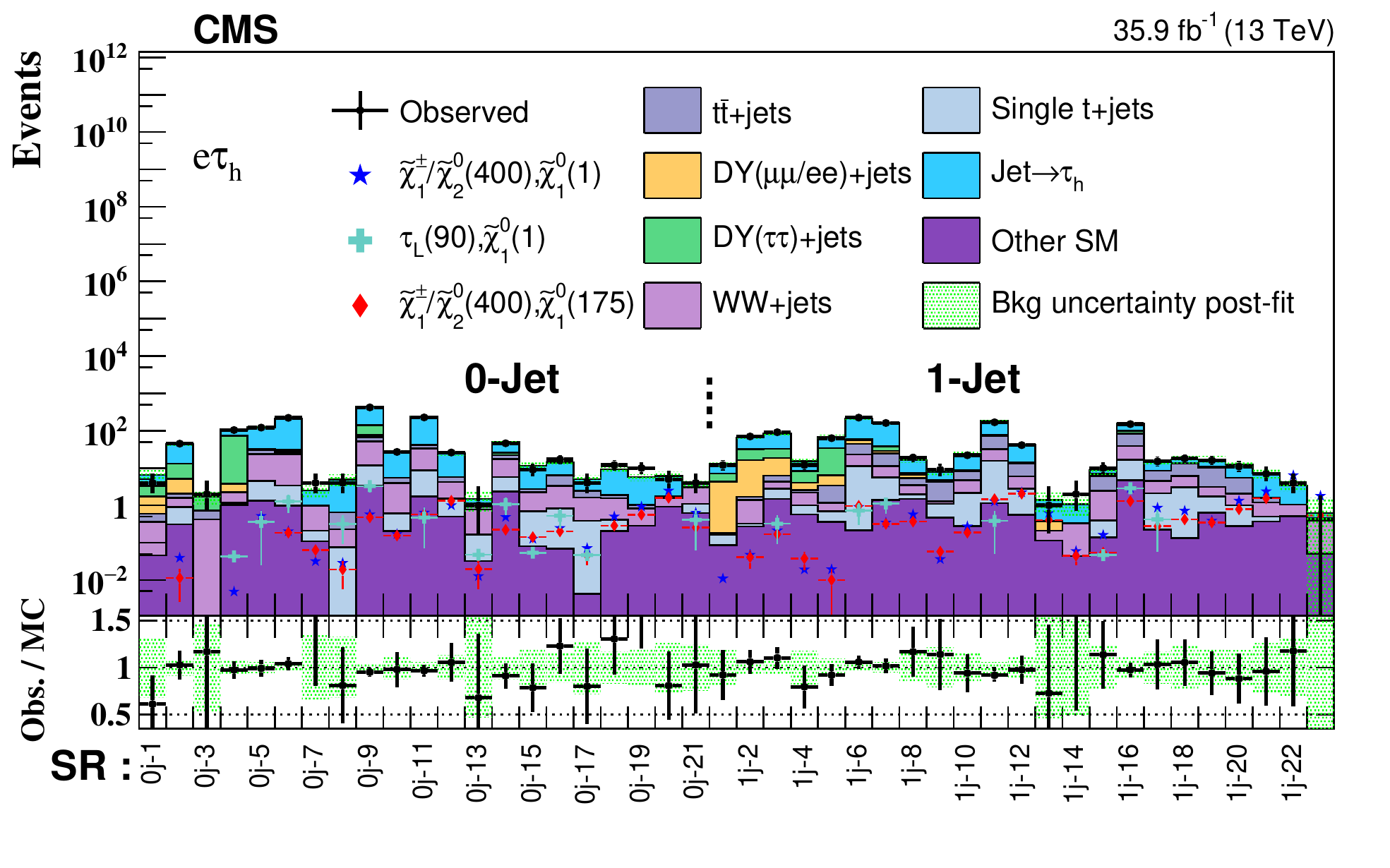}
\caption{Pre-fit (upper) and post-fit (lower) results for the SRs used for the final signal extraction in the \etau final state. Distributions for two benchmark models of chargino-neutralino production, and one of direct left-handed \PSGt pair production, are overlaid. The numbers within parentheses in the legend correspond to the masses of the parent SUSY particle and the \PSGczDo in GeV for these benchmark models. In the ratio panels, the black markers indicate the ratio of the observed data in each SR to the corresponding pre-fit or post-fit SM background prediction.
}
\label{fig:SR_ResultsElTau}
\end{figure}

\begin{figure}[htbp]
\centering
\includegraphics[width=0.95\textwidth]{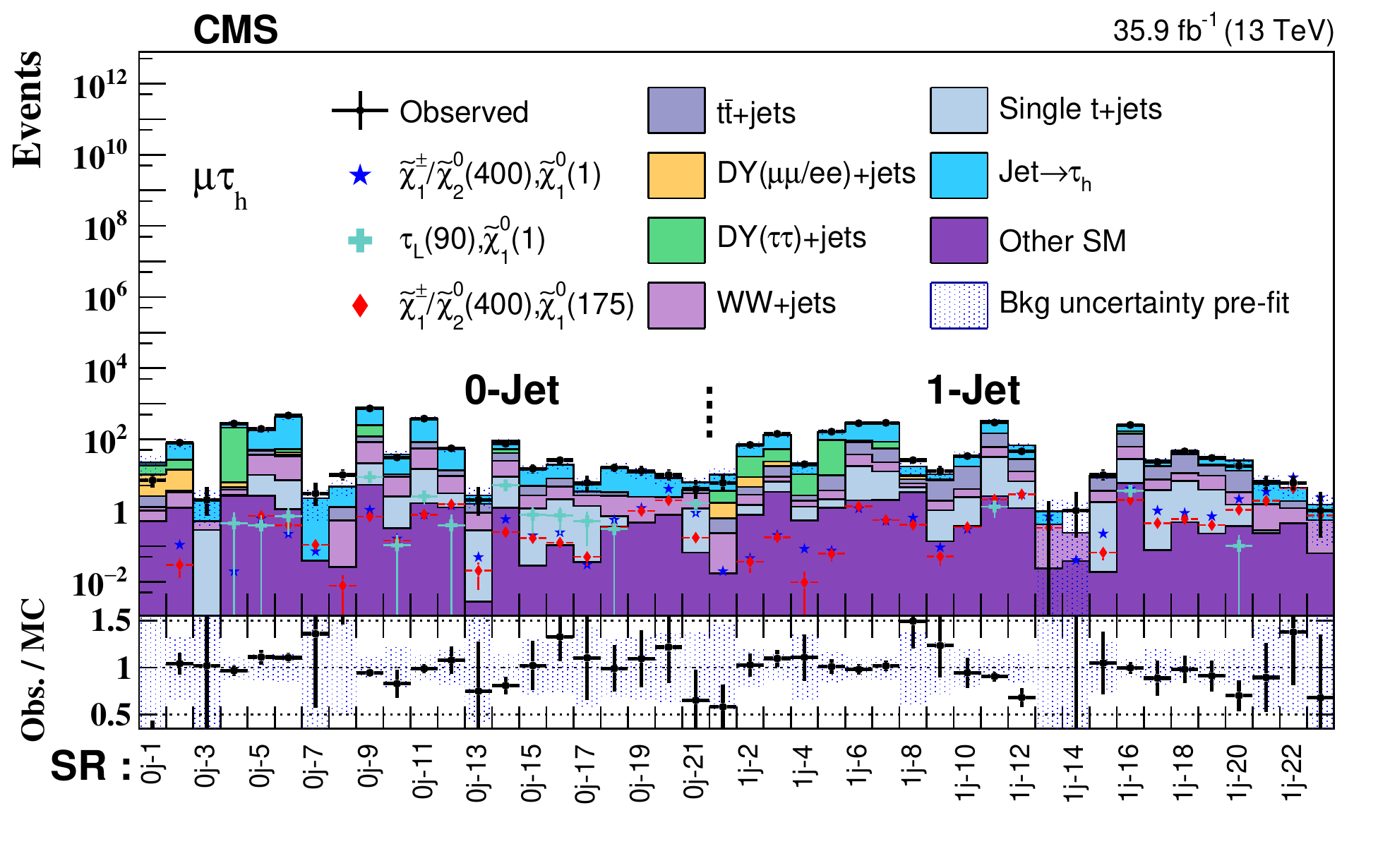} \\
\includegraphics[width=0.95\textwidth]{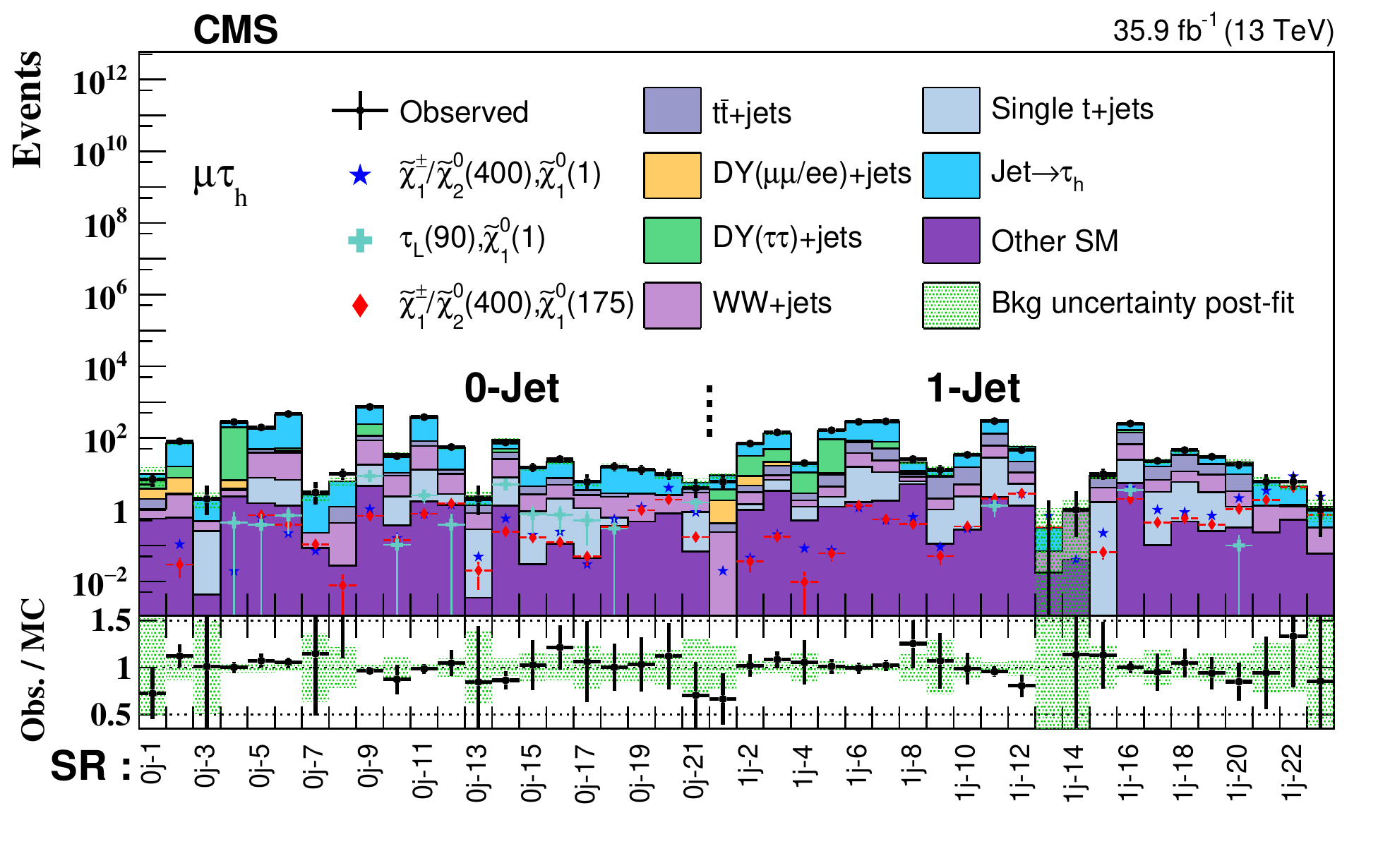}

\caption{Pre-fit (upper) and post-fit (lower) results for the SRs used for the final signal extraction in the \mutau final state. Distributions for two benchmark models of chargino-neutralino production, and one of direct left-handed \PSGt pair production, are overlaid. The numbers within parentheses in the legend correspond to the masses of the parent SUSY particle and the \PSGczDo in GeV for these benchmark models. In the ratio panels, the black markers indicate the ratio of the observed data in each SR to the corresponding pre-fit or post-fit SM background prediction.
}
\label{fig:SR_ResultsMuTau}
\end{figure}

\begin{figure}[htbp]
\centering
\includegraphics[width=0.95\textwidth]{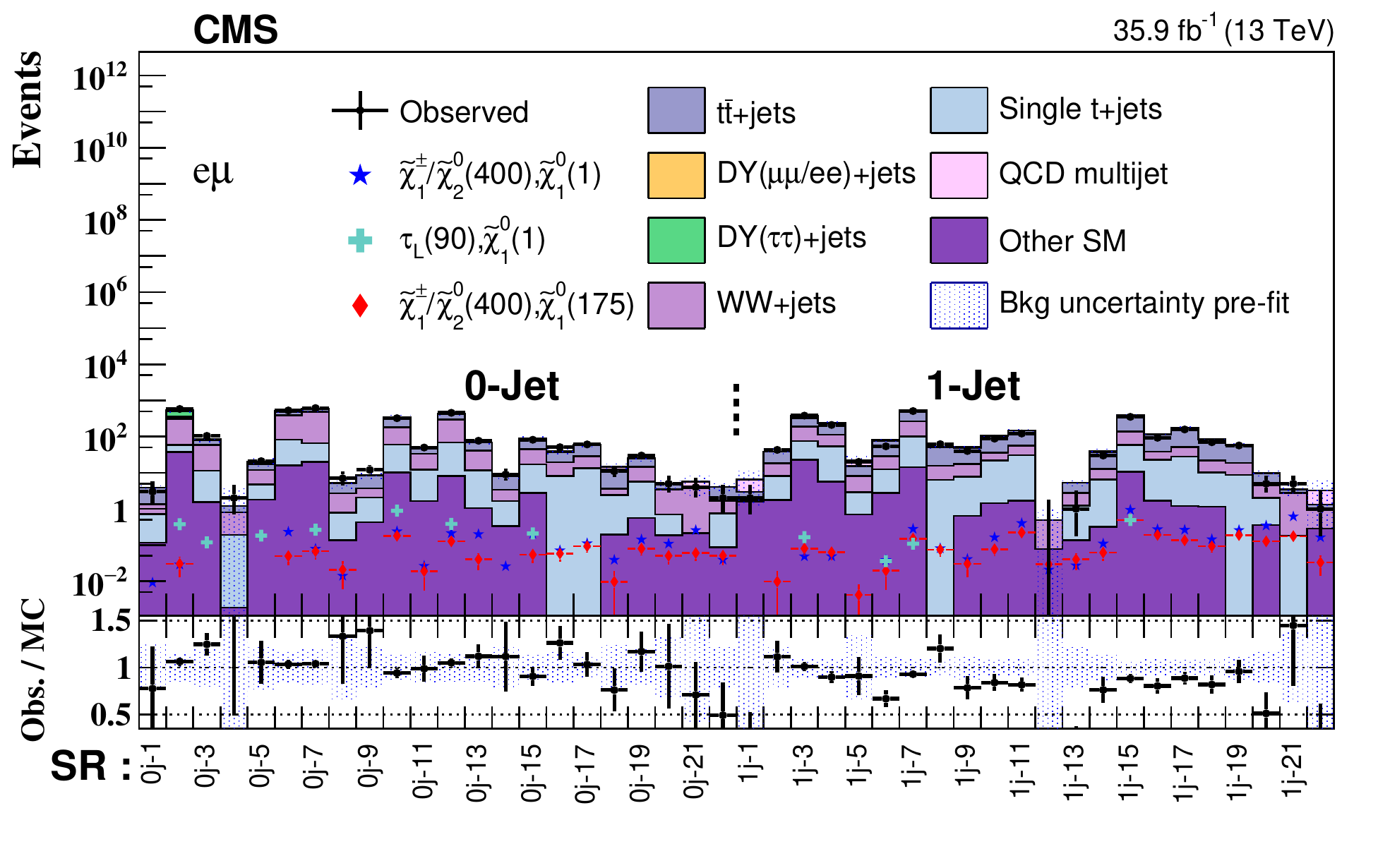}\\
\includegraphics[width=0.95\textwidth]{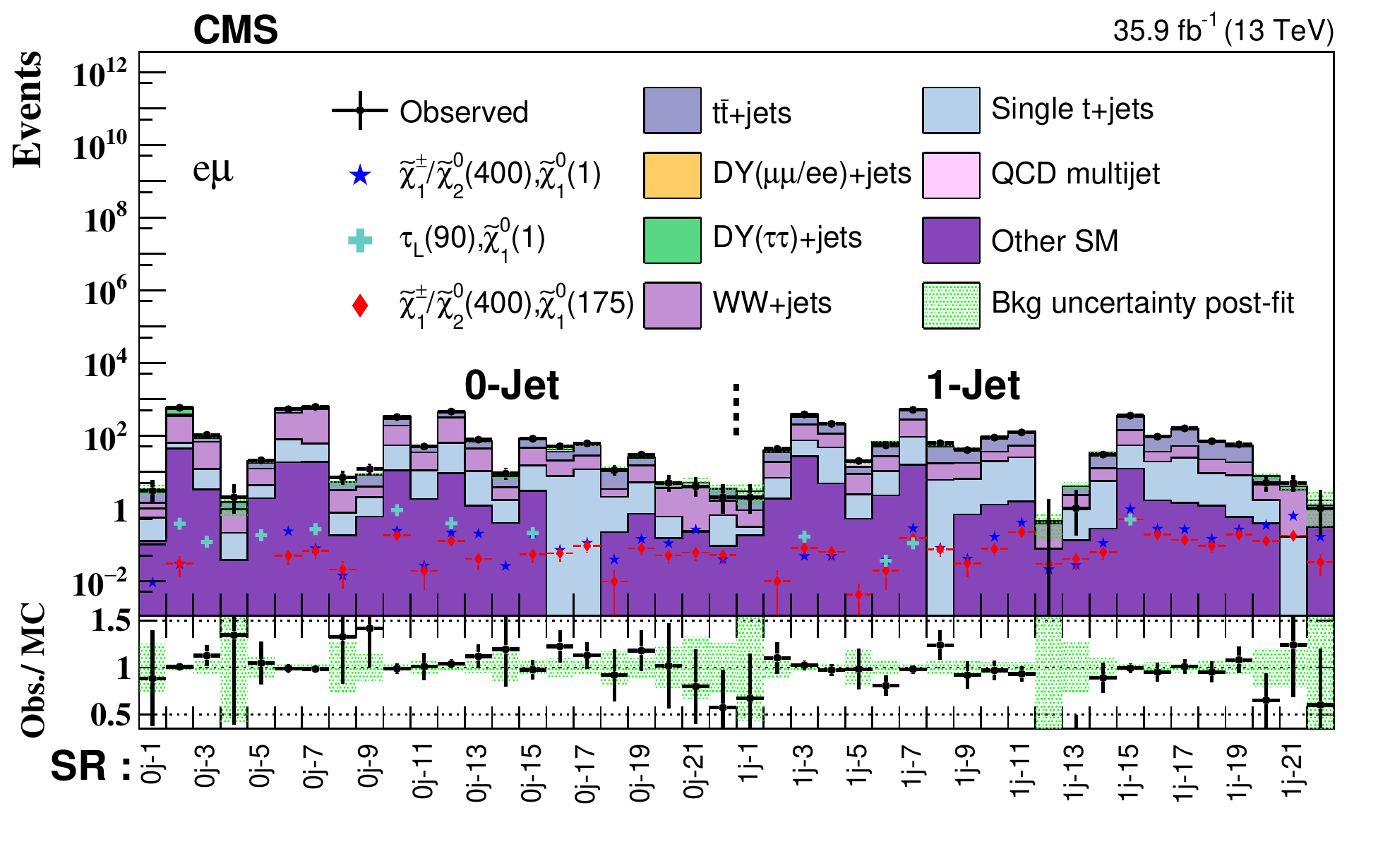}
\caption{Pre-fit (upper) and post-fit (lower) results for the SRs used for the final signal extraction in the \emu final state. Distributions for two benchmark models of chargino-neutralino production, and one of direct left-handed \PSGt pair production, are overlaid. The numbers within parentheses in the legend correspond to the masses of the parent SUSY particle and the \PSGczDo in GeV for these benchmark models. In the ratio panels, the black markers indicate the ratio of the observed data in each SR to the corresponding pre-fit or post-fit SM background prediction.
}
\label{fig:SR_ResultsMuEl}
\end{figure}

A comparison of the observed data with the background prediction for the search variables \ptmiss and \sumMT is shown for the all-hadronic final state in Fig.~\ref{fig:search_variables_tautau} after the baseline selection. Similar comparisons are shown for the three search variables \ptmiss, \mttwo, and \DZ used in the leptonic final states (\etau, \mutau, and \emu) in Figs.~\ref{fig:search_variables_etau}--\ref{fig:search_variables_emu}. The background estimates derived for all the SRs in the leptonic final states, as defined in Tables~\ref{tab:srdefinition_0jet} and~\ref{tab:srdefinition_1jet}, together with their uncertainties, are used as inputs to a simultaneous maximum likelihood fit to the observed data. The results for the SR bins that are used for the signal extraction in the final statistical interpretation procedure are shown in Figs.~\ref{fig:SR_ResultsElTau}--\ref{fig:SR_ResultsMuEl}. Both  histograms before the simultaneous fitting of all SRs (pre-fit) and after fitting (post-fit) are shown. The numbers of expected and observed events in each SR are also reported in Tables~\ref{tab:res_elehad}--\ref{tab:res_elemu} in Appendix~\ref{app:yieldtables}.

\begin{figure}[htbp]
\centering
\includegraphics[width=0.48\textwidth]{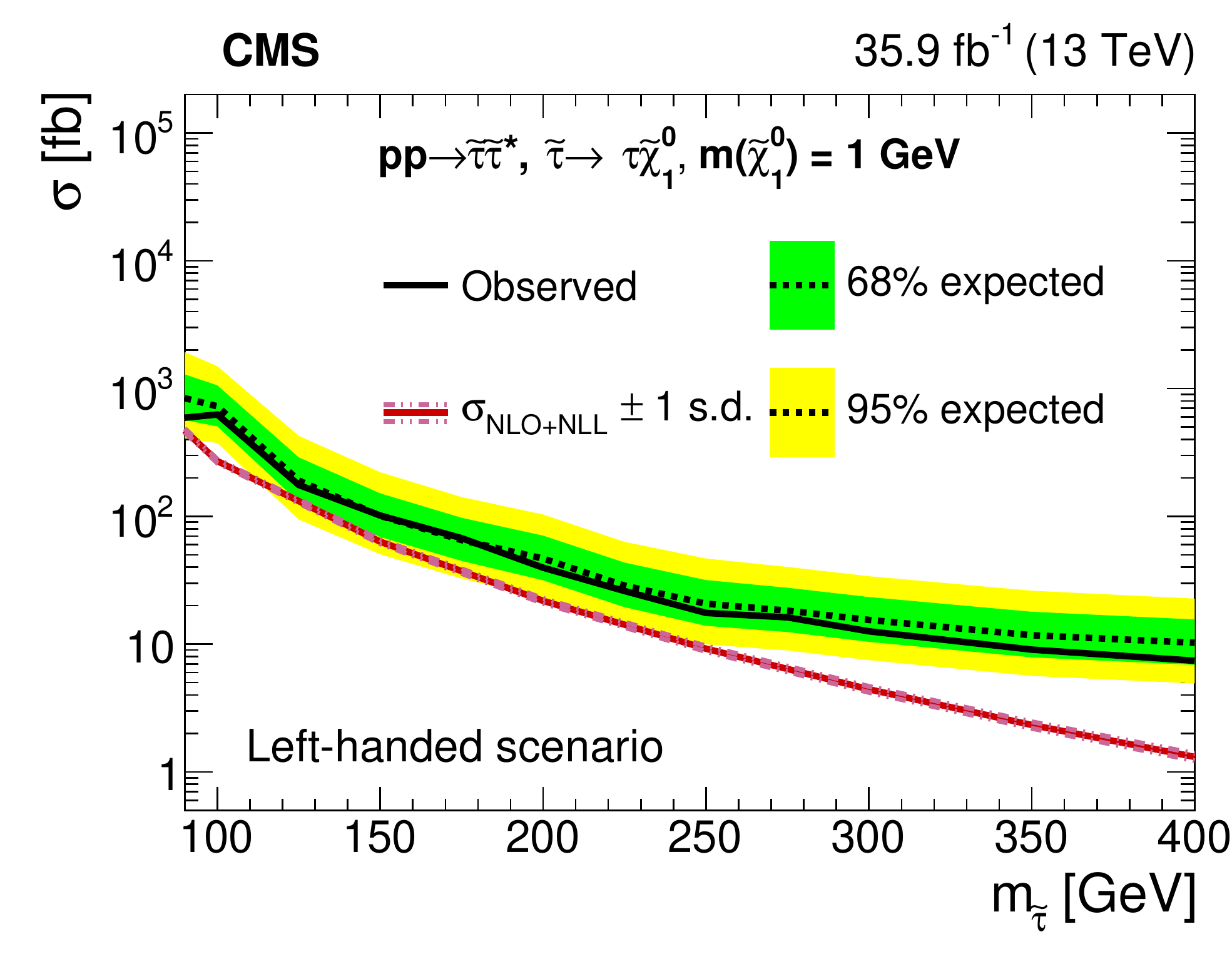}
\includegraphics[width=0.48\textwidth]{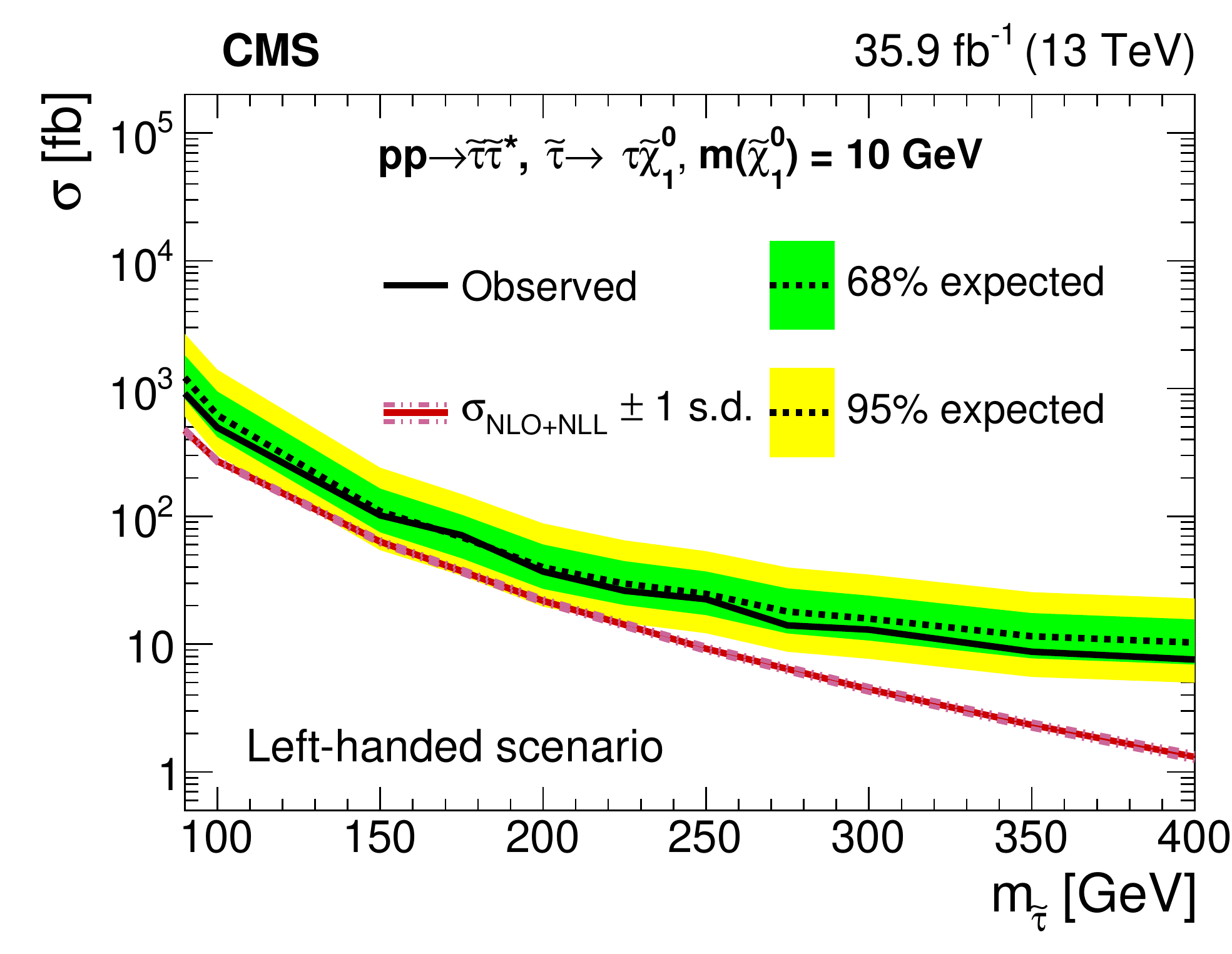}
\\
\includegraphics[width=0.48\textwidth]{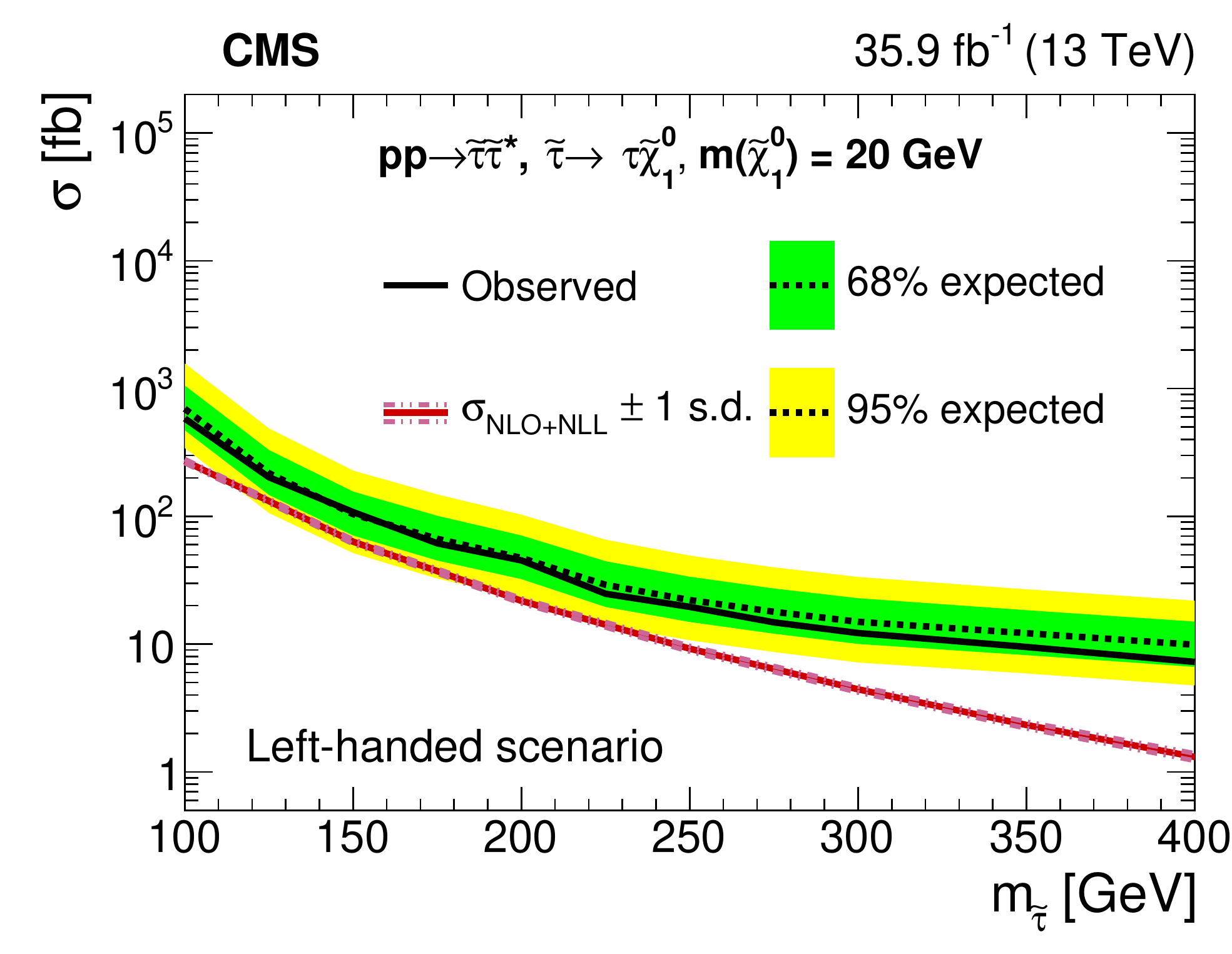}
\includegraphics[width=0.48\textwidth]{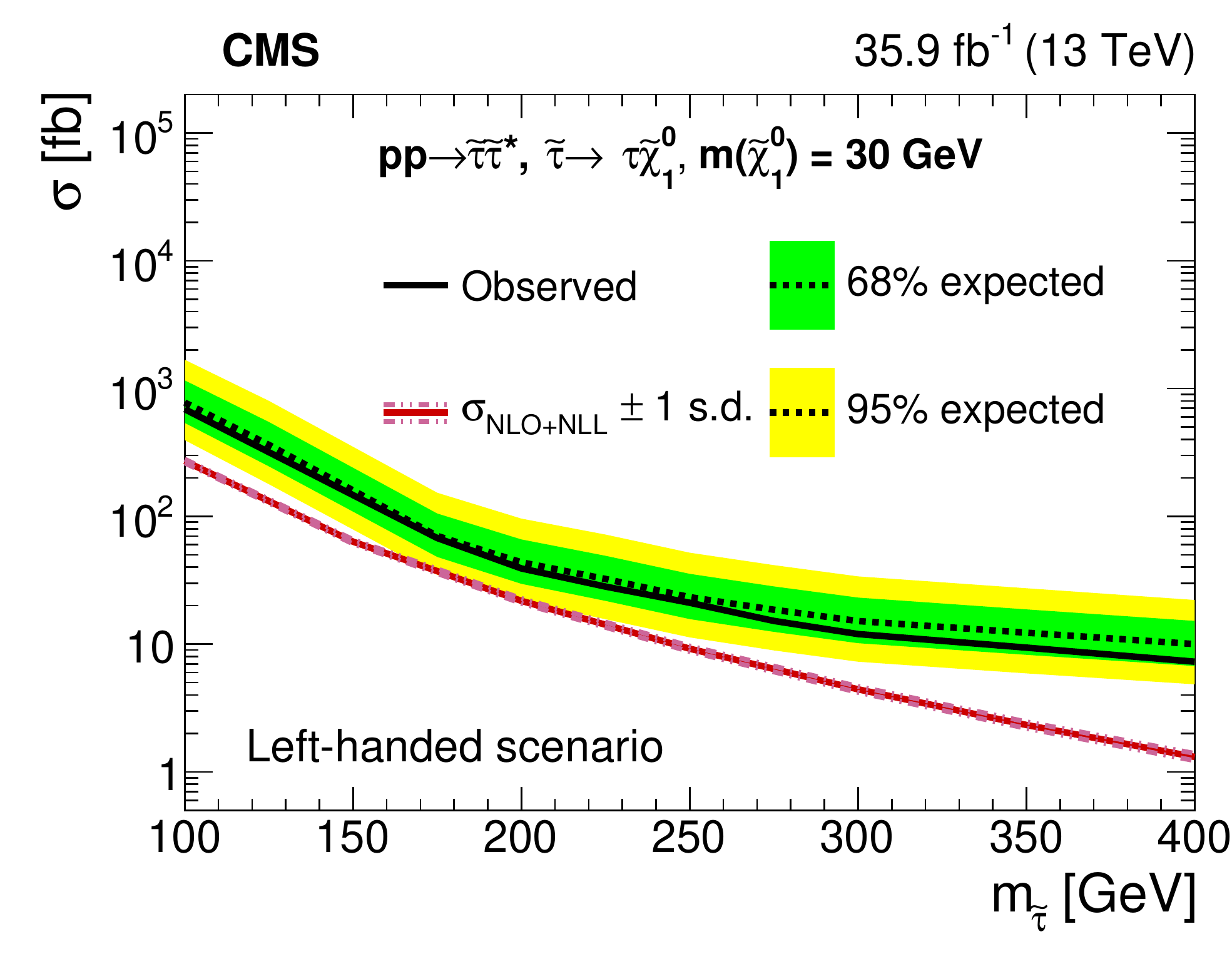}
\\
\includegraphics[width=0.48\textwidth]{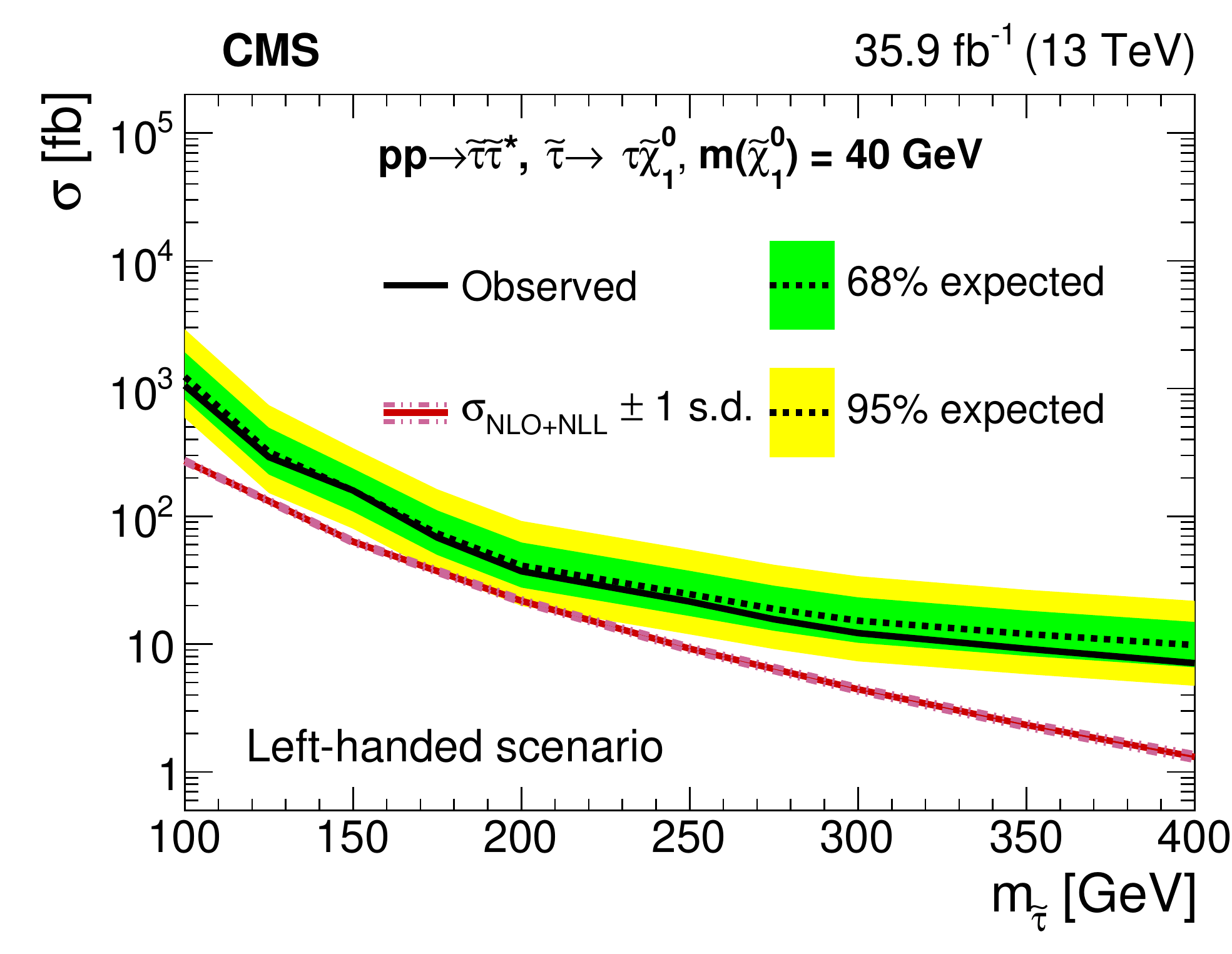}
\includegraphics[width=0.48\textwidth]{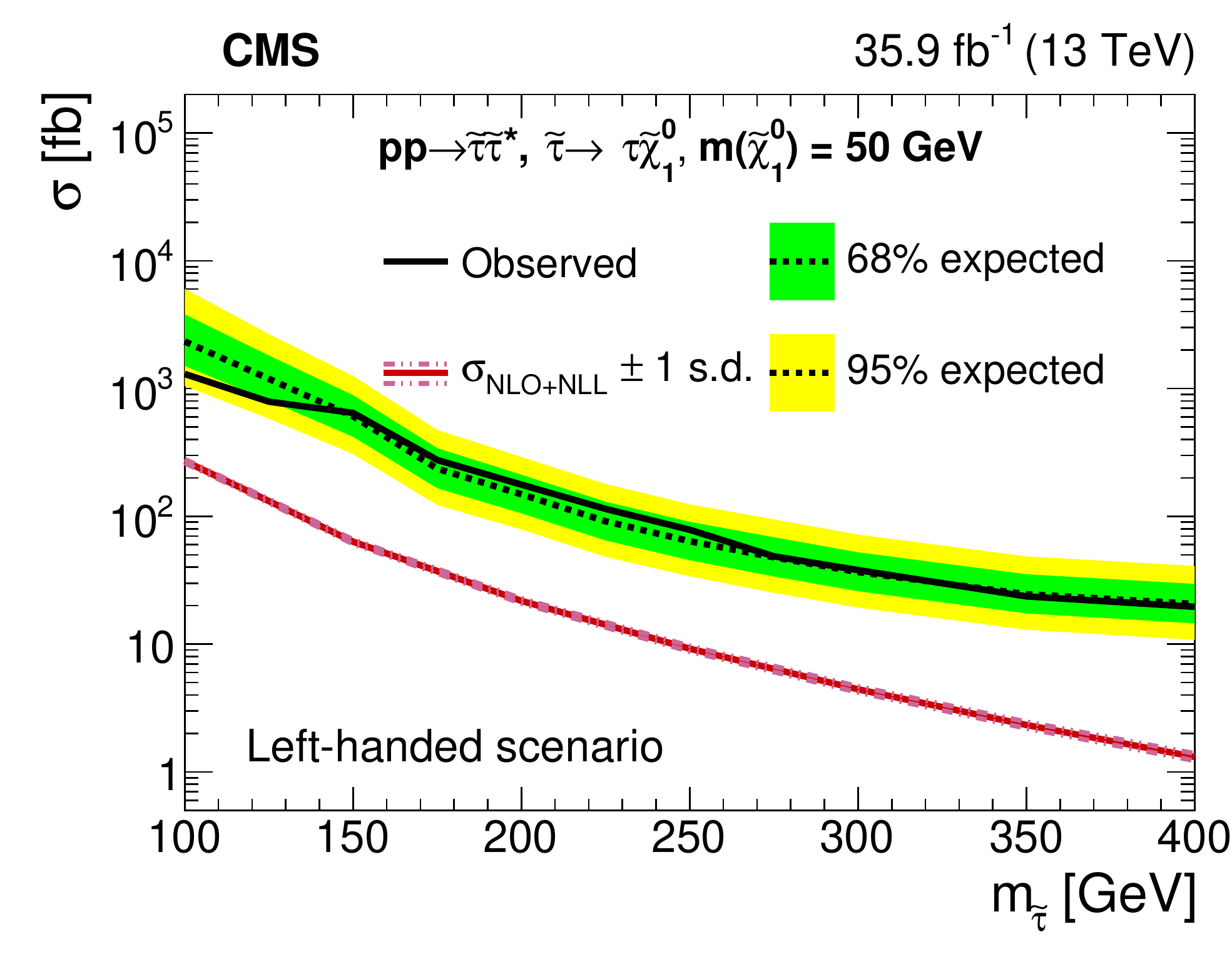}

\caption{\label{fig:ULs1D_left}Excluded \PSGt pair production cross section as a function of the \PSGt mass for the left-handed \PSGt scenario, and for different  \PSGczDo masses of   1, 10, 20, 30, 40, and 50\GeV from upper left to lower right, respectively. The inner (green) band and the outer (yellow) band indicate the regions containing 68 and 95\%, respectively, of the distribution of limits expected under the background-only hypothesis. The red line indicates the NLO+NLL prediction for the signal production cross section calculated with \textsc{Resummino}~\cite{Fuks:2013lya}, while the red hatched band represents the uncertainty in the prediction.}
\end{figure}

\begin{figure}[htbp]
\centering
\includegraphics[width=0.48\textwidth]{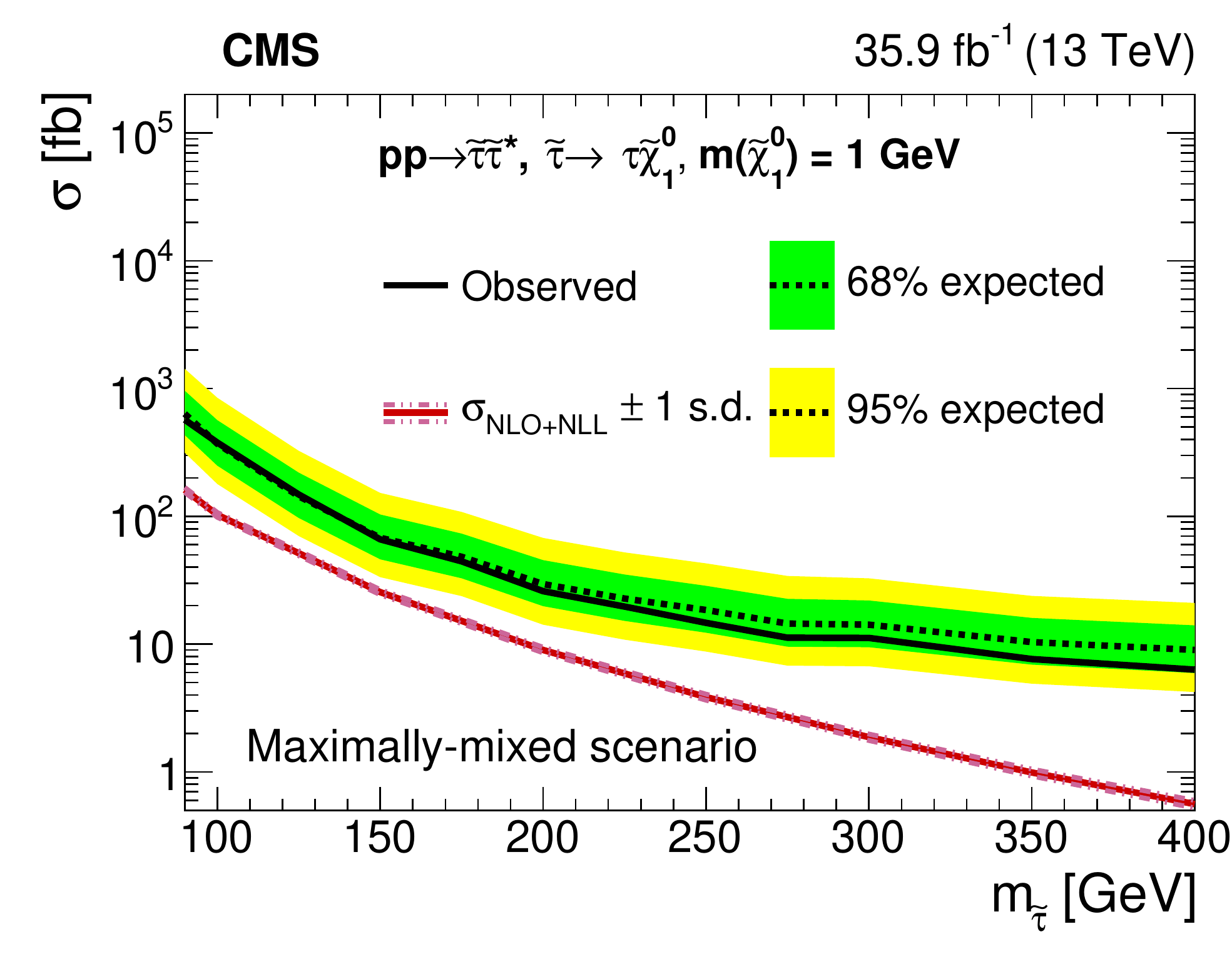}
\includegraphics[width=0.48\textwidth]{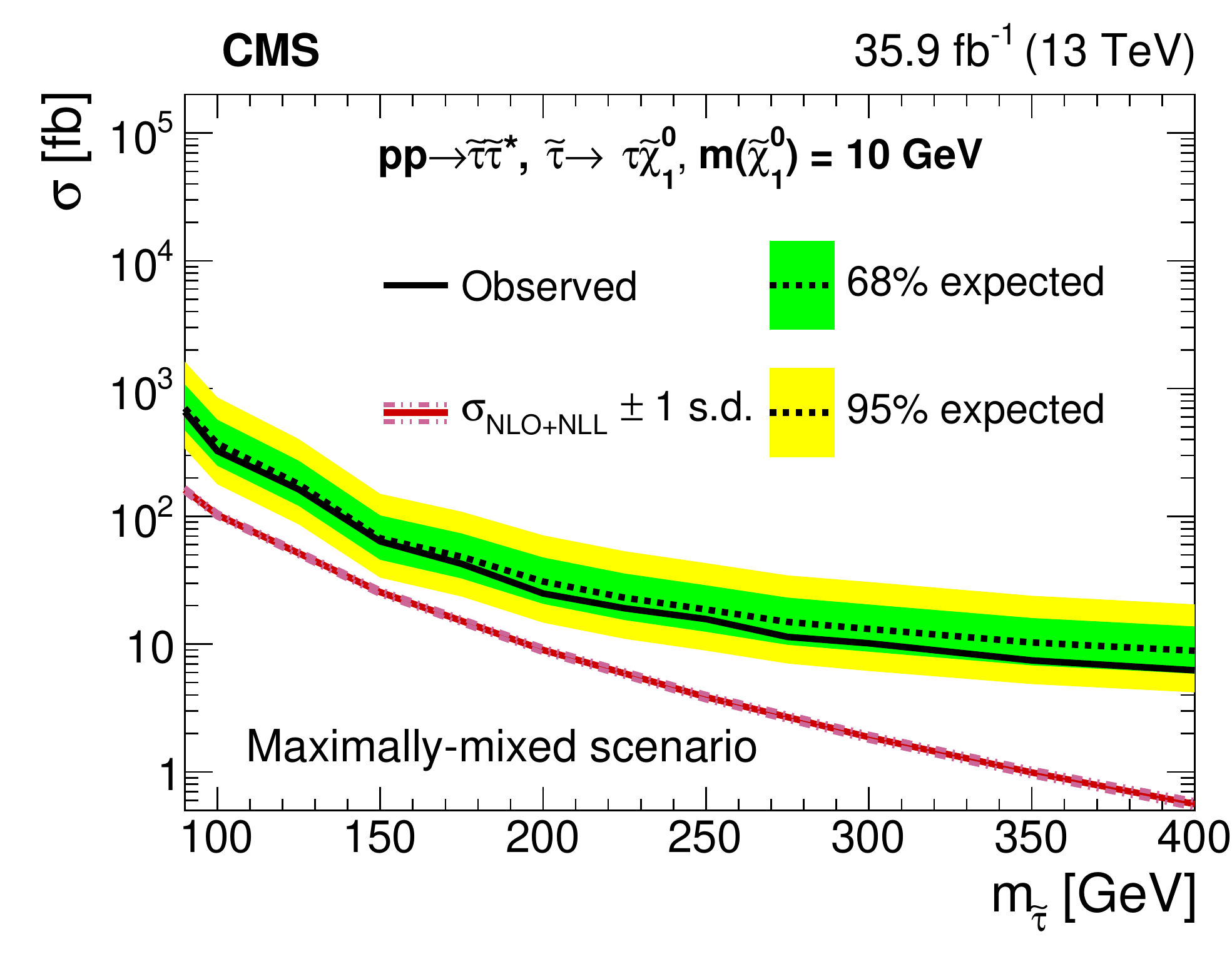}
\\
\includegraphics[width=0.48\textwidth]{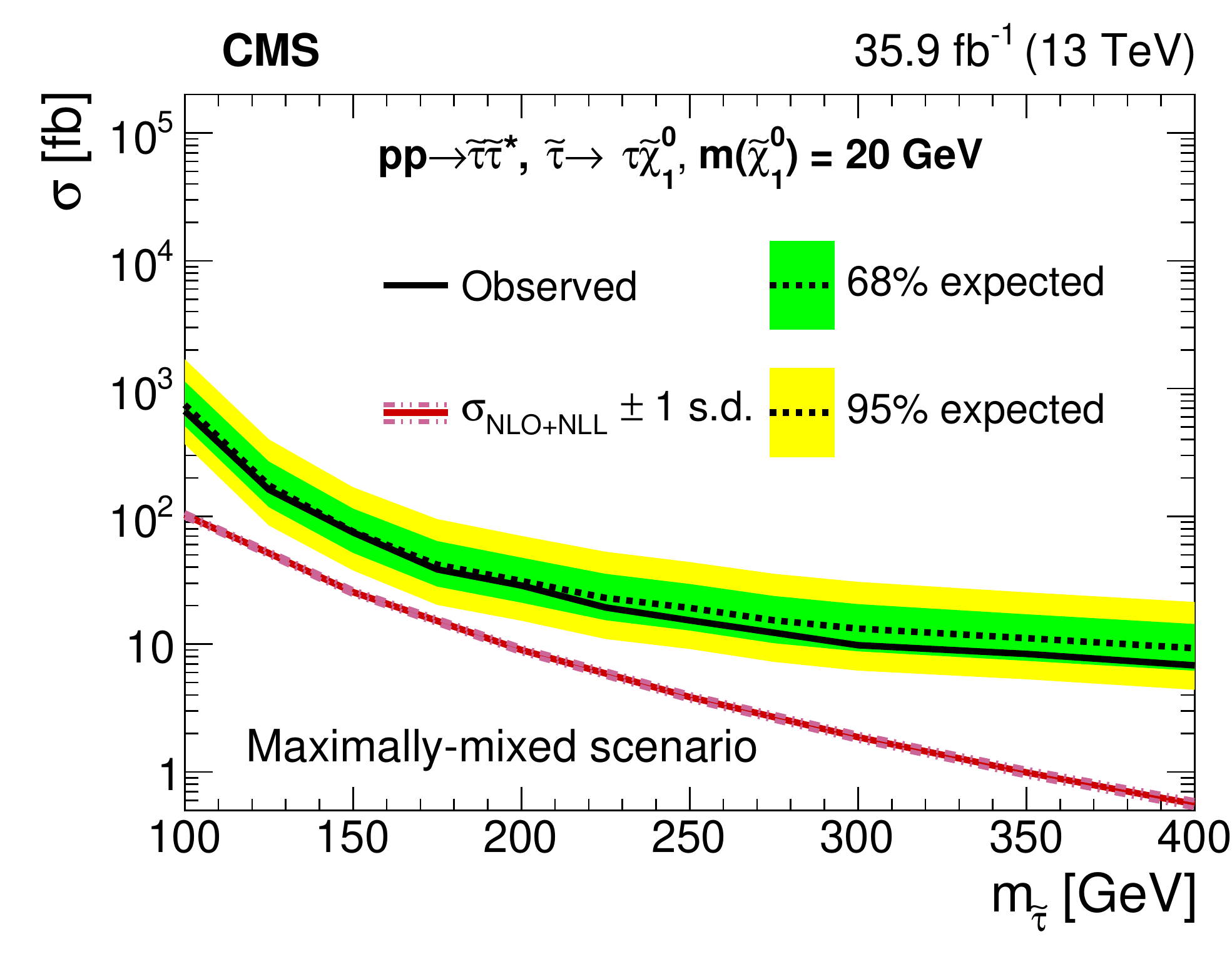}
\includegraphics[width=0.48\textwidth]{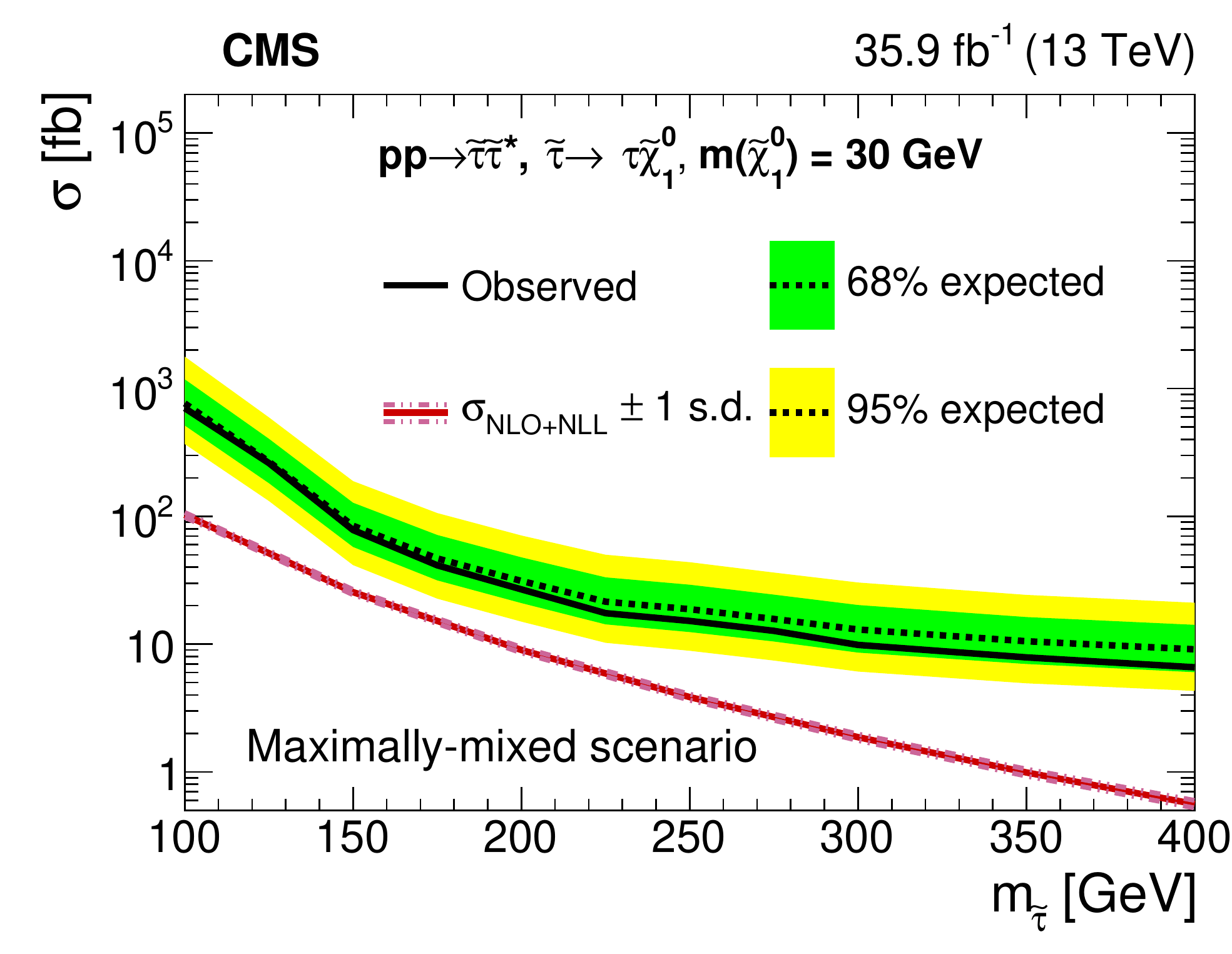}
\\
\includegraphics[width=0.48\textwidth]{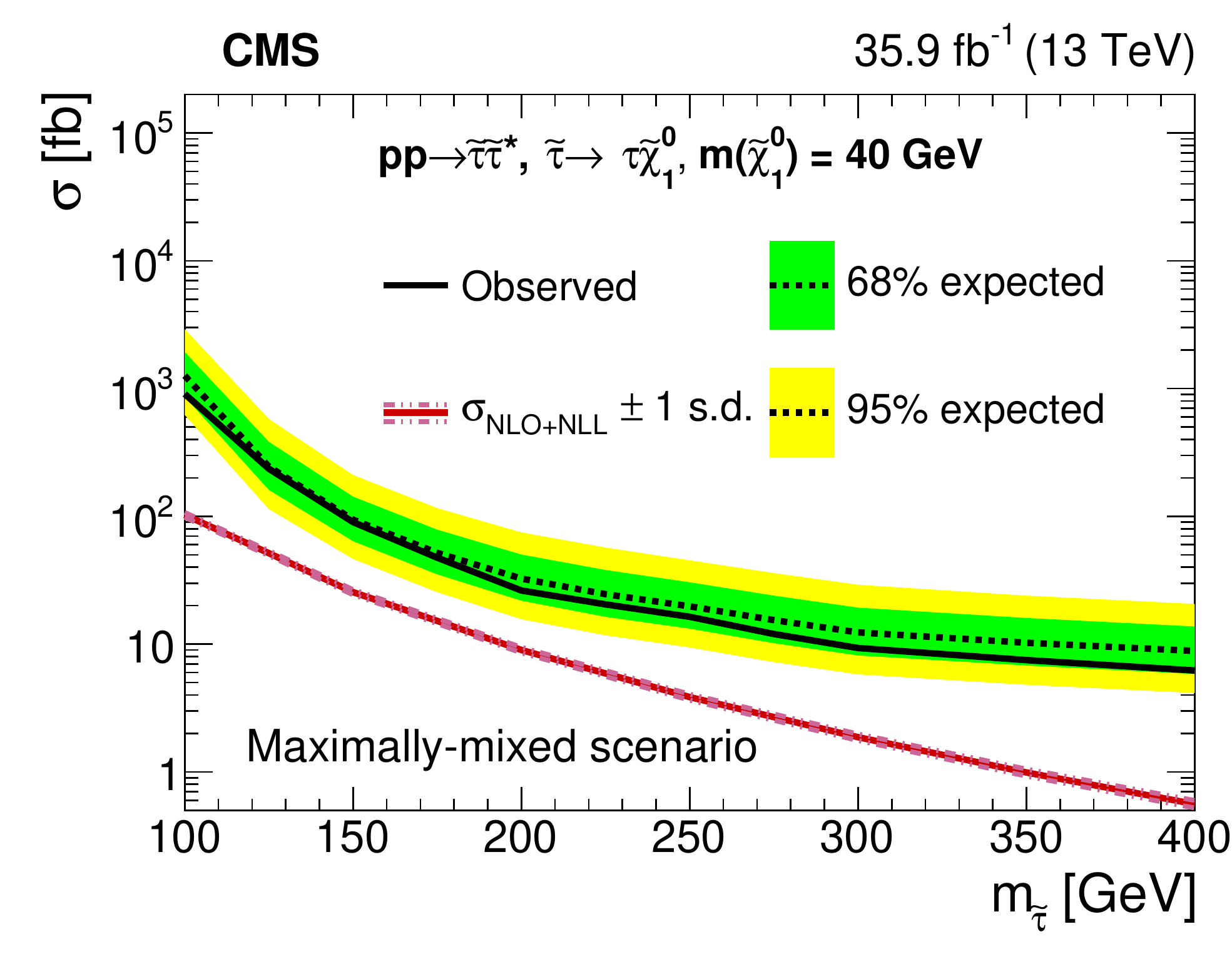}
\includegraphics[width=0.48\textwidth]{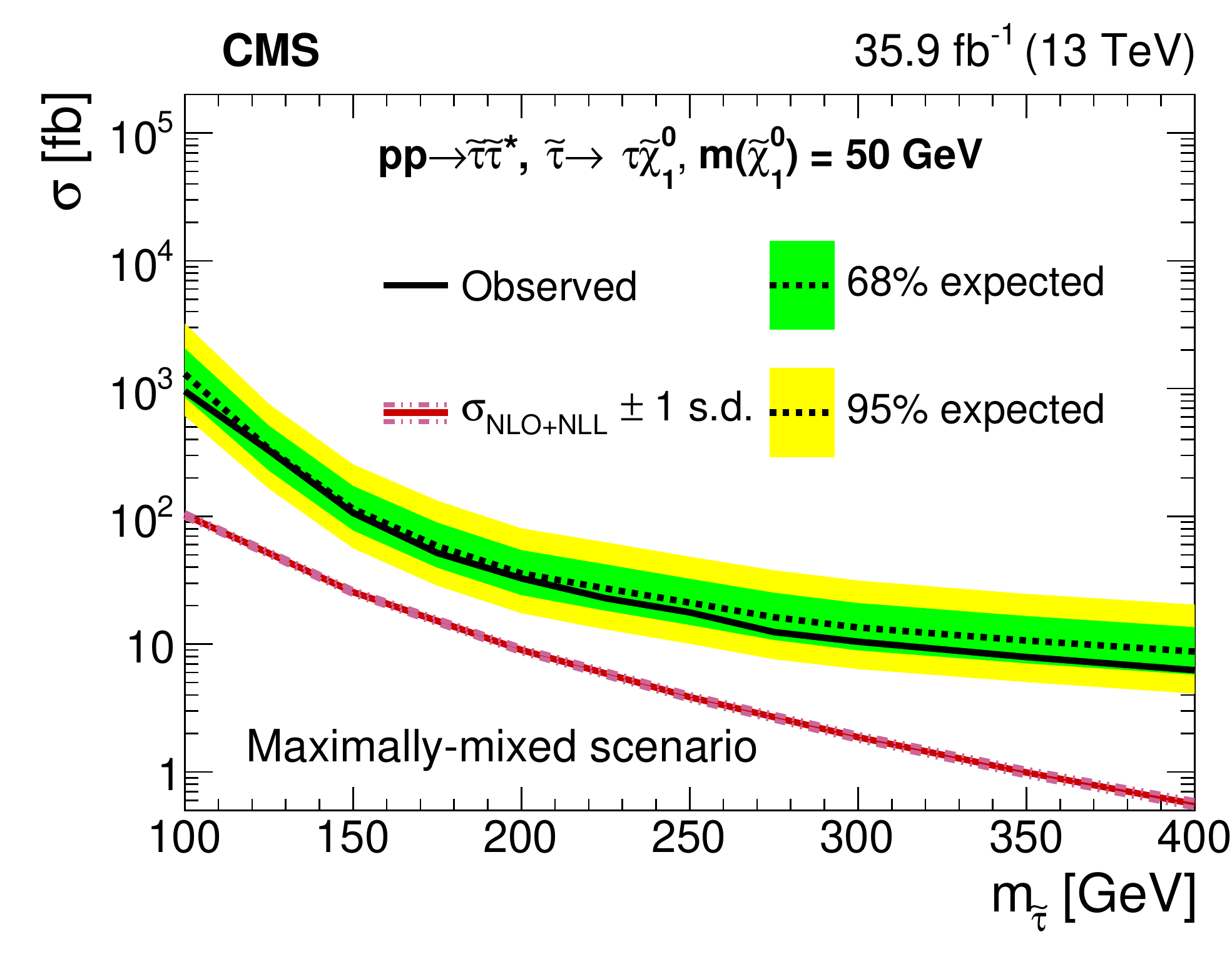}

\caption{\label{fig:ULs1D_max}Excluded \PSGt pair production cross section as a function of the \PSGt mass for the maximally-mixed \PSGt scenario, and for different  \PSGczDo masses of   1, 10, 20, 30, 40, and 50\GeV from upper left to lower right, respectively. The inner (green) band and the outer (yellow) band indicate the regions containing 68 and 95\%, respectively, of the distribution of limits expected under the background-only hypothesis. The red line indicates the NLO+NLL prediction for the signal production cross section calculated with \textsc{Resummino}~\cite{Fuks:2013lya}, while the red hatched band represents the uncertainty in the prediction.}
\end{figure}

\begin{figure}[htbp]
\centering
\includegraphics[width=0.48\textwidth]{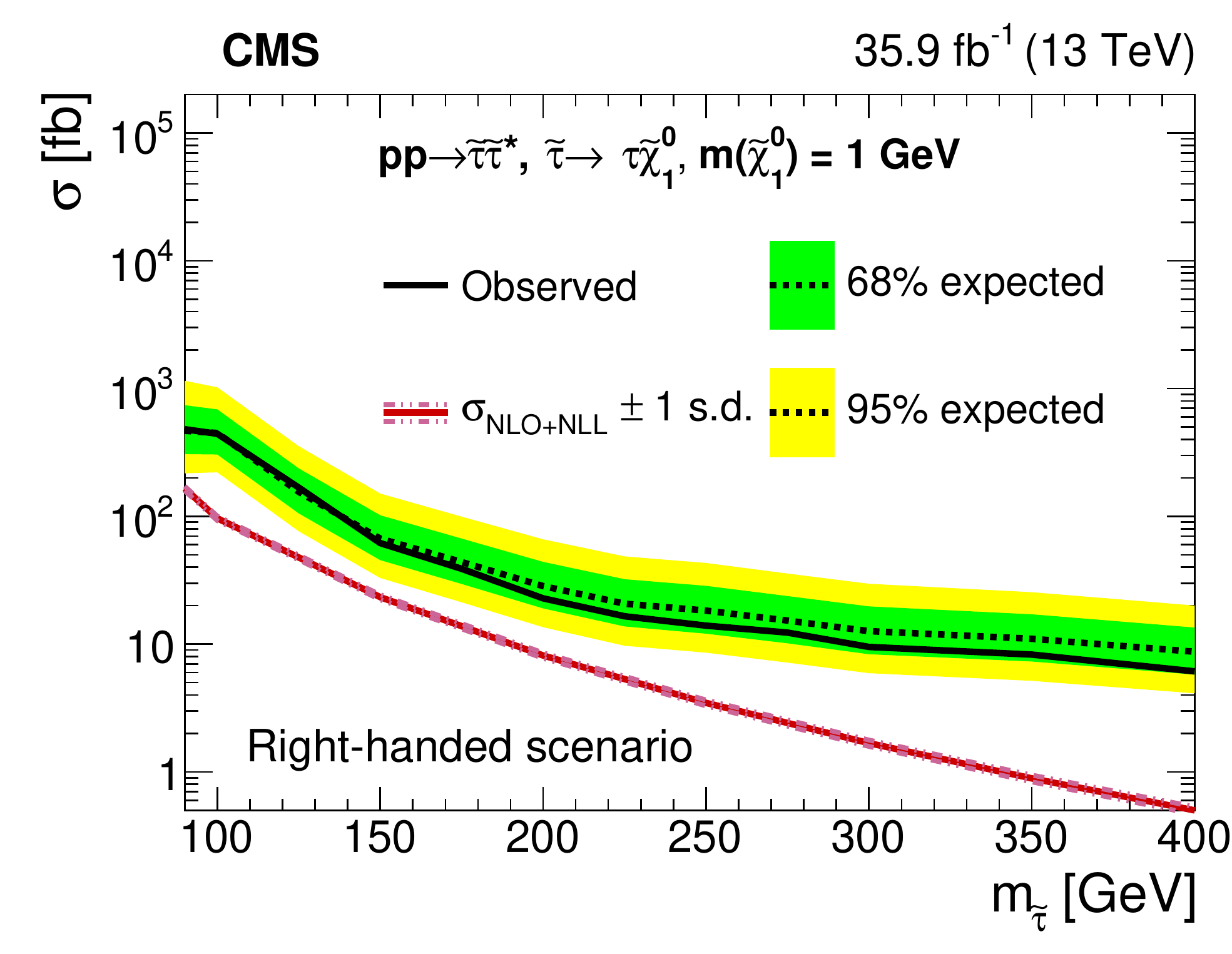}
\includegraphics[width=0.48\textwidth]{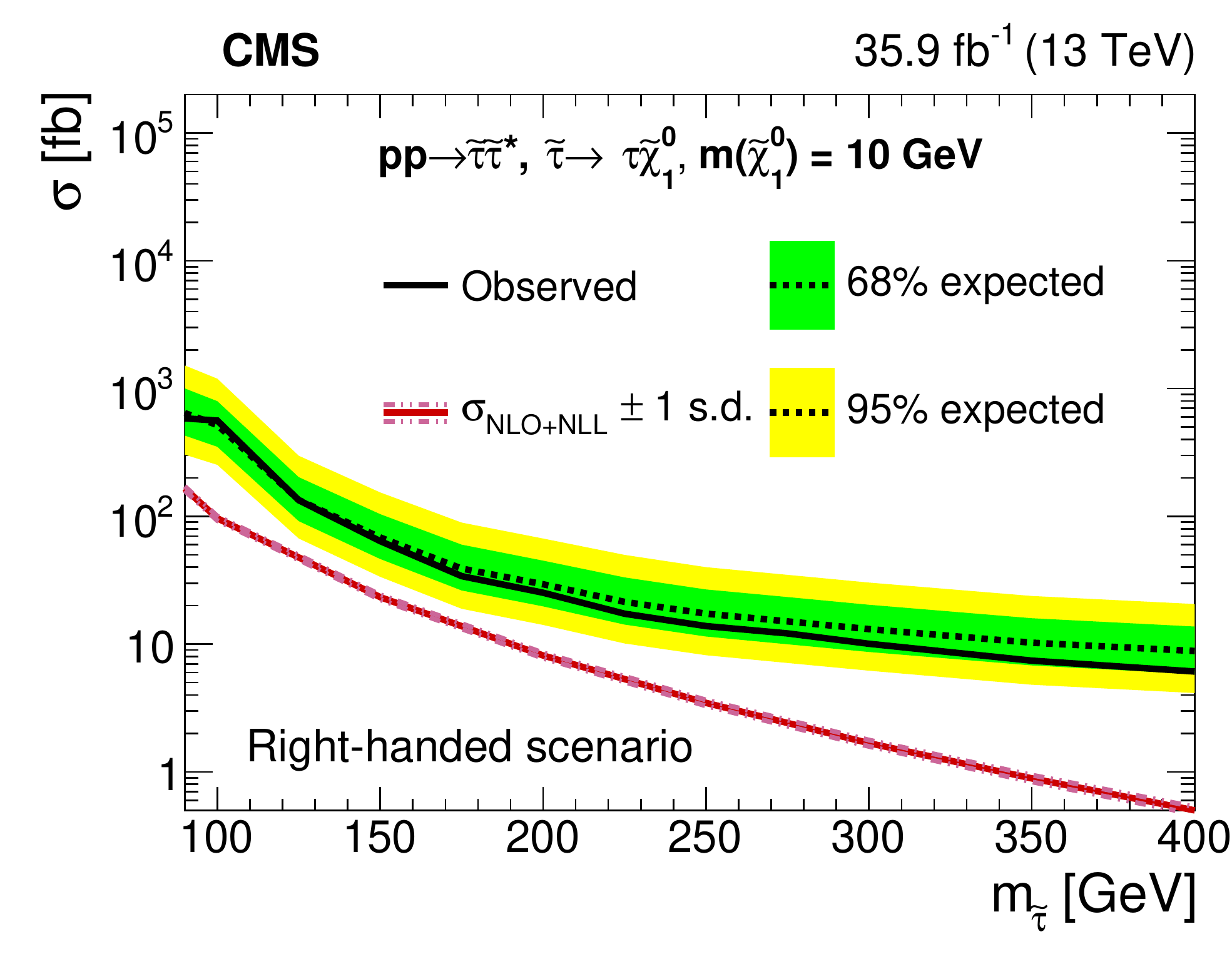}
\\
\includegraphics[width=0.48\textwidth]{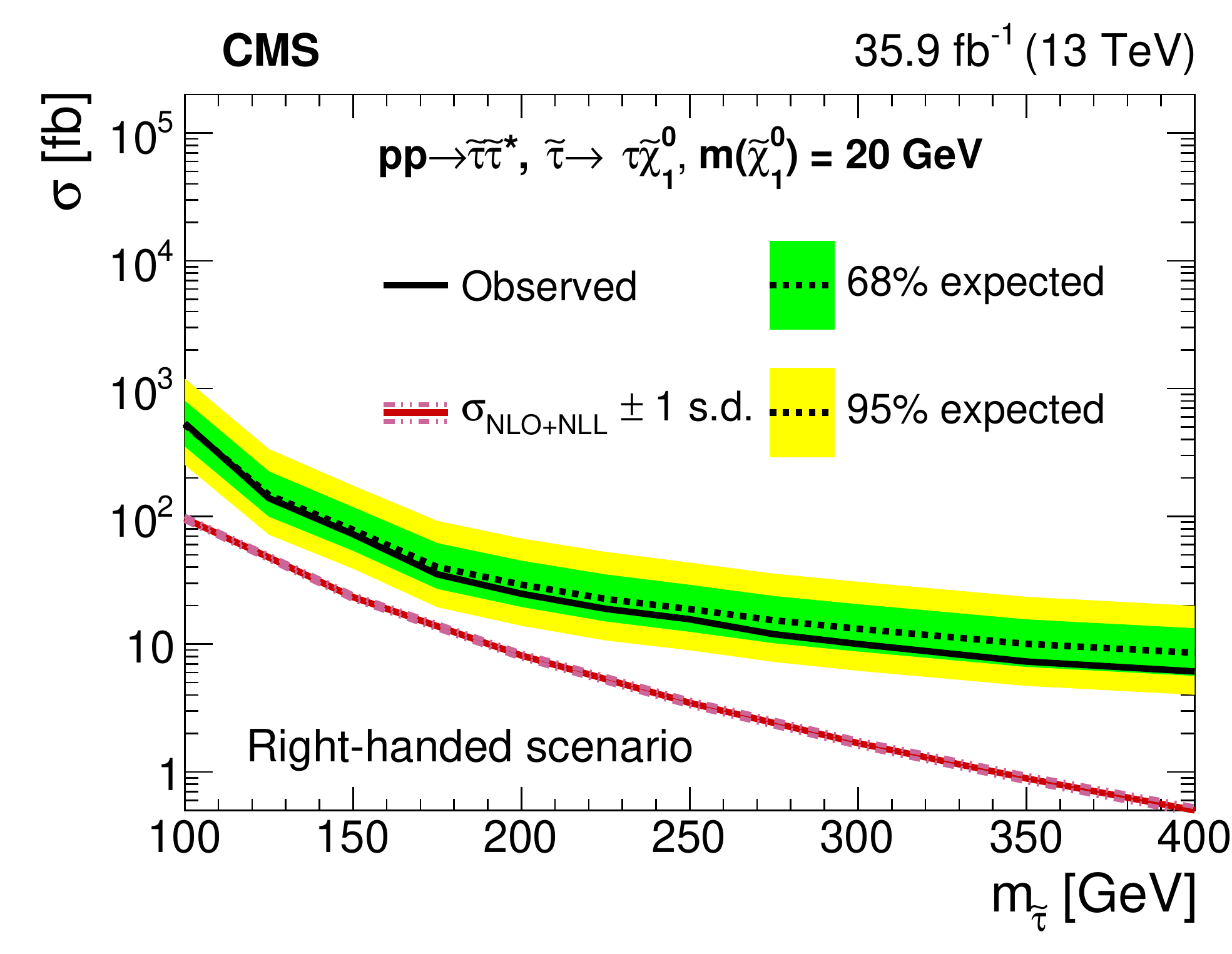}
\includegraphics[width=0.48\textwidth]{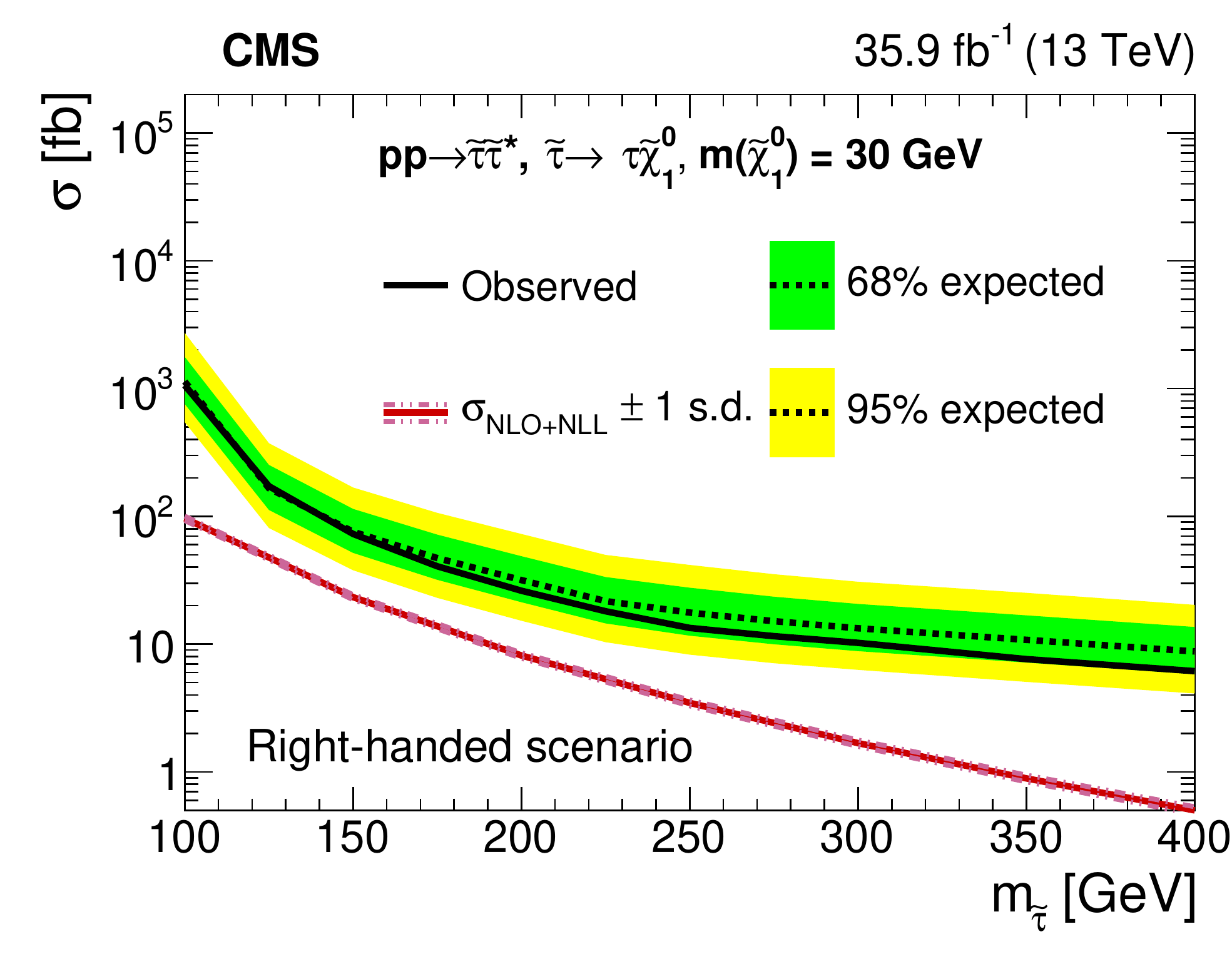}
\\
\includegraphics[width=0.48\textwidth]{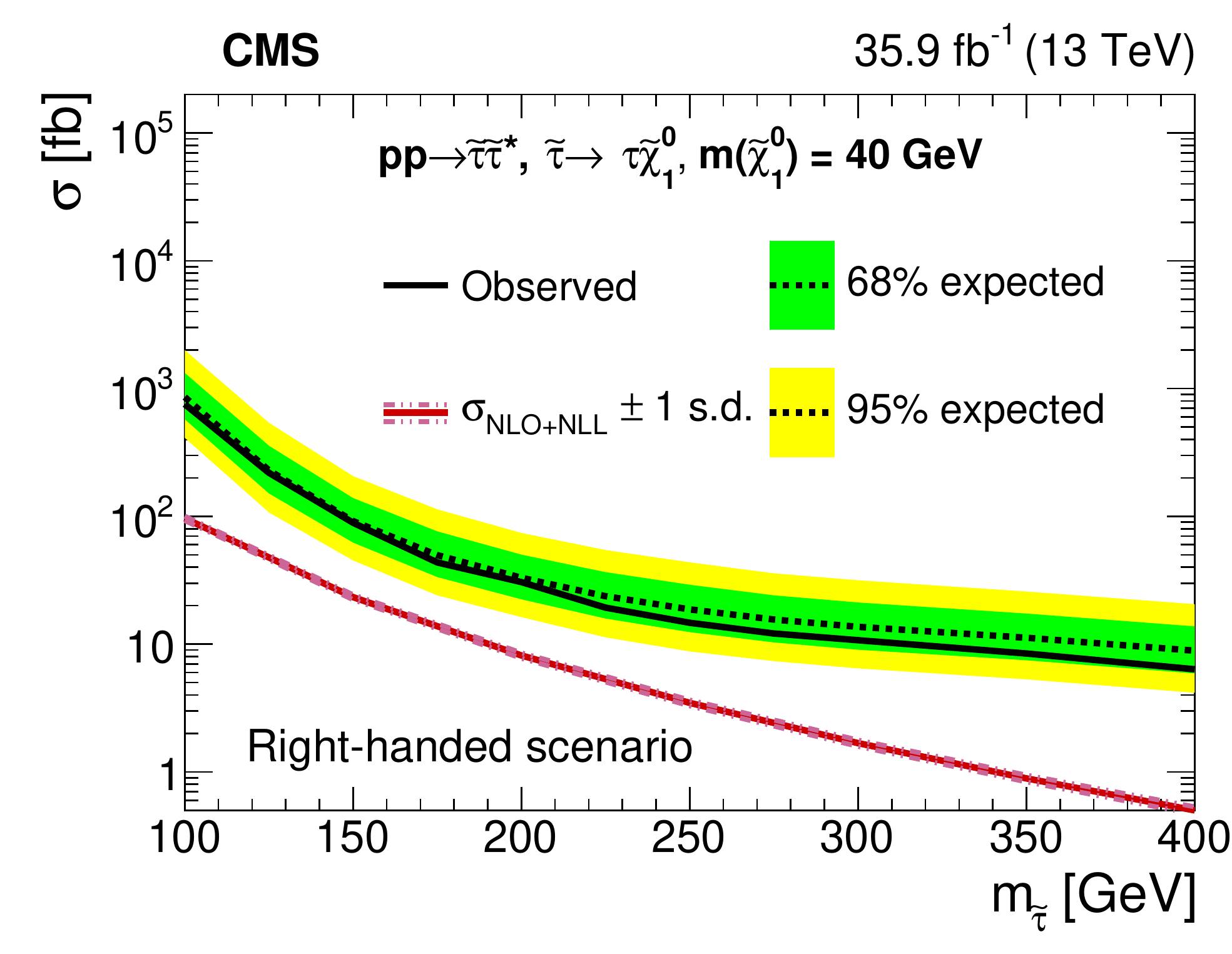}
\includegraphics[width=0.48\textwidth]{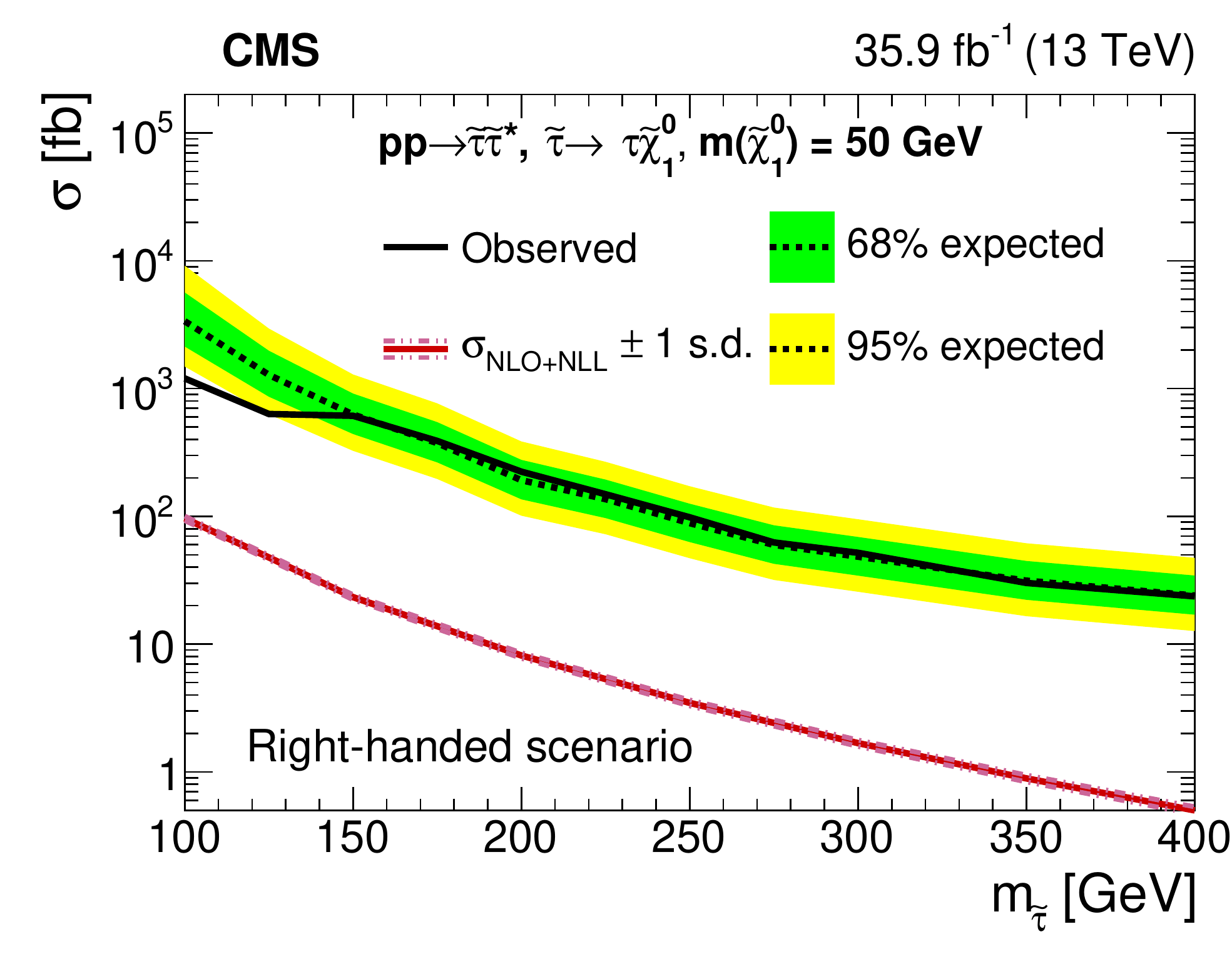}

\caption{\label{fig:ULs1D_right}Excluded \PSGt pair production cross section as a function of the \PSGt mass for the right-handed \PSGt scenario, and for different  \PSGczDo masses of   1, 10, 20, 30, 40, and 50\GeV from upper right to lower right, respectively. The inner (green) band and the outer (yellow) band indicate the regions containing 68 and 95\%, respectively, of the distribution of limits expected under the background-only hypothesis. The red line indicates the NLO+NLL prediction for the signal production cross section calculated with \textsc{Resummino}~\cite{Fuks:2013lya}, while the red hatched band represents the uncertainty in the prediction.}
\end{figure}

No significant deviations from the expected SM background are observed in this search. The results are interpreted as limits on the cross section for the production of \PSGt pairs in the context of simplified models.  The produced \PSGt is assumed to always decay to a \PGt lepton and a \PSGczDo.  The 95\% confidence level (\CL) upper limits on SUSY production cross sections are calculated using a modified frequentist approach  with the \CLs criterion~\cite{Junk:1999kv,Read:2002hq} and asymptotic approximation for the test statistic~\cite{CMS-NOTE-2011-005,Cowan:2010js}. Since the cross section of direct \PSGt pair production and the \PGt lepton decay are strongly dependent on chirality, the results are shown for three different scenarios.   Figures~\ref{fig:ULs1D_left}-\ref{fig:ULs1D_right} show the cross section upper limits obtained for \distau production for the left-handed, maximally mixed, and right-handed scenarios as a function of the \PSGt mass for different \PSGczDo mass hypotheses, namely 1, 10, 20, 30, 40, and 50\GeV.  It can be seen that the constraints are reduced for higher \PSGczDo masses due to the smaller experimental acceptance. The stronger than expected limits observed at low \PSGt mass values for a \PSGczDo mass of 50\GeV in the purely left- and right-handed scenarios are driven by a deficit in the \mutau final state in the 0--jet category, leading to strong constraints on the predicted background contribution in SRs sensitive to these signal models. The extremely small \distau production cross sections make this scenario in general very challenging. This analysis is most sensitive to scenarios with a left-handed \PSGt and a nearly massless \PSGczDo, in which we exclude production rates larger than 1.26 (1.34) times the expected SUSY cross section for a \PSGt mass of 90 (125)\GeV.

\begin{figure}[htbp]
\centering
\includegraphics[width=.7\textwidth]{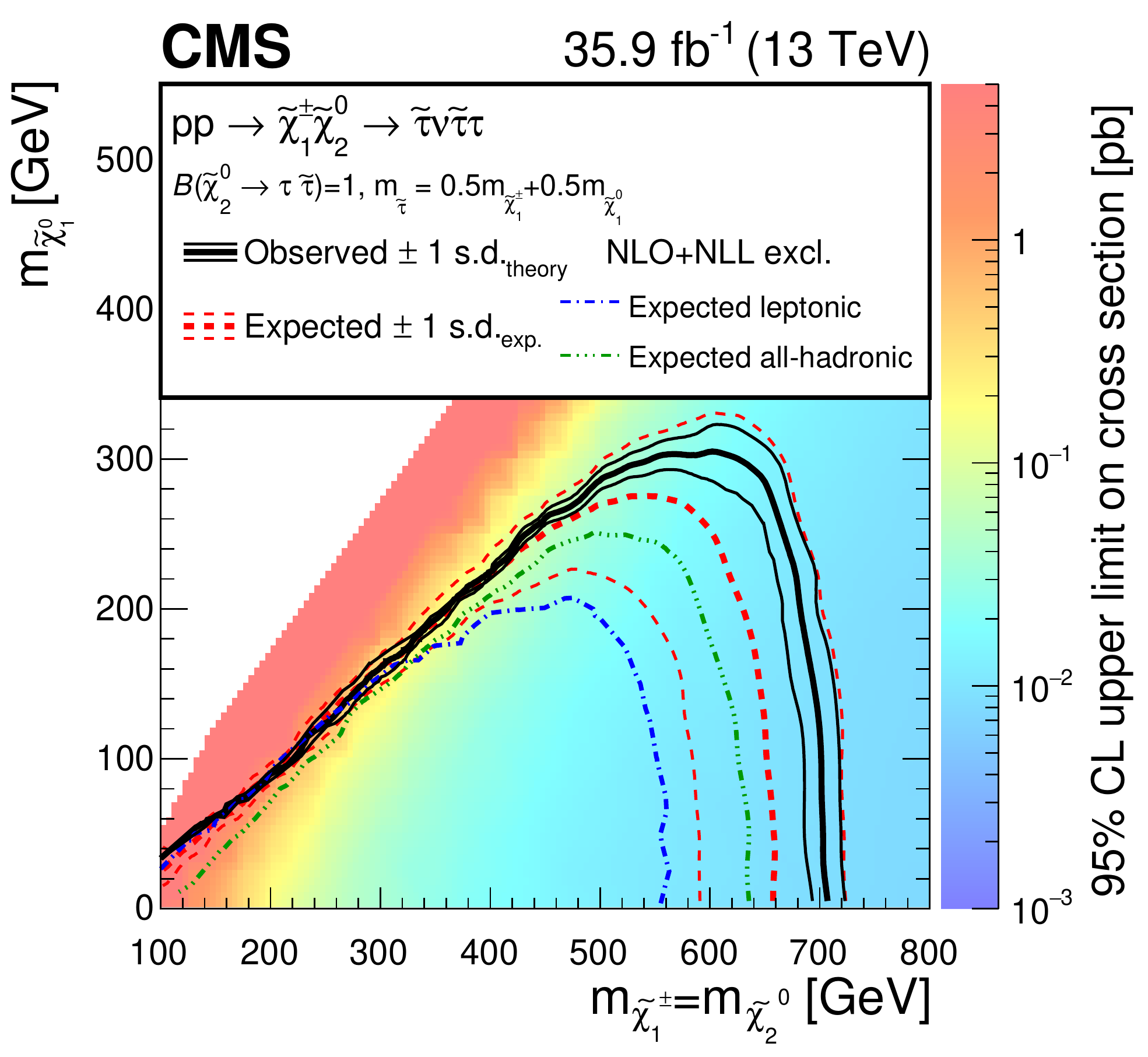}
\caption{\label{fig:c1n2limit} Exclusion limits at 95\% \CL for chargino-neutralino production with decays through \PSGt to final states with \PGt leptons. The production cross sections are computed at NLO+NLL precision assuming mass-degenerate wino \cone and \ntwo, light bino \PSGczDo, and with all the other sparticles assumed to be heavy and decoupled~\cite{Fuks:2012qx, Fuks:2013vua}. The regions enclosed by the thick black curves represent the observed exclusion at 95\% \CL, while the thick dashed red line indicates the expected exclusion at 95\% \CL. The thin black lines show the effect of variations of the signal cross sections within theoretical uncertainties on the observed exclusion. The thin red dashed lines indicate the region containing 68\% of the distribution of limits expected under the background-only hypothesis. The green and blue dashed lines show separately the expected exclusion regions for the analyses in the all-hadronic and leptonic final states, respectively.}
\end{figure}

\begin{figure}[htbp]
\centering
\includegraphics[width=.7\textwidth]{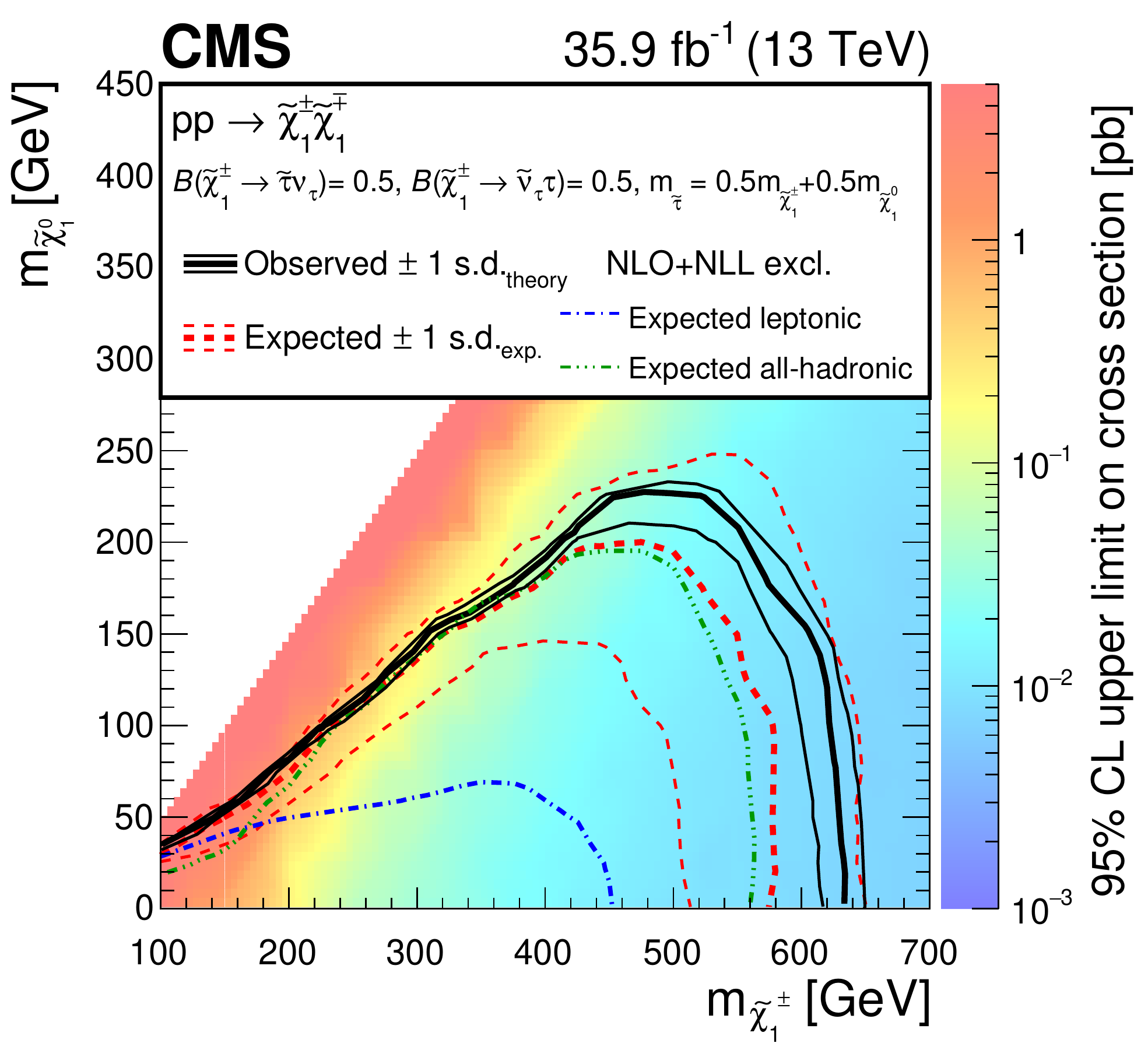}
\caption{\label{fig:c1c1limit} Exclusion limits at 95\% \CL for chargino pair production with decays through \PSGt to final states with \PGt leptons. The production cross sections are computed at NLO+NLL precision assuming a wino-like \cone, light bino \PSGczDo, and with all the other sparticles assumed to be heavy and decoupled~\cite{Fuks:2012qx, Fuks:2013vua}. The regions enclosed by the thick black curves represent the observed exclusion at 95\% \CL, while the thick dashed red line indicates the expected exclusion at 95\% \CL. The thin black lines show the effect of variations of the signal cross sections within theoretical uncertainties on the observed exclusion. The thin red dashed lines indicate the region containing 68\% of the distribution of limits expected under the background-only hypothesis. The green and blue dashed lines show separately the expected exclusion regions for the analyses in the all-hadronic and leptonic final states, respectively.}
\end{figure}

We also interpret the results as exclusion limits in simplified models of mass-degenerate chargino-neutralino (\cone\ntwo) and chargino pair (\cone\conemp) production with decays to \PGt leptons in the final state via the decay chains $\cone \to \PSGt \PGnGt / \tausneu \PGt \to \PGt \PGnGt \PSGczDo$, $\ntwo \to \PGt \PSGt \to \PGt \PGt \PSGczDo$. Equal branching fractions are assumed for each of the two possible \cone decay chains considered. The \PSGt and \tausneu masses are assumed to be degenerate in these models and to have a value halfway between the mass of the parent sparticles and the \PSGczDo mass. Figure~\ref{fig:c1n2limit} shows the 95\% \CL exclusion limits in the mass plane of \cone/\ntwo versus \PSGczDo mass obtained for the \cone\ntwo scenario. We exclude \cone/\ntwo masses up to around 710\GeV for a nearly massless \PSGczDo hypothesis in this scenario. Figure~\ref{fig:c1c1limit} shows the corresponding limits for the \cone\conemp signal scenario in the plane of \cone versus \PSGczDo mass. In this scenario, we exclude \cone masses up to around 630\GeV for a nearly massless \PSGczDo hypothesis.

In order to simplify the reinterpretation of the results obtained in the leptonic final states using other signal models, we define a small set of aggregate SRs by combining subsets of the SRs. These aggregate SRs are chosen to have sensitivity to a range of signal models. Since they are not exclusive, the results obtained for these aggregate SRs cannot be statistically combined. These results are tabulated in Table~\ref{tab:aggregated}.

\begin{table}[htbp]
\centering
\topcaption{Definition of the aggregate SRs to be used for easier reinterpretation of the results in the \etau, \mutau, and \emu final states. In all of these regions, a selection requirement of $\DZ>-500\GeV$ is applied.}
\label{tab:aggregated}
\begin{tabular}{cccccc}
\hline
Channel & \njet & \ptmiss [\GeVns{}] & \mttwo [\GeVns{}] & Background & Observed \\\hline
\etau & 0 & ${>}120$ & ${>}100$ & $10.8 \pm 2.1$  $\pm 2.5$ & 9 \\
\etau & 1 & ${>}120$ & ${>}120$ & $4.9 \pm 1.5$ $\pm 1.9$ & 4 \\
\etau & 1 & ${>}250$ &  ${>}80$ & $1.6 \pm 0.9$ $\pm 1.2$ & 0 \\
\mutau & 0 & ${>}120$ & ${>}100$ & $14.4 \pm 2.5$ $\pm  3.1$ & 14 \\
\mutau & 1 & ${>}120$ & ${>}120$ & $5.8 \pm 1.8$ $\pm 2.7$ & 7 \\
\mutau & 1 & ${>}250$ &  ${>}80$ & $1.5 \pm 0.9$ $\pm 1.1$ & 1 \\
\emu & 0 & ${>}120$ & ${>}100$ & $9.7 \pm 2.4$ $\pm 3.0$ & 6 \\
\emu & 1 & ${>}120$ & ${>}120$ & $6.8 \pm 2.2$  $\pm 2.7$ & 6 \\
\emu & 1 & ${>}250$ &  ${>}80$ & $3.3 \pm 2.0$ $\pm 2.3$ & 1 \\
\hline
\end{tabular}
\end{table}

\clearpage

\section {Summary}
\label{sec:summary}
A search for the direct and indirect production of \PGt sleptons has been performed in proton-proton collisions at a center-of-mass energy of 13\TeV in events with a \PGt lepton pair and significant missing transverse momentum in the final state. Both leptonic and hadronic decay modes of the \PGt leptons are considered. Search regions are defined using discriminating kinematic observables that exploit expected differences between signal and background. The data sample used for this search corresponds to an integrated luminosity of 35.9\fbinv. No excess above the expected standard model background has been observed. Upper limits on the cross section of direct \PSGt pair production are derived for simplified models in which each \PSGt decays to a \PGt lepton and the lightest neutralino, with the latter being assumed to be the lightest supersymmetric particle (LSP). The analysis is most sensitive to a \PSGt that is purely left-handed.  For a left-handed \PSGt of 90\GeV decaying to a nearly massless LSP, the observed limit is 1.26 times the expected production cross section in the simplified model. The limits obtained for direct \PSGt pair production represent a considerable improvement in sensitivity for this production mechanism with respect to previous ATLAS and CMS measurements. Exclusion limits are also derived for simplified models of chargino-neutralino and chargino pair production with decays to \PGt leptons that involve indirect \PSGt production via the chargino and neutralino decay chains. In the chargino-neutralino production model, in which the parent chargino and second-lightest neutralino are assumed to have the same mass, we exclude chargino masses up to 710\GeV under the hypothesis of a nearly massless LSP. In the chargino pair production model, we exclude chargino masses up to 630\GeV under the same hypothesis. In both cases, we significantly extend the exclusion limits with respect to previous CMS measurements.

\begin{acknowledgments}
\hyphenation{Bundes-ministerium Forschungs-gemeinschaft Forschungs-zentren Rachada-pisek} We congratulate our colleagues in the CERN accelerator departments for the excellent performance of the LHC and thank the technical and administrative staffs at CERN and at other CMS institutes for their contributions to the success of the CMS effort. In addition, we gratefully acknowledge the computing centres and personnel of the Worldwide LHC Computing Grid for delivering so effectively the computing infrastructure essential to our analyses. Finally, we acknowledge the enduring support for the construction and operation of the LHC and the CMS detector provided by the following funding agencies: the Austrian Federal Ministry of Science, Research and Economy and the Austrian Science Fund; the Belgian Fonds de la Recherche Scientifique, and Fonds voor Wetenschappelijk Onderzoek; the Brazilian Funding Agencies (CNPq, CAPES, FAPERJ, FAPERGS, and FAPESP); the Bulgarian Ministry of Education and Science; CERN; the Chinese Academy of Sciences, Ministry of Science and Technology, and National Natural Science Foundation of China; the Colombian Funding Agency (COLCIENCIAS); the Croatian Ministry of Science, Education and Sport, and the Croatian Science Foundation; the Research Promotion Foundation, Cyprus; the Secretariat for Higher Education, Science, Technology and Innovation, Ecuador; the Ministry of Education and Research, Estonian Research Council via IUT23-4 and IUT23-6 and European Regional Development Fund, Estonia; the Academy of Finland, Finnish Ministry of Education and Culture, and Helsinki Institute of Physics; the Institut National de Physique Nucl\'eaire et de Physique des Particules~/~CNRS, and Commissariat \`a l'\'Energie Atomique et aux \'Energies Alternatives~/~CEA, France; the Bundesministerium f\"ur Bildung und Forschung, Deutsche Forschungsgemeinschaft, and Helmholtz-Gemeinschaft Deutscher Forschungszentren, Germany; the General Secretariat for Research and Technology, Greece; the National Research, Development and Innovation Fund, Hungary; the Department of Atomic Energy and the Department of Science and Technology, India; the Institute for Studies in Theoretical Physics and Mathematics, Iran; the Science Foundation, Ireland; the Istituto Nazionale di Fisica Nucleare, Italy; the Ministry of Science, ICT and Future Planning, and National Research Foundation (NRF), Republic of Korea; the Ministry of Education and Science of the Republic of Latvia; the Lithuanian Academy of Sciences; the Ministry of Education, and University of Malaya (Malaysia); the Ministry of Science of Montenegro; the Mexican Funding Agencies (BUAP, CINVESTAV, CONACYT, LNS, SEP, and UASLP-FAI); the Ministry of Business, Innovation and Employment, New Zealand; the Pakistan Atomic Energy Commission; the Ministry of Science and Higher Education and the National Science Centre, Poland; the Funda\c{c}\~ao para a Ci\^encia e a Tecnologia, Portugal; JINR, Dubna; the Ministry of Education and Science of the Russian Federation, the Federal Agency of Atomic Energy of the Russian Federation, Russian Academy of Sciences, the Russian Foundation for Basic Research, and the National Research Center ``Kurchatov Institute"; the Ministry of Education, Science and Technological Development of Serbia; the Secretar\'{\i}a de Estado de Investigaci\'on, Desarrollo e Innovaci\'on, Programa Consolider-Ingenio 2010, Plan Estatal de Investigaci\'on Cient\'{\i}fica y T\'ecnica y de Innovaci\'on 2013-2016, Plan de Ciencia, Tecnolog\'{i}a e Innovaci\'on 2013-2017 del Principado de Asturias, and Fondo Europeo de Desarrollo Regional, Spain; the Ministry of Science, Technology and Research, Sri Lanka; the Swiss Funding Agencies (ETH Board, ETH Zurich, PSI, SNF, UniZH, Canton Zurich, and SER); the Ministry of Science and Technology, Taipei; the Thailand Center of Excellence in Physics, the Institute for the Promotion of Teaching Science and Technology of Thailand, Special Task Force for Activating Research and the National Science and Technology Development Agency of Thailand; the Scientific and Technical Research Council of Turkey, and Turkish Atomic Energy Authority; the National Academy of Sciences of Ukraine, and State Fund for Fundamental Researches, Ukraine; the Science and Technology Facilities Council, UK; the US Department of Energy, and the US National Science Foundation.

Individuals have received support from the Marie-Curie programme and the European Research Council and Horizon 2020 Grant, contract No. 675440 (European Union); the Leventis Foundation; the A. P. Sloan Foundation; the Alexander von Humboldt Foundation; the Belgian Federal Science Policy Office; the Fonds pour la Formation \`a la Recherche dans l'Industrie et dans l'Agriculture (FRIA-Belgium); the Agentschap voor Innovatie door Wetenschap en Technologie (IWT-Belgium); the F.R.S.-FNRS and FWO (Belgium) under the ``Excellence of Science - EOS" - be.h project n. 30820817; the Ministry of Education, Youth and Sports (MEYS) of the Czech Republic; the Lend\"ulet (``Momentum") Programme and the J\'anos Bolyai Research Scholarship of the Hungarian Academy of Sciences, the New National Excellence Program \'UNKP, the NKFIA research grants 123842, 123959, 124845, 124850 and 125105 (Hungary); the Council of Scientific and Industrial Research, India; the HOMING PLUS programme of the Foundation for Polish Science, cofinanced from European Union, Regional Development Fund, the Mobility Plus programme of the Ministry of Science and Higher Education, the National Science Center (Poland), contracts Harmonia 2014/14/M/ST2/00428, Opus 2014/13/B/ST2/02543, 2014/15/B/ST2/03998, and 2015/19/B/ST2/02861, Sonata-bis 2012/07/E/ST2/01406; the National Priorities Research Program by Qatar National Research Fund; the Programa de Excelencia Mar\'{i}a de Maeztu, and the Programa Severo Ochoa del Principado de Asturias; the Thalis and Aristeia programmes cofinanced by EU-ESF, and the Greek NSRF; the Rachadapisek Sompot Fund for Postdoctoral Fellowship, Chulalongkorn University, and the Chulalongkorn Academic into Its 2nd Century Project Advancement Project (Thailand); the Welch Foundation, contract C-1845; and the Weston Havens Foundation (USA).
\end{acknowledgments}

\newpage

\bibliography{auto_generated}

\clearpage
\appendix
\section{Event yields for the analysis in the \texorpdfstring{\etau, \mutau, and \emu}{e-tau, mu-tau, and e-mu} final states}
\label{app:yieldtables}

\begin{sidewaystable}
\topcaption{Numbers of expected and observed events in the \etau channel. The total background includes the total uncertainty, while for each process the statistical and systematic uncertainties are quoted separately. The two numbers that are quoted for the benchmark signal models are the masses of the parent SUSY particle and the \PSGczDo, respectively, in GeV. In the case of the chargino-neutralino signal models, the first number within parentheses indicates the common \cone and \ntwo mass in GeV.} \label{tab:res_elehad}
\centering
\cmsTable{
\begin{tabular}{l|c|c|c|c|c|c|c|c|c|c|c}  \hline 
SR label &   \ttbar     & DY+jets &WW+jets &WW+jets &  Rest  & Jet$\rightarrow \tau_h$  & Total Bkg    & \cone\ntwo(400,1) &\cone\ntwo(400,175) &$\PSGt_L$ (90,1) & Observed \\ \hline

$\mathrm{0j}-{1}$  & $0.4 \pm 0.4$ $\pm 0.4$ & $4.8 \pm 2.2$ $\pm 2.4$ & $  {<}0.1$  & $ 0.4$ $\pm$ $0.3 ^{+0.6}_{-0.4}$  &  ${<}0.1$  & $4.2 \pm 1.6$ $\pm 1.7$ & $9.8 \pm 2.8$ $\pm 3.0$ & $ {<}0.1$  & $ {<}0.1$  & $ {<}0.1$  & 4 \\\hline
$\mathrm{0j}-{2}$  & $0.7 \pm 0.5$ $\pm 0.7$ & $11.4 \pm 4.1$ $\pm 4.4$ & $ 0.7$ $\pm$ $0.4 ^{+0.9}_{-0.7}$  & $0.6 \pm 0.4$ $\pm 0.6$ & $ 0.4$ $\pm$ $0.4 ^{+0.9}_{-0.4}$  & $29.8 \pm 3.2$ $\pm 5.5$ & $43.7 \pm 5.2$ $\pm 7.2$ & $ {<}0.1$  &  $ {<}0.1$  & $ {<}0.1$  & 45 \\\hline
$\mathrm{0j}-{3}$  & $0.4 \pm 0.4$ $\pm 0.4$ & $1.0 \pm 0.7$ $\pm 0.8$ & $  {<}0.1$  & $ 0.4$ $\pm$ $0.3 ^{+0.5}_{-0.4}$  & $ {<}0.1$  & $ {<}0.1$  & $1.8 \pm 0.8$ $\pm 1.0$ & $ {<}0.1$  & $ {<}0.1$  & $ {<}0.1$  & 2 \\\hline
$\mathrm{0j}-{4}$  & $ {<}0.1$  & $80.3 \pm 9.1$ $\pm 17.3$ & $ 0.2$ $\pm$ $0.2 ^{+0.4}_{-0.2}$  & $0.9 \pm 0.4$ $\pm 0.7$ & $ 1.1$ $\pm$ $0.8 ^{+3.7}_{-1.1}$  & $33.3 \pm 6.3$ $\pm 8.0$ & $115.8 \pm 11.1$ $\pm 19.4$ & $ {<}0.1$  & $ {<}0.1$  & $ 0.0$ $\pm$ $0.0 ^{+0.7}_{-0.0}$  & 104 \\\hline
$\mathrm{0j}-{5}$  & $7.4 \pm 1.7$ $\pm 2.3$ & $2.5 \pm 1.8$ $\pm 1.8$ & $3.6 \pm 0.8$ $\pm 1.5$ & $17.8 \pm 1.9$ $\pm 6.2$ & $1.4 \pm 0.3$ $\pm 0.9$ & $88.7 \pm 7.8$ $\pm 15.4$ & $121.5 \pm 8.4$ $\pm 16.9$ & $0.5 \pm 0.1$ $\pm 0.2$ & $0.4 \pm 0.1$ $\pm 0.3$ & $ 0.4$ $\pm$ $0.3 ^{+1.0}_{-0.4}$  & 121 \\\hline
$\mathrm{0j}-{6}$  & $3.9 \pm 1.2$ $\pm 2.4$ & $4.8 \pm 1.9$ $\pm 2.0$ & $2.6 \pm 0.7$ $\pm 1.3$ & $16.3 \pm 1.8$ $\pm 6.5$ & $ 1.7$ $\pm$ $0.9 ^{+2.6}_{-1.7}$  & $174.9 \pm 10.4$ $\pm 28.2$ & $204.2 \pm 10.8$ $\pm 29.3$ & $0.2 \pm 0.0$ $\pm 0.1$ & $0.2 \pm 0.0$ $\pm 0.2$ & $ 1.3$ $\pm$ $0.6 ^{+2.0}_{-1.3}$  & 221 \\\hline
$\mathrm{0j}-{7}$  & $ {<}0.5$  & $ {<}0.1$  & $ 0.2$ $\pm$ $0.2 ^{+0.3}_{-0.2}$  & $ 0.8$ $\pm$ $0.4 ^{+1.2}_{-0.8}$  & $0.1 \pm 0.1$ $\pm 0.1$ & $1.1 \pm 0.8$ $\pm 0.8$ & $2.2 \pm 0.9$ $\pm 1.6$ & $ {<}0.1$  &  $ {<}0.1$  & $ {<}0.1$  & 4 \\\hline
$\mathrm{0j}-{8}$  &  0.4 $\pm$ $0.4 ^{+0.9}_{-0.4}$  & $ {<}0.1$  & $ 0.0$ $\pm$ $0.0 ^{+0.3}_{-0.0}$  & $ 0.1$ $\pm$ $0.1 ^{+0.4}_{-0.1}$  & $ {<}1.9$  & $4.8 \pm 1.8$ $\pm 1.9$ & $5.4 \pm 1.8$ $\pm 2.9$ &  ${<}0.1$  & $ {<}0.1$  & $ 0.3$ $\pm$ $0.2 ^{+0.7}_{-0.3}$  & 4 \\\hline
$\mathrm{0j}-{9}$  & $19.4 \pm 2.7$ $\pm 5.0$ & $82.0 \pm 9.3$ $\pm 17.7$ & $9.9 \pm 1.3$ $\pm 3.0$ & $36.9 \pm 2.7$ $\pm 11.5$ & $3.1 \pm 1.4$ $\pm 3.1$ & $308.8 \pm 14.6$ $\pm 48.5$ & $460.1 \pm 17.8$ $\pm 53.4$ & $0.6 \pm 0.1$ $\pm 0.2$ & $0.5 \pm 0.1$ $\pm 0.2$ & $ 3.3$ $\pm$ $1.0 ^{+5.5}_{-3.3}$  & 421 \\\hline
$\mathrm{0j}-{10}$  &  1.7 $\pm$ $0.8 ^{+2.3}_{-1.7}$  & $ {<}0.1$  & $ 0.5$ $\pm$ $0.3 ^{+0.7}_{-0.5}$  & $3.0 \pm 0.8$ $\pm 1.5$ & $0.2 \pm 0.1$ $\pm 0.1$ & $22.1 \pm 4.0$ $\pm 5.2$ & $27.5 \pm 4.1$ $\pm 5.9$ & $ 0.2$ $\pm$ $0.0 ^{+0.2}_{-0.2}$  &  $ {<}0.2$  & $ {<}0.1$  & 27 \\\hline
$\mathrm{0j}-{11}$  & $9.2 \pm 1.8$ $\pm 4.8$ & $ 1.3$ $\pm$ $1.3 ^{+1.4}_{-1.3}$  & $8.0 \pm 1.2$ $\pm 3.5$ & $21.8 \pm 2.1$ $\pm 9.1$ & $ 0.6$ $\pm$ $0.1 ^{+4.8}_{-0.6}$  & $194.2 \pm 11.5$ $\pm 31.3$ & $235.2 \pm 12.0$ $\pm 33.5$ & $0.6 \pm 0.1$ $\pm 0.2$ & $0.6 \pm 0.1$ $\pm 0.3$ & $ 0.5$ $\pm$ $0.4 ^{+1.5}_{-0.5}$  & 227 \\\hline
$\mathrm{0j}-{12}$  & $2.3 \pm 0.9$ $\pm 2.0$ & $ {<}0.1$  & $1.1 \pm 0.4$ $\pm 0.7$ & $2.4 \pm 0.7$ $\pm 1.4$ & $0.5 \pm 0.1$ $\pm 0.2$ & $18.0 \pm 3.5$ $\pm 4.4$ & $24.2 \pm 3.7$ $\pm 5.1$ & $1.0 \pm 0.1$ $\pm 0.2$ & $1.3 \pm 0.1$ $\pm 0.3$ & $ {<}0.1$  & 26 \\\hline
$\mathrm{0j}-{13}$  &  0.6 $\pm$ $0.4 ^{+0.7}_{-0.6}$  & $ {<}0.1$  & $ 0.2$ $\pm$ $0.2 ^{+0.2}_{-0.2}$  & $0.6 \pm 0.4$ $\pm 0.5$ & $ {<}0.1$  & $0.5^{+0.8}_{-0.5}$ $\pm 0.8$ & $2.0 \pm 1.0$ $\pm 1.2$ & $ {<}0.1$  & $ {<}0.1$  & $ 0.0$ $\pm$ $0.0 ^{+0.7}_{-0.0}$  & 1 \\\hline
$\mathrm{0j}-{14}$  & $5.2 \pm 1.3$ $\pm 3.5$ & $4.3 \pm 2.1$ $\pm 2.3$ & $4.0 \pm 0.8$ $\pm 1.5$ & $9.5 \pm 1.4$ $\pm 5.5$ & $2.2 \pm 1.5$ $\pm 1.5$ & $26.1 \pm 4.3$ $\pm 5.8$ & $51.2 \pm 5.4$ $\pm 9.3$ & $0.5 \pm 0.1$ $\pm 0.2$ & $0.2 \pm 0.0$ $\pm 0.1$ & $ 1.1$ $\pm$ $0.5 ^{+1.2}_{-1.1}$  & 46 \\\hline
$\mathrm{0j}-{15}$  &  0.7 $\pm$ $0.5 ^{+1.1}_{-0.7}$  & $ {<}0.1$  & $0.7 \pm 0.3$ $\pm 0.5$ & $1.6 \pm 0.6$ $\pm 1.0$ & $ 0.1$ $\pm$ $0.0 ^{+0.1}_{-0.1}$  & $10.1 \pm 2.6$ $\pm 3.0$ & $13.1 \pm 2.7$ $\pm 3.4$ &  ${<}0.1$  & $0.1 \pm 0.0$ $\pm 0.1$ & $ 0.1$ $\pm$ $0.0 ^{+0.8}_{-0.1}$  & 9 \\\hline
$\mathrm{0j}-{16}$  & $3.1 \pm 1.0$ $\pm 2.1$ & $ {<}0.1$  & $0.8 \pm 0.4$ $\pm 0.5$ & $2.1 \pm 0.7$ $\pm 1.0$ & $ {<}0.1$  & $6.2 \pm 2.3$ $\pm 2.5$ & $12.3 \pm 2.6$ $\pm 3.4$ &  ${<}0.2$  & $0.2 \pm 0.0$ $\pm 0.1$ & $ 0.5$ $\pm$ $0.3 ^{+1.2}_{-0.5}$  & 17 \\\hline
$\mathrm{0j}-{17}$  &  1.1 $\pm$ $0.6 ^{+1.2}_{-1.1}$  & $ {<}0.1$  & $0.5 \pm 0.3$ $\pm 0.4$ & $1.2 \pm 0.5$ $\pm 0.7$ & $ {<}0.1$  & $2.9 \pm 1.3$ $\pm 1.3$ & $5.7 \pm 1.5$ $\pm 1.9$ & $ 0.1$ $\pm$ $0.0 ^{+0.1}_{-0.1}$  &  $ {<}0.1$  & $ 0.0$ $\pm$ $0.0 ^{+1.4}_{-0.0}$  & 4 \\\hline
$\mathrm{0j}-{18}$  & $ 0.4$ $\pm$ $0.4 ^{+0.6}_{-0.4}$  & $ {<}0.1$  & $ 0.3$ $\pm$ $0.2 ^{+0.5}_{-0.3}$  & $0.8 \pm 0.4$ $\pm 0.7$ &  ${<}0.2$  & $6.0 \pm 2.0$ $\pm 2.2$ & $7.8 \pm 2.1$ $\pm 2.4$ &  ${<}0.5$  & $0.3 \pm 0.1$ $\pm 0.1$ & $ {<}0.1$  & 12 \\\hline
$\mathrm{0j}-{19}$  & $ 0.3$ $\pm$ $0.3 ^{+0.8}_{-0.3}$  & $ {<}0.1$  & $ {<}0.2$  & $ 0.6$ $\pm$ $0.4 ^{+0.7}_{-0.6}$  &  ${<}0.3$  & $3.0 \pm 1.5$ $\pm 1.5$ & $4.2 \pm 1.5$ $\pm 1.9$ & $0.9 \pm 0.1$ $\pm 0.3$ & $0.5 \pm 0.1$ $\pm 0.3$ & $ {<}0.1$  & 10 \\\hline
$\mathrm{0j}-{20}$  & $ 0.4$ $\pm$ $0.4 ^{+1.0}_{-0.4}$  & $ {<}0.1$  & $  {<}0.1$  & $ 0.6$ $\pm$ $0.4 ^{+0.7}_{-0.6}$  & $0.9 \pm 0.2$ $\pm 0.3$ & $5.0 \pm 1.7$ $\pm 1.9$ & $6.9 \pm 1.8$ $\pm 2.3$ & $2.4 \pm 0.1$ $\pm 0.5$ & $1.6 \pm 0.1$ $\pm 0.2$ & $ {<}0.1$  & 5 \\\hline
$\mathrm{0j}-{21}$  & $1.1 \pm 0.6$ $\pm 0.6$ & $ {<}0.1$  & $0.5 \pm 0.3$ $\pm 0.3$ & $1.7 \pm 0.6$ $\pm 0.8$ & $0.6 \pm 0.4$ $\pm 0.4$ & $ {<}0.1$  & $3.9 \pm 1.0$ $\pm 1.2$ & $0.6 \pm 0.1$ $\pm 0.2$ & $0.2 \pm 0.0$ $\pm 0.1$ & $ 0.4$ $\pm$ $0.3 ^{+0.5}_{-0.4}$  & 4 \\\hline
$\mathrm{1j}-{1}$  & $ {<}0.1$  & $8.0 \pm 2.4$ $\pm 2.6$ & $ 0.2$ $\pm$ $0.2 ^{+0.4}_{-0.2}$  & $ {<}0.3$  & $ 0.2$ $\pm$ $0.1 ^{+0.2}_{-0.2}$  & $6.1 \pm 2.0$ $\pm 2.2$ & $14.4 \pm 3.1$ $\pm 3.5$ &  ${<}0.1$  & $ {<}0.1$  & $ {<}0.1$  & 12 \\\hline
$\mathrm{1j}-{2}$  & $ 0.3$ $\pm$ $0.3 ^{+0.9}_{-0.3}$  & $31.2 \pm 4.5$ $\pm 7.1$ & $ {<}0.2$  & $0.8 \pm 0.4$ $\pm 0.7$ & $0.4^{+0.9}_{-0.4}$ $\pm 1.3$ & $32.5 \pm 4.8$ $\pm 6.8$ & $65.2 \pm 6.7$ $\pm 10.0$ &  ${<}0.1$  & $ {<}0.1$  & $ {<}0.1$  & 70 \\\hline
$\mathrm{1j}-{3}$  & $2.8 \pm 1.0$ $\pm 2.8$ & $26.8 \pm 4.2$ $\pm 6.0$ & $1.4 \pm 0.5$ $\pm 0.8$ & $ 1.2$ $\pm$ $0.6 ^{+1.4}_{-1.2}$  & $0.7^{+1.2}_{-0.7}$ $\pm 2.1$ & $46.1 \pm 5.5$ $\pm 8.8$ & $79.1 \pm 7.1$ $\pm 11.3$ & $0.2 \pm 0.0$ $\pm 0.2$ & $ 0.2$ $\pm$ $0.0 ^{+0.2}_{-0.2}$  & $ 0.3$ $\pm$ $0.2 ^{+0.8}_{-0.3}$  & 91 \\\hline
$\mathrm{1j}-{4}$  &  0.4 $\pm$ $0.4 ^{+0.6}_{-0.4}$  & $7.1 \pm 1.9$ $\pm 2.3$ & $0.5 \pm 0.3$ $\pm 0.5$ & $ 0.8$ $\pm$ $0.4 ^{+1.4}_{-0.8}$  & $0.7 \pm 0.4$ $\pm 0.5$ & $7.8 \pm 2.4$ $\pm 2.6$ & $17.3 \pm 3.1$ $\pm 3.9$ & $ {<}0.1$  & $ {<}0.1$  & $ {<}0.1$  & 12 \\\hline
$\mathrm{1j}-{5}$  & $2.7 \pm 1.0$ $\pm 1.9$ & $36.1 \pm 5.4$ $\pm 8.3$ & $ 0.3$ $\pm$ $0.2 ^{+1.3}_{-0.3}$  & $0.4 \pm 0.3$ $\pm 0.4$ & $ 0.4$ $\pm$ $0.3 ^{+0.9}_{-0.4}$  & $36.6 \pm 5.6$ $\pm 7.8$ & $76.4 \pm 7.8$ $\pm 11.6$ & $ {<}0.1$  & $ {<}0.1$  & $ {<}0.1$  & 63 \\\hline
$\mathrm{1j}-{6}$  & $25.8 \pm 3.1$ $\pm 5.9$ & $16.3 \pm 3.4$ $\pm 4.4$ & $12.6 \pm 1.5$ $\pm 4.1$ & $10.5 \pm 1.5$ $\pm 3.7$ & $0.1^{+1.0}_{-0.1}$ $\pm 1.1$ & $143.3 \pm 10.1$ $\pm 23.7$ & $208.7 \pm 11.3$ $\pm 25.5$ & $0.9 \pm 0.1$ $\pm 0.6$ & $0.9 \pm 0.1$ $\pm 0.4$ & $ 0.7$ $\pm$ $0.4 ^{+0.9}_{-0.7}$  & 224 \\\hline
$\mathrm{1j}-{7}$  & $16.4 \pm 2.4$ $\pm 3.9$ & $15.3 \pm 2.9$ $\pm 4.1$ & $4.5 \pm 0.9$ $\pm 2.8$ & $4.9 \pm 1.0$ $\pm 2.5$ & $ 1.3$ $\pm$ $0.6 ^{+1.6}_{-1.3}$  & $116.7 \pm 8.7$ $\pm 19.6$ & $159.1 \pm 9.6$ $\pm 20.8$ & $0.3 \pm 0.1$ $\pm 0.3$ & $0.3 \pm 0.1$ $\pm 0.2$ & $ 1.1$ $\pm$ $0.5 ^{+1.2}_{-1.1}$  & 161 \\\hline
$\mathrm{1j}-{8}$  & $2.2 \pm 1.0$ $\pm 1.4$ & $1.8 \pm 1.1$ $\pm 1.1$ & $ 0.7$ $\pm$ $0.4 ^{+1.2}_{-0.7}$  & $1.3 \pm 0.5$ $\pm 0.8$ & $ 1.0$ $\pm$ $0.8 ^{+1.1}_{-1.0}$  & $8.2 \pm 2.1$ $\pm 2.5$ & $15.1 \pm 2.8$ $\pm 3.6$ & $0.6 \pm 0.1$ $\pm 0.2$ & $0.4 \pm 0.1$ $\pm 0.1$ & $ {<}0.1$  & 19 \\\hline
$\mathrm{1j}-{9}$  & $3.3 \pm 1.1$ $\pm 1.3$ & $ 0.4$ $\pm$ $0.4 ^{+0.5}_{-0.4}$  & $0.7 \pm 0.4$ $\pm 0.6$ & $ 0.2$ $\pm$ $0.2 ^{+0.3}_{-0.2}$  & $0.4^{+0.4}_{-0.4}$ $\pm 0.5$ & $2.8 \pm 1.5$ $\pm 1.5$ & $7.8 \pm 2.0$ $\pm 2.2$ &  ${<}0.1$  &  $ {<}0.1$  & $ {<}0.1$  & 9 \\\hline
$\mathrm{1j}-{10}$  & $4.4 \pm 1.3$ $\pm 3.9$ & $0.9 \pm 0.8$ $\pm 0.8$ & $2.5 \pm 0.7$ $\pm 1.2$ & $2.5 \pm 0.7$ $\pm 1.8$ & $0.3 \pm 0.1$ $\pm 0.2$ & $15.1 \pm 3.5$ $\pm 4.1$ & $25.5 \pm 3.9$ $\pm 6.2$ & $0.3 \pm 0.1$ $\pm 0.1$ & $0.2 \pm 0.0$ $\pm 0.1$ & $ {<}0.1$  & 22 \\\hline
$\mathrm{1j}-{11}$  & $51.5 \pm 4.3$ $\pm 9.2$ & $2.7 \pm 1.2$ $\pm 1.4$ & $17.0 \pm 1.7$ $\pm 5.3$ & $14.0 \pm 1.7$ $\pm 7.4$ & $1.0 \pm 0.2$ $\pm 0.7$ & $113.2 \pm 9.4$ $\pm 19.4$ & $199.3 \pm 10.7$ $\pm 23.3$ & $1.2 \pm 0.1$ $\pm 0.5$ & $1.4 \pm 0.1$ $\pm 0.3$ & $ 0.4$ $\pm$ $0.3 ^{+0.6}_{-0.4}$  & 168 \\\hline
$\mathrm{1j}-{12}$  & $10.1 \pm 1.9$ $\pm 6.7$ & $ 0.8$ $\pm$ $0.8 ^{+0.8}_{-0.8}$  & $2.7 \pm 0.7$ $\pm 1.4$ & $3.0 \pm 0.8$ $\pm 1.8$ & $0.5 \pm 0.2$ $\pm 0.2$ & $28.2 \pm 4.3$ $\pm 6.0$ & $45.3 \pm 4.9$ $\pm 9.3$ & $2.2 \pm 0.1$ $\pm 0.6$ & $2.0 \pm 0.1$ $\pm 0.7$ & $ {<}0.1$  & 41 \\\hline
$\mathrm{1j}-{13}$  & $ {<}0.1$  & $0.2 \pm 0.2$ $\pm 0.2$ & $  {<}0.1$  & $ {<}0.2$  & $ 0.2$ $\pm$ $0.1 ^{+0.8}_{-0.2}$  & $1.2 \pm 0.8$ $\pm 0.8$ & $1.5 \pm 0.9$ $\pm 1.2$ & $0.6 \pm 0.1$ $\pm 0.1$ & $0.4 \pm 0.1$ $\pm 0.3$ & $ {<}0.1$  & 1 \\\hline
$\mathrm{1j}-{14}$  & $ {<}0.4$  & $ {<}0.1$  & $ {<}0.2$  & $0.2 \pm 0.2$ $\pm 0.2$ &  ${<}0.1$  & $0.4^{+0.6}_{-0.4}$ $\pm 0.6$ & $0.7^{+0.7}_{-0.7}$ $\pm 0.8$ & $ 0.1$ $\pm$ $0.0 ^{+0.1}_{-0.1}$  &  $ {<}0.1$  & $ {<}0.1$  & 2 \\\hline
$\mathrm{1j}-{15}$  & $3.7 \pm 1.2$ $\pm 3.0$ & $0.9 \pm 0.6$ $\pm 0.6$ & $ 0.2$ $\pm$ $0.2 ^{+0.7}_{-0.2}$  & $1.8 \pm 0.6$ $\pm 0.9$ & $0.1 \pm 0.1$ $\pm 0.1$ & $1.5 \pm 1.5$ $\pm 1.5$ & $8.3 \pm 2.1$ $\pm 3.6$ &  ${<}0.2$  &  $ {<}0.1$  & $ 0.0$ $\pm$ $0.0 ^{+0.7}_{-0.0}$  & 10 \\\hline
$\mathrm{1j}-{16}$  & $52.4 \pm 4.3$ $\pm 18.7$ & $19.8 \pm 3.2$ $\pm 6.6$ & $12.6 \pm 1.5$ $\pm 7.9$ & $18.9 \pm 1.9$ $\pm 9.3$ & $4.7 \pm 1.9$ $\pm 2.0$ & $54.2 \pm 6.9$ $\pm 10.6$ & $162.5 \pm 9.2$ $\pm 25.7$ & $2.8 \pm 0.2$ $\pm 0.7$ & $1.3 \pm 0.1$ $\pm 0.5$ & $2.9 \pm 0.9$ $\pm 2.4$ & 150 \\\hline
$\mathrm{1j}-{17}$  & $ 3.8$ $\pm$ $1.2 ^{+4.2}_{-3.8}$  & $ {<}0.1$  & $2.5 \pm 0.7$ $\pm 1.6$ & $2.4 \pm 0.7$ $\pm 1.3$ & $0.2 \pm 0.1$ $\pm 0.2$ & $5.8 \pm 2.6$ $\pm 2.7$ & $14.7 \pm 3.0$ $\pm 5.4$ & $0.8 \pm 0.1$ $\pm 0.2$ & $0.3 \pm 0.1$ $\pm 0.1$ & $ 0.4$ $\pm$ $0.4 ^{+0.6}_{-0.4}$  & 15 \\\hline
$\mathrm{1j}-{18}$  & $8.5 \pm 1.7$ $\pm 5.4$ & $ {<}0.1$  & $3.3 \pm 0.7$ $\pm 1.5$ & $2.5 \pm 0.7$ $\pm 2.2$ & $0.1^{+0.1}_{-0.1}$ $\pm 0.2$ & $3.2 \pm 2.5$ $\pm 2.6$ & $17.5 \pm 3.2$ $\pm 6.6$ & $0.7 \pm 0.1$ $\pm 0.2$ & $0.4 \pm 0.1$ $\pm 0.1$ & $ {<}0.1$  & 18 \\\hline
$\mathrm{1j}-{19}$  & $8.3 \pm 1.7$ $\pm 4.9$ & $0.4 \pm 0.4$ $\pm 0.4$ & $1.5 \pm 0.5$ $\pm 0.9$ & $ 1.0$ $\pm$ $0.4 ^{+1.5}_{-1.0}$  & $0.6 \pm 0.4$ $\pm 0.5$ & $6.5 \pm 2.1$ $\pm 2.3$ & $18.2 \pm 2.9$ $\pm 5.7$ & $0.4 \pm 0.1$ $\pm 0.1$ & $0.3 \pm 0.1$ $\pm 0.2$ & $ {<}0.1$  & 16 \\\hline
$\mathrm{1j}-{20}$  & $4.0 \pm 1.3$ $\pm 2.8$ & $ {<}0.1$  & $ 1.2$ $\pm$ $0.4 ^{+1.3}_{-1.2}$  & $ 1.1$ $\pm$ $0.5 ^{+1.2}_{-1.1}$  & $0.3 \pm 0.1$ $\pm 0.1$ & $7.7 \pm 2.5$ $\pm 2.7$ & $14.2 \pm 2.8$ $\pm 4.3$ & $1.3 \pm 0.1$ $\pm 0.2$ & $0.8 \pm 0.1$ $\pm 0.2$ & $ {<}0.1$  & 11 \\\hline
$\mathrm{1j}-{21}$  & $ 1.1$ $\pm$ $0.6 ^{+2.5}_{-1.1}$  & $ {<}0.1$  & $0.2 \pm 0.2$ $\pm 0.2$ & $ 0.8$ $\pm$ $0.4 ^{+1.0}_{-0.8}$  & $ 0.5$ $\pm$ $0.3 ^{+0.8}_{-0.5}$  & $5.8 \pm 2.0$ $\pm 2.1$ & $8.3 \pm 2.1$ $\pm 3.5$ & $2.3 \pm 0.1$ $\pm 0.3$ & $1.5 \pm 0.1$ $\pm 0.2$ & $ {<}0.1$  & 7 \\\hline
$\mathrm{1j}-{22}$  & $ {<}0.1$  & $ {<}0.1$  & $  {<}0.1$  & $ 0.7$ $\pm$ $0.4 ^{+0.7}_{-0.7}$  & $0.5 \pm 0.1$ $\pm 0.2$ & $2.1 \pm 1.2$ $\pm 1.2$ & $3.2 \pm 1.2$ $\pm 1.4$ & $6.2 \pm 0.2$ $\pm 0.6$ & $3.9 \pm 0.2$ $\pm 0.6$ & $ {<}0.1$  & 4 \\\hline
$\mathrm{1j}-{23}$  & $ 0.3$ $\pm$ $0.3 ^{+0.8}_{-0.3}$  & $ {<}0.1$  & $  {<}0.1$  & $0.4 \pm 0.3$ $\pm 0.3$ & $ 0.1$ $\pm$ $0.0 ^{+0.4}_{-0.1}$  & $0.9 \pm 0.8$ $\pm 0.8$ & $1.6 \pm 0.9$ $\pm 1.2$ & $1.8 \pm 0.1$ $\pm 0.5$ & $0.5 \pm 0.1$ $\pm 0.4$ & $ {<}0.1$  & 0 \\
 \hline
\end{tabular}}
\end{sidewaystable}

\begin{sidewaystable}
\topcaption{Numbers of expected and observed events in the \mutau channel. The total background includes the total uncertainty, while for each process the statistical and systematic uncertainties are quoted separately. The two numbers that are quoted for the benchmark signal models are the masses of the parent SUSY particle and the \PSGczDo, respectively, in GeV. In the case of the chargino-neutralino signal models, the first number within parentheses indicates the common \cone and \ntwo mass in GeV.} \label{tab:res_muhad}
\centering
\cmsTable{
\begin{tabular}{l|c|c|c|c|c|c|c|c|c|c|c}  \hline 
  SR label & \ttbar    & DY+jets &WW+jets &WW+jets &  Rest  & Jet$\rightarrow \tau_h$  & Total Bkg   & \cone\ntwo(400,1) &\cone\ntwo(400,175) &$\PSGt_L$(90,1) & Observed \\ \hline
$\mathrm{0j}-{1}$  & $1.3 \pm 0.8$ $\pm 1.2$ & $16.2 \pm 4.4$ $\pm 14.2$ & $  {<}0.1$  & $ 0.7$ $\pm$ $0.4 ^{+0.8}_{-0.7}$  & $ 0.5$ $\pm$ $0.5 ^{+1.1}_{-0.5}$  & $3.5 \pm 1.6$ $\pm 1.7$ & $22.2 \pm 4.8$ $\pm 14.5$ & $ {<}0.1$  & $ {<}0.1$  & $ {<}0.1$  & 7 \\\hline
$\mathrm{0j}-{2}$& $ 0.4$ $\pm$ $0.4 ^{+1.4}_{-0.4}$  & $23.1 \pm 5.8$ $\pm 22.5$ & $ {<}0.2$  & $2.0 \pm 0.7$ $\pm 1.5$ & $ 1.2$ $\pm$ $0.6 ^{+1.5}_{-1.2}$  & $51.1 \pm 5.7$ $\pm 9.6$ & $77.7 \pm 8.2$ $\pm 24.5$ &  ${<}0.1$  &$ {<}0.1$  & $ {<}0.1$  & 81 \\\hline
$\mathrm{0j}-{3}$& $ {<}0.1$  & $ {<}0.1$  & $0.3 \pm 0.2$ $\pm 0.3$ & $ 0.2$ $\pm$ $0.2 ^{+0.4}_{-0.2}$  & $ {<}0.1$  & $1.5 \pm 1.0$ $\pm 1.0$ & $2.0 \pm 1.1$ $\pm 1.2$ & $ {<}0.1$  & $ {<}0.1$  & $ {<}0.1$  & 2 \\\hline
$\mathrm{0j}-{4}$& $ 0.7$ $\pm$ $0.5 ^{+1.0}_{-0.7}$  & $208.3 \pm 15.7$ $\pm 27.2$ & $ 0.1$ $\pm$ $0.1 ^{+0.5}_{-0.1}$  & $1.2 \pm 0.5$ $\pm 0.8$ & $2.6 \pm 1.1$ $\pm 2.2$ & $76.1 \pm 9.6$ $\pm 14.9$ & $288.9 \pm 18.5$ $\pm 31.1$ &  ${<}0.1$  & $ {<}0.1$  & $ 0.4$ $\pm$ $0.4 ^{+0.8}_{-0.4}$  & 279 \\\hline
$\mathrm{0j}-{5}$& $12.3 \pm 2.4$ $\pm 5.0$ & $3.3 \pm 1.6$ $\pm 3.2$ & $7.3 \pm 1.2$ $\pm 3.3$ & $26.3 \pm 2.5$ $\pm 7.6$ & $ 2.6$ $\pm$ $1.7 ^{+4.1}_{-2.6}$  & $125.5 \pm 9.3$ $\pm 21.0$ & $177.4 \pm 10.2$ $\pm 23.7$ & $0.4 \pm 0.1$ $\pm 0.2$ & $0.6 \pm 0.1$ $\pm 0.2$ & $ 0.4$ $\pm$ $0.4 ^{+0.6}_{-0.4}$  & 197 \\\hline
$\mathrm{0j}-{6}$& $4.1 \pm 1.3$ $\pm 3.4$ & $15.5 \pm 4.1$ $\pm 10.8$ & $6.0 \pm 1.1$ $\pm 2.7$ & $25.8 \pm 2.5$ $\pm 9.5$ & $ 1.1$ $\pm$ $0.3 ^{+1.3}_{-1.1}$  & $372.0 \pm 15.2$ $\pm 57.8$ & $424.5 \pm 16.0$ $\pm 59.8$ & $0.2 \pm 0.1$ $\pm 0.1$ & $0.3 \pm 0.1$ $\pm 0.1$ & $ 0.7$ $\pm$ $0.5 ^{+1.8}_{-0.7}$  & 469 \\\hline
$\mathrm{0j}-{7}$& $ {<}0.1$  & $ {<}0.1$  & $  {<}0.1$  & $ {<}0.7$  & $ {<}0.1$  & $2.2 \pm 1.1$ $\pm 1.1$ & $2.2 \pm 1.1$ $\pm 1.3$ &  ${<}0.1$  &  $ {<}0.1$  & $ {<}0.1$  & 3 \\\hline
$\mathrm{0j}-{8}$& $ 0.7$ $\pm$ $0.5 ^{+1.0}_{-0.7}$  & $ {<}0.1$  & $ {<}0.2$  & $0.5 \pm 0.3$ $\pm 0.5$ &  ${<}0.1$  & $3.4 \pm 1.6$ $\pm 1.6$ & $4.7 \pm 1.7$ $\pm 2.0$ & $ {<}0.1$  &  $ {<}0.1$  & $ {<}0.1$  & 10 \\\hline
$\mathrm{0j}-{9}$& $35.3 \pm 3.7$ $\pm 10.4$ & $133.9 \pm 11.9$ $\pm 23.8$ & $16.0 \pm 1.7$ $\pm 4.6$ & $61.8 \pm 3.7$ $\pm 17.7$ & $5.3 \pm 1.3$ $\pm 2.3$ & $531.1 \pm 19.4$ $\pm 82.0$ & $783.4 \pm 23.4$ $\pm 88.0$ & $1.2 \pm 0.1$ $\pm 0.2$ & $0.8 \pm 0.1$ $\pm 0.1$ & $8.7 \pm 1.8$ $\pm 2.3$ & 739 \\\hline
$\mathrm{0j}-{10}$& $ 1.6$ $\pm$ $1.0 ^{+1.7}_{-1.6}$  & $ {<}0.1$  & $ 2.2$ $\pm$ $0.8 ^{+2.3}_{-2.2}$  & $6.3 \pm 1.3$ $\pm 2.9$ & $0.3 \pm 0.1$ $\pm 0.3$ & $27.0 \pm 4.4$ $\pm 6.0$ & $37.5 \pm 4.7$ $\pm 7.3$ & $0.2 \pm 0.1$ $\pm 0.1$ & $0.1 \pm 0.1$ $\pm 0.1$ & $ 0.1$ $\pm$ $0.1 ^{+0.4}_{-0.1}$  & 31 \\\hline
$\mathrm{0j}-{11}$& $26.8 \pm 3.4$ $\pm 5.1$ & $ 1.0$ $\pm$ $0.7 ^{+1.1}_{-1.0}$  & $13.0 \pm 1.7$ $\pm 4.1$ & $40.4 \pm 3.1$ $\pm 11.9$ & $1.6 \pm 0.3$ $\pm 0.6$ & $305.0 \pm 14.3$ $\pm 47.9$ & $387.8 \pm 15.1$ $\pm 49.8$ & $0.8 \pm 0.1$ $\pm 0.2$ & $0.7 \pm 0.1$ $\pm 0.2$ & $2.5 \pm 1.0$ $\pm 1.2$ & 383 \\\hline
$\mathrm{0j}-{12}$& $3.9 \pm 1.4$ $\pm 2.6$ & $ {<}0.1$  & $1.8 \pm 0.6$ $\pm 1.4$ & $6.3 \pm 1.3$ $\pm 2.4$ & $1.2 \pm 0.2$ $\pm 0.5$ & $38.7 \pm 4.9$ $\pm 7.6$ & $52.0 \pm 5.3$ $\pm 8.5$ & $1.2 \pm 0.2$ $\pm 0.2$ & $1.2 \pm 0.2$ $\pm 0.3$ & $ 0.4$ $\pm$ $0.4 ^{+0.6}_{-0.4}$  & 56 \\\hline
$\mathrm{0j}-{13}$& $ 0.7$ $\pm$ $0.5 ^{+0.8}_{-0.7}$  & $ {<}0.1$  & $ 0.3$ $\pm$ $0.2 ^{+0.5}_{-0.3}$  & $0.6 \pm 0.3$ $\pm 0.6$ & $ {<}0.1$  & $1.1 \pm 0.9$ $\pm 0.9$ & $2.7 \pm 1.1$ $\pm 1.5$ &  ${<}0.1$  & $ {<}0.1$  & $ {<}0.1$  & 2 \\\hline
$\mathrm{0j}-{14}$& $16.1 \pm 2.5$ $\pm 6.5$ & $11.6 \pm 3.8$ $\pm 5.3$ & $7.7 \pm 1.2$ $\pm 2.5$ & $16.2 \pm 1.8$ $\pm 5.0$ & $1.2 \pm 0.5$ $\pm 0.6$ & $40.0 \pm 5.2$ $\pm 8.0$ & $92.8 \pm 7.3$ $\pm 12.9$ & $0.7 \pm 0.1$ $\pm 0.1$ &  $ {<}0.4$  & $5.1 \pm 1.4$ $\pm 1.9$ & 75 \\\hline
$\mathrm{0j}-{15}$& $2.3 \pm 0.9$ $\pm 1.1$ & $ {<}0.1$  & $1.1 \pm 0.5$ $\pm 0.7$ & $1.5 \pm 0.6$ $\pm 1.2$ & $ {<}0.1$  & $9.8 \pm 2.7$ $\pm 3.1$ & $14.7 \pm 3.0$ $\pm 3.5$ &  ${<}0.2$  &  $ {<}0.2$  & $0.8 \pm 0.5$ $\pm 0.6$ & 15 \\\hline
$\mathrm{0j}-{16}$& $3.3 \pm 1.1$ $\pm 1.2$ & $ {<}0.1$  & $2.0 \pm 0.6$ $\pm 1.0$ & $2.7 \pm 0.8$ $\pm 1.9$ &  ${<}0.1$  & $11.6 \pm 3.0$ $\pm 3.5$ & $19.6 \pm 3.3$ $\pm 4.2$ & $0.2 \pm 0.1$ $\pm 0.1$ &  $ {<}0.1$  & $0.7 \pm 0.5$ $\pm 0.7$ & 26 \\\hline
$\mathrm{0j}-{17}$& $ 0.7$ $\pm$ $0.5 ^{+1.3}_{-0.7}$  & $ {<}0.1$  & $1.3 \pm 0.5$ $\pm 0.8$ & $1.3 \pm 0.5$ $\pm 0.8$ & $ {<}0.1$ & $2.0 \pm 1.3$ $\pm 1.4$ & $5.4 \pm 1.6$ $\pm 2.2$ &  ${<}0.1$  &  $ {<}0.1$  & $ 0.5$ $\pm$ $0.4 ^{+0.6}_{-0.5}$  & 6 \\\hline
$\mathrm{0j}-{18}$& $0.7 \pm 0.5$ $\pm 0.6$ & $ {<}0.1$  & $0.3 \pm 0.2$ $\pm 0.3$ & $1.9 \pm 0.6$ $\pm 1.5$ & $0.4 \pm 0.1$ $\pm 0.2$ & $12.9 \pm 3.0$ $\pm 3.5$ & $16.2 \pm 3.1$ $\pm 3.9$ & $0.6 \pm 0.1$ $\pm 0.1$ & $0.4 \pm 0.1$ $\pm 0.2$ & $ 0.3$ $\pm$ $0.3 ^{+0.3}_{-0.3}$  & 16 \\\hline
$\mathrm{0j}-{19}$& $ {<}0.1$  & $ {<}0.1$  & $  {<}0.1$  & $2.0 \pm 0.7$ $\pm 0.9$ & $0.5 \pm 0.1$ $\pm 0.2$ & $9.4 \pm 2.5$ $\pm 2.8$ & $11.9 \pm 2.6$ $\pm 3.0$ & $1.3 \pm 0.2$ $\pm 0.2$ & $1.1 \pm 0.1$ $\pm 0.2$ & $ {<}0.1$  & 13 \\\hline
$\mathrm{0j}-{20}$& $ {<}0.1$  & $ {<}0.1$  & $  {<}0.1$  & $1.6 \pm 0.7$ $\pm 0.9$ & $0.8 \pm 0.1$ $\pm 0.3$ & $5.8 \pm 1.8$ $\pm 2.0$ & $8.2 \pm 2.0$ $\pm 2.3$ & $4.1 \pm 0.3$ $\pm 0.4$ & $2.0 \pm 0.2$ $\pm 0.3$ & $ {<}0.1$  & 10 \\\hline
$\mathrm{0j}-{21}$& $ 0.8$ $\pm$ $0.5 ^{+1.3}_{-0.8}$  & $ {<}0.1$  & $1.1 \pm 0.4$ $\pm 0.6$ & $1.9 \pm 0.7$ $\pm 1.0$ &  ${<}0.1$  & $2.3 \pm 1.2$ $\pm 1.3$ & $6.1 \pm 1.6$ $\pm 2.2$ & $1.0 \pm 0.1$ $\pm 0.2$ &  $ {<}0.2$  & $1.6 \pm 0.8$ $\pm 1.0$ & 4 \\\hline
$\mathrm{1j}-{1}$& $ 0.4$ $\pm$ $0.4 ^{+0.5}_{-0.4}$  & $ 3.0$ $\pm$ $1.4 ^{+3.1}_{-3.0}$  & $  {<}0.1$  & $ 0.2$ $\pm$ $0.2 ^{+0.3}_{-0.2}$  & $0.0^{+0.0}_{-0.0}$ $\pm 0.6$ & $6.7 \pm 2.1$ $\pm 2.3$ & $10.3 \pm 2.6$ $\pm 4.0$ &  ${<}0.1$  & $ {<}0.1$  & $ {<}0.1$  & 6 \\\hline
$\mathrm{1j}-{2}$& $ 1.8$ $\pm$ $0.9 ^{+1.9}_{-1.8}$  & $28.0 \pm 3.9$ $\pm 9.3$ & $ 0.7$ $\pm$ $0.4 ^{+0.9}_{-0.7}$  & $1.4 \pm 0.6$ $\pm 0.8$ & $0.8^{+0.9}_{-0.8}$ $\pm 1.0$ & $35.6 \pm 5.0$ $\pm 7.3$ & $68.2 \pm 6.5$ $\pm 12.1$ &  ${<}0.1$  & $ {<}0.1$  & $ {<}0.1$  & 70 \\\hline
$\mathrm{1j}-{3}$& $9.0 \pm 2.1$ $\pm 5.1$ & $35.7 \pm 4.4$ $\pm 8.1$ & $3.2 \pm 0.9$ $\pm 2.6$ & $2.3 \pm 0.7$ $\pm 1.8$ & $3.4 \pm 2.1$ $\pm 3.4$ & $77.0 \pm 7.0$ $\pm 13.5$ & $130.5 \pm 8.8$ $\pm 17.2$ &  ${<}0.2$  &  $ {<}0.1$  & $ {<}0.1$  & 143 \\\hline
$\mathrm{1j}-{4}$& $ 0.7$ $\pm$ $0.5 ^{+2.2}_{-0.7}$  & $7.9 \pm 1.9$ $\pm 4.9$ & $0.8 \pm 0.4$ $\pm 0.6$ & $0.6 \pm 0.4$ $\pm 0.6$ & $ 0.5$ $\pm$ $0.4 ^{+0.8}_{-0.5}$  & $7.5 \pm 2.2$ $\pm 2.5$ & $18.1 \pm 3.0$ $\pm 6.0$ &  ${<}0.1$  & $ {<}0.1$  & $ {<}0.1$  & 20 \\\hline
$\mathrm{1j}-{5}$& $5.9 \pm 1.6$ $\pm 2.3$ & $86.1 \pm 8.9$ $\pm 18.4$ & $ 0.9$ $\pm$ $0.4 ^{+1.9}_{-0.9}$  & $1.6 \pm 0.6$ $\pm 1.4$ & $ 1.2$ $\pm$ $0.6 ^{+1.4}_{-1.2}$  & $67.0 \pm 8.3$ $\pm 13.0$ & $162.6 \pm 12.3$ $\pm 22.8$ &$ {<}0.1$ & $ {<}0.1$  & $ {<}0.1$  & 164 \\\hline
$\mathrm{1j}-{6}$& $44.8 \pm 4.3$ $\pm 9.5$ & $9.3 \pm 2.3$ $\pm 4.8$ & $15.8 \pm 1.8$ $\pm 5.5$ & $19.9 \pm 2.2$ $\pm 6.3$ & $ 1.9$ $\pm$ $0.7 ^{+1.9}_{-1.9}$  & $197.5 \pm 11.9$ $\pm 31.9$ & $289.2 \pm 13.2$ $\pm 34.7$ & $0.7 \pm 0.1$ $\pm 0.2$ & $0.9 \pm 0.1$ $\pm 0.2$ & $ {<}0.1$  & 283 \\\hline
$\mathrm{1j}-{7}$& $31.7 \pm 3.7$ $\pm 7.2$ & $31.4 \pm 3.8$ $\pm 6.7$ & $10.5 \pm 1.5$ $\pm 3.8$ & $10.2 \pm 1.6$ $\pm 5.4$ & $2.0 \pm 0.7$ $\pm 1.4$ & $201.1 \pm 11.5$ $\pm 32.3$ & $286.9 \pm 12.9$ $\pm 34.4$ & $0.5 \pm 0.1$ $\pm 0.1$ & $0.3 \pm 0.1$ $\pm 0.1$ & $ {<}0.1$  & 292 \\\hline
$\mathrm{1j}-{8}$& $1.8 \pm 0.8$ $\pm 1.4$ & $ 2.3$ $\pm$ $1.7 ^{+2.9}_{-2.3}$  & $1.3 \pm 0.5$ $\pm 0.8$ & $ 1.2$ $\pm$ $0.6 ^{+1.9}_{-1.2}$  & $3.3 \pm 3.0$ $\pm 3.0$ & $7.6 \pm 2.2$ $\pm 2.4$ & $17.4 \pm 4.2$ $\pm 5.5$ & $0.6 \pm 0.1$ $\pm 0.2$ & $0.3 \pm 0.1$ $\pm 0.1$ & $ {<}0.1$  & 26 \\\hline
$\mathrm{1j}-{9}$& $ 5.2$ $\pm$ $1.6 ^{+6.1}_{-5.2}$  & $ 0.5$ $\pm$ $0.4 ^{+0.9}_{-0.5}$  & $0.9 \pm 0.4$ $\pm 0.8$ & $ 0.9$ $\pm$ $0.4 ^{+0.9}_{-0.9}$  & $ 0.1$ $\pm$ $0.1 ^{+0.2}_{-0.1}$  & $3.0 \pm 1.7$ $\pm 1.7$ & $10.5 \pm 2.5$ $\pm 6.5$ &  ${<}0.1$  &  $ {<}0.1$  & $ {<}0.1$  & 13 \\\hline
$\mathrm{1j}-{10}$& $13.2 \pm 2.6$ $\pm 3.6$ & $ {<}0.1$  & $2.0 \pm 0.7$ $\pm 1.5$ & $ 1.6$ $\pm$ $0.6 ^{+3.1}_{-1.6}$  & $ 0.4$ $\pm$ $0.2 ^{+0.8}_{-0.4}$  & $18.8 \pm 3.9$ $\pm 4.8$ & $36.0 \pm 4.8$ $\pm 7.0$ & $0.3 \pm 0.1$ $\pm 0.1$ & $0.3 \pm 0.1$ $\pm 0.2$ & $ {<}0.1$  & 34 \\\hline
$\mathrm{1j}-{11}$& $86.2 \pm 6.1$ $\pm 24.6$ & $2.3 \pm 1.0$ $\pm 1.4$ & $29.7 \pm 2.5$ $\pm 10.4$ & $28.5 \pm 2.6$ $\pm 9.7$ & $2.5 \pm 0.6$ $\pm 1.1$ & $178.0 \pm 11.6$ $\pm 29.1$ & $327.2 \pm 13.6$ $\pm 40.7$ & $1.9 \pm 0.2$ $\pm 0.3$ & $1.9 \pm 0.2$ $\pm 0.3$ & $1.3 \pm 0.7$ $\pm 0.8$ & 296 \\\hline
$\mathrm{1j}-{12}$& $15.4 \pm 2.8$ $\pm 4.5$ & $ {<}0.1$  & $5.4 \pm 1.1$ $\pm 3.1$ & $6.0 \pm 1.2$ $\pm 2.6$ & $1.2 \pm 0.3$ $\pm 0.4$ & $39.6 \pm 5.1$ $\pm 7.8$ & $67.6 \pm 6.0$ $\pm 9.9$ & $2.5 \pm 0.2$ $\pm 0.3$ & $3.0 \pm 0.2$ $\pm 0.4$ & $ {<}0.1$  & 46 \\\hline
$\mathrm{1j}-{13}$& $ {<}0.8$  & $ {<}0.1$  & $  {<}0.1$  & $ 0.4$ $\pm$ $0.4 ^{+0.6}_{-0.4}$  & $ {<}0.1$  & $0.6^{+0.6}_{-0.6}$ $\pm 0.6$ & $1.0 \pm 0.7$ $\pm 1.1$ & $0.8 \pm 0.1$ $\pm 0.2$ & $0.3 \pm 0.1$ $\pm 0.1$ & $ {<}0.1$  & 0 \\\hline
$\mathrm{1j}-{14}$& $ {<}0.1$  & $ {<}0.1$  & $  {<}0.1$  & $ 0.2$ $\pm$ $0.2 ^{+1.5}_{-0.2}$  & $ {<}0.1$  & $ {<}0.1$  & $0.2 \pm 0.2$ $\pm 1.5$ &  ${<}0.1$  &  $ {<}0.1$  & $ {<}0.1$  & 1 \\\hline
$\mathrm{1j}-{15}$& $5.0 \pm 1.5$ $\pm 2.6$ & $0.5 \pm 0.5$ $\pm 0.5$ & $1.7 \pm 0.6$ $\pm 1.0$ & $1.8 \pm 0.7$ $\pm 1.4$ & $ {<}0.1$  & $0.6^{+1.0}_{-0.6}$ $\pm 1.0$ & $9.5 \pm 2.1$ $\pm 3.3$ &  ${<}0.2$  &  $ {<}0.1$  & $ {<}0.1$  & 10 \\\hline
$\mathrm{1j}-{16}$& $81.0 \pm 5.6$ $\pm 30.0$ & $25.2 \pm 3.6$ $\pm 7.8$ & $22.0 \pm 2.0$ $\pm 7.9$ & $34.5 \pm 2.7$ $\pm 14.8$ & $5.9 \pm 1.3$ $\pm 1.9$ & $86.3 \pm 8.5$ $\pm 15.5$ & $255.0 \pm 11.4$ $\pm 38.5$ & $4.4 \pm 0.3$ $\pm 0.6$ & $2.3 \pm 0.2$ $\pm 0.3$ & $3.6 \pm 1.3$ $\pm 1.6$ & 254 \\\hline
$\mathrm{1j}-{17}$& $10.2 \pm 2.0$ $\pm 6.8$ & $ {<}0.1$  & $3.7 \pm 0.9$ $\pm 1.5$ & $3.6 \pm 0.9$ $\pm 1.5$ & $ 0.1$ $\pm$ $0.0 ^{+0.1}_{-0.1}$  & $8.4 \pm 2.9$ $\pm 3.2$ & $25.9 \pm 3.8$ $\pm 7.9$ & $0.9 \pm 0.1$ $\pm 0.2$ & $0.4 \pm 0.1$ $\pm 0.2$ & $ {<}0.1$  & 23 \\\hline
$\mathrm{1j}-{18}$& $26.8 \pm 3.6$ $\pm 6.3$ & $ {<}0.1$  & $6.3 \pm 1.1$ $\pm 2.6$ & $4.3 \pm 1.0$ $\pm 2.5$ & $0.5 \pm 0.2$ $\pm 0.2$ & $9.0 \pm 3.0$ $\pm 3.3$ & $46.9 \pm 4.9$ $\pm 8.0$ & $0.8 \pm 0.1$ $\pm 0.2$ & $0.6 \pm 0.1$ $\pm 0.1$ & $ {<}0.1$  & 46 \\\hline
$\mathrm{1j}-{19}$& $9.3 \pm 1.9$ $\pm 7.9$ & $ 1.1$ $\pm$ $0.8 ^{+1.7}_{-1.1}$  & $3.6 \pm 0.9$ $\pm 1.5$ & $5.7 \pm 1.2$ $\pm 2.5$ & $0.2 \pm 0.1$ $\pm 0.1$ & $13.1 \pm 3.0$ $\pm 3.6$ & $32.9 \pm 4.0$ $\pm 9.3$ & $0.7 \pm 0.1$ $\pm 0.1$ & $0.4 \pm 0.1$ $\pm 0.1$ & $ {<}0.1$  & 30 \\\hline
$\mathrm{1j}-{20}$& $9.9 \pm 2.2$ $\pm 4.7$ & $ {<}0.1$  & $1.1 \pm 0.4$ $\pm 0.9$ & $1.8 \pm 0.6$ $\pm 1.1$ & $0.4 \pm 0.1$ $\pm 0.2$ & $12.6 \pm 3.0$ $\pm 3.5$ & $25.7 \pm 3.8$ $\pm 6.1$ & $2.0 \pm 0.2$ $\pm 0.3$ & $1.1 \pm 0.1$ $\pm 0.2$ & $ 0.1$ $\pm$ $0.1 ^{+0.2}_{-0.1}$  & 18 \\\hline
$\mathrm{1j}-{21}$& $ 0.0$ $\pm$ $0.0 ^{+0.4}_{-0.0}$  & $ {<}0.1$  & $ 0.0$ $\pm$ $0.0 ^{+0.4}_{-0.0}$  & $1.2 \pm 0.5$ $\pm 0.9$ &  ${<}0.2$  & $5.2 \pm 1.9$ $\pm 2.0$ & $6.7 \pm 2.0$ $\pm 2.3$ & $3.6 \pm 0.3$ $\pm 0.6$ & $1.9 \pm 0.2$ $\pm 0.3$ & $ {<}0.1$  & 6 \\\hline
$\mathrm{1j}-{22}$& $ 0.7$ $\pm$ $0.7 ^{+1.9}_{-0.7}$  & $ {<}0.1$  & $  {<}0.1$  & $0.7 \pm 0.4$ $\pm 0.7$ & $0.4 \pm 0.1$ $\pm 0.3$ & $2.5 \pm 1.3$ $\pm 1.4$ & $4.4 \pm 1.6$ $\pm 2.5$ & $8.7 \pm 0.4$ $\pm 0.8$ & $4.4 \pm 0.3$ $\pm 0.5$ & $ {<}0.1$  & 6 \\\hline
$\mathrm{1j}-{23}$& $ {<}0.1$  & $ {<}0.1$  & $  {<}0.1$  & $ 0.5$ $\pm$ $0.3 ^{+0.7}_{-0.5}$  &  ${<}0.1$  & $0.9 \pm 0.8$ $\pm 0.8$ & $1.5 \pm 0.9$ $\pm 1.1$ & $2.7 \pm 0.2$ $\pm 0.3$ & $0.9 \pm 0.1$ $\pm 0.2$ & $ {<}0.1$  & 1 \\

\hline
\end{tabular}}
\end{sidewaystable}

\begin{sidewaystable}
\topcaption{Numbers of expected and observed events in the \emu channel. The total background includes the total uncertainty, while for each process the statistical and systematic uncertainties are quoted separately. The two numbers that are quoted for the benchmark signal models are the masses of the parent SUSY particle and the \PSGczDo, respectively, in GeV. In the case of the chargino-neutralino signal models, the first number within parentheses indicates the common \cone and \ntwo mass in GeV.} \label{tab:res_elemu}
\centering
\cmsTable{
\begin{tabular}{l|c|c|c|c|c|c|c|c|c|c|c}  \hline 

 SR label  &  \ttbar    & DY+jets &WW+jets &WW+jets &  Rest  & QCD      & Total Bkg   & \cone\ntwo(400,1) &\cone\ntwo(400,175) &$\PSGt_L$(90,1) & Observed \\ \hline

$\mathrm{0j}-{1}$  & $2.5 \pm 1.0$ $\pm 1.6$ & $ {<}0.1$  & $0.6 \pm 0.3$ $\pm 0.4$ & $ 0.6$ $\pm$ $0.4 ^{+0.7}_{-0.6}$  & $0.1 \pm 0.1$ $\pm 0.1$ & $ {<}0.1$  & $3.9 \pm 1.1$ $\pm 1.8$ & $ {<}0.1$  & $ {<}0.1$  & $ {<}0.1$  & 3 \\\hline
$\mathrm{0j}-{2}$  & $40.0 \pm 3.8$ $\pm 12.9$ & $155.4 \pm 13.5$ $\pm 20.7$ & $21.1 \pm 1.9$ $\pm 6.0$ & $248.7 \pm 7.1$ $\pm 64.4$ & $37.3 \pm 11.6$ $\pm 22.4$ & $35.0 \pm 16.2$ $\pm 23.8$ & $537.5 \pm 25.4$ $\pm 76.4$ & $ {<}0.1$  &  $ {<}0.1$  & $ 0.4$ $\pm$ $0.0 ^{+2.5}_{-0.4}$  & 584 \\\hline
$\mathrm{0j}-{3}$ & $21.3 \pm 2.8$ $\pm 7.1$ & $ {<}0.1$  & $9.9 \pm 1.3$ $\pm 3.8$ & $47.2 \pm 3.1$ $\pm 13.3$ & $1.6^{+1.6}_{-1.6}$ $\pm 3.9$ & $4.3^{+5.1}_{-4.3}$ $\pm 5.5$ & $84.2 \pm 6.9$ $\pm 16.9$ & $ {<}0.1$  & $ {<}0.1$  & $ 0.1$ $\pm$ $0.0 ^{+0.8}_{-0.1}$  & 105 \\\hline
$\mathrm{0j}-{4}$ & $ 0.4$ $\pm$ $0.4 ^{+0.8}_{-0.4}$  & $ {<}0.1$  & $ 0.2$ $\pm$ $0.2 ^{+0.6}_{-0.2}$  & $0.6 \pm 0.4$ $\pm 0.6$ & $0.0^{+0.0}_{-0.0}$ $\pm 2.3$ & $ {<}0.1$  & $1.2 \pm 0.6$ $\pm 2.5$ & $ {<}0.1$  & $ {<}0.1$  & $ {<}0.1$  & 2 \\\hline
$\mathrm{0j}-{5}$ & $5.7 \pm 1.4$ $\pm 2.8$ & $2.4 \pm 1.5$ $\pm 1.6$ & $2.9 \pm 0.7$ $\pm 1.2$ & $7.1 \pm 1.2$ $\pm 2.2$ & $ 1.8$ $\pm$ $1.5 ^{+2.4}_{-1.8}$  & $ {<}0.1$  & $20.0 \pm 2.9$ $\pm 4.8$ & $ {<}0.1$  & $ {<}0.1$  & $ 0.2$ $\pm$ $0.0 ^{+1.2}_{-0.2}$  & 21 \\\hline
$\mathrm{0j}-{6}$ & $105.3 \pm 6.2$ $\pm 33.2$ & $ {<}0.1$  & $66.2 \pm 3.4$ $\pm 18.8$ & $302.9 \pm 7.8$ $\pm 79.8$ & $16.1 \pm 5.6$ $\pm 10.7$ & $22.6 \pm 11.2$ $\pm 15.9$ & $513.1 \pm 16.4$ $\pm 90.6$ & $ 0.2$ $\pm$ $0.0 ^{+0.6}_{-0.2}$  &  $ {<}0.1$  & $ {<}0.1$  & 531 \\\hline
$\mathrm{0j}-{7}$ & $82.9 \pm 5.5$ $\pm 29.4$ & $ 1.4$ $\pm$ $1.4 ^{+1.5}_{-1.4}$  & $46.0 \pm 2.8$ $\pm 13.1$ & $424.6 \pm 9.3$ $\pm 110.0$ & $19.9 \pm 6.2$ $\pm 16.1$ & $19.6 \pm 13.8$ $\pm 16.9$ & $594.4 \pm 18.8$ $\pm 116.9$ & $ 0.1$ $\pm$ $0.0 ^{+0.2}_{-0.1}$  &  $ {<}0.1$  & $ 0.3$ $\pm$ $0.0 ^{+1.8}_{-0.3}$  & 618 \\\hline
$\mathrm{0j}-{8}$ & $ 2.6$ $\pm$ $0.9 ^{+2.9}_{-2.6}$  & $ {<}0.1$  & $0.6 \pm 0.3$ $\pm 0.6$ & $1.9 \pm 0.6$ $\pm 1.5$ & $ 0.1$ $\pm$ $0.1 ^{+0.2}_{-0.1}$  & $ {<}0.1$  & $5.3 \pm 1.1$ $\pm 3.4$ & $ {<}0.1$  &  $ {<}0.1$  & $ {<}0.1$  & 7 \\\hline
$\mathrm{0j}-{9}$ & $4.9 \pm 1.3$ $\pm 1.9$ & $ {<}0.1$  & $1.6 \pm 0.5$ $\pm 0.8$ & $1.7 \pm 0.6$ $\pm 1.4$ & $ 0.4$ $\pm$ $0.3 ^{+0.7}_{-0.4}$  & $ {<}0.1$  & $8.6 \pm 1.5$ $\pm 2.6$ & $ {<}0.1$  & $ {<}0.1$  & $ {<}0.1$  & 12 \\\hline
$\mathrm{0j}-{10}$ & $119.2 \pm 6.5$ $\pm 33.4$ & $28.3 \pm 5.9$ $\pm 8.0$ & $49.7 \pm 2.9$ $\pm 13.2$ & $123.9 \pm 5.0$ $\pm 36.1$ & $10.2 \pm 3.9$ $\pm 8.8$ & $13.0 \pm 10.3$ $\pm 12.2$ & $344.2 \pm 15.2$ $\pm 53.7$ & $ 0.2$ $\pm$ $0.0 ^{+0.6}_{-0.2}$  &  $ {<}0.4$  & $ 0.9$ $\pm$ $0.0 ^{+6.1}_{-0.9}$  & 324 \\\hline
$\mathrm{0j}-{11}$ & $17.0 \pm 2.5$ $\pm 6.6$ & $ {<}0.1$  & $10.5 \pm 1.3$ $\pm 3.4$ & $21.4 \pm 2.1$ $\pm 6.3$ & $ 1.6$ $\pm$ $1.0 ^{+3.3}_{-1.6}$  & $ {<}0.1$  & $50.7 \pm 3.6$ $\pm 10.3$ & $ {<}0.1$  &  $ {<}0.1$  & $ {<}0.1$  & 50 \\\hline
$\mathrm{0j}-{12}$ & $129.0 \pm 6.8$ $\pm 36.9$ & $0.5 \pm 0.5$ $\pm 0.5$ & $61.3 \pm 3.2$ $\pm 16.5$ & $224.7 \pm 6.7$ $\pm 58.9$ & $8.2 \pm 3.2$ $\pm 3.4$ & $11.6 \pm 7.9$ $\pm 9.8$ & $435.3 \pm 13.2$ $\pm 72.2$ & $ 0.2$ $\pm$ $0.0 ^{+0.6}_{-0.2}$  &  $ {<}0.2$  & $ 0.4$ $\pm$ $0.0 ^{+2.4}_{-0.4}$  & 457 \\\hline
$\mathrm{0j}-{13}$ & $27.9 \pm 3.2$ $\pm 8.8$ & $ {<}0.1$  & $10.7 \pm 1.3$ $\pm 3.7$ & $29.2 \pm 2.4$ $\pm 8.9$ & $ 1.0$ $\pm$ $0.2 ^{+1.5}_{-1.0}$  & $ {<}0.1$  & $68.8 \pm 4.2$ $\pm 13.1$ & $ 0.2$ $\pm$ $0.0 ^{+0.5}_{-0.2}$  &  $ {<}0.1$  & $ {<}0.1$  & 77 \\\hline
$\mathrm{0j}-{14}$ & $4.6 \pm 1.2$ $\pm 2.1$ & $ {<}0.1$  & $1.3 \pm 0.5$ $\pm 1.1$ & $1.8 \pm 0.6$ $\pm 1.0$ & $ 0.3$ $\pm$ $0.3 ^{+0.5}_{-0.3}$  & $ {<}0.1$  & $8.1 \pm 1.5$ $\pm 2.6$ & $ {<}0.1$ & $ {<}0.1$  & $ {<}0.1$  & 9 \\\hline
$\mathrm{0j}-{15}$ & $40.2 \pm 3.7$ $\pm 12.7$ & $4.8 \pm 2.3$ $\pm 2.4$ & $14.3 \pm 1.5$ $\pm 4.0$ & $27.8 \pm 2.3$ $\pm 7.6$ & $2.8 \pm 1.4$ $\pm 1.9$ & $0.7^{+4.1}_{-0.7}$ $\pm 4.2$ & $90.5 \pm 6.8$ $\pm 16.2$ & $ 0.2$ $\pm$ $0.0 ^{+0.5}_{-0.2}$  &  $ {<}0.1$  & $ 0.2$ $\pm$ $0.0 ^{+1.6}_{-0.2}$  & 82 \\\hline
$\mathrm{0j}-{16}$ & $18.0 \pm 2.5$ $\pm 5.6$ & $ {<}0.1$  & $8.1 \pm 1.2$ $\pm 2.8$ & $11.4 \pm 1.5$ $\pm 3.4$ & $ {<}0.1$  & $2.9^{+3.4}_{-2.9}$ $\pm 3.7$ & $40.5 \pm 4.7$ $\pm 8.1$ & $ 0.1$ $\pm$ $0.0 ^{+0.2}_{-0.1}$  &  $ {<}0.1$  & $ {<}0.1$  & 51 \\\hline
$\mathrm{0j}-{17}$ & $30.5 \pm 3.2$ $\pm 10.4$ & $ {<}0.1$  & $13.5 \pm 1.5$ $\pm 4.0$ & $15.2 \pm 1.7$ $\pm 4.9$ & $ {<}0.1$  & $ {<}0.1$  & $59.3 \pm 4.0$ $\pm 12.2$ & $ 0.1$ $\pm$ $0.0 ^{+0.3}_{-0.1}$  &  $ {<}0.2$  & $ {<}0.1$  & 61 \\\hline
$\mathrm{0j}-{18}$ & $9.0 \pm 1.8$ $\pm 3.7$ & $ {<}0.1$  & $2.2 \pm 0.6$ $\pm 1.0$ & $ 1.1$ $\pm$ $0.5 ^{+1.2}_{-1.1}$  & $0.2 \pm 0.1$ $\pm 0.1$ & $1.9^{+1.9}_{-1.9}$ $\pm 2.1$ & $14.5 \pm 2.7$ $\pm 4.5$ & $ {<}0.1$  & $ {<}0.1$  & $ {<}0.1$  & 11 \\\hline
$\mathrm{0j}-{19}$ & $10.5 \pm 1.9$ $\pm 3.7$ & $ {<}0.1$  & $5.1 \pm 0.9$ $\pm 1.7$ & $8.7 \pm 1.3$ $\pm 3.2$ & $0.6 \pm 0.4$ $\pm 0.5$ & $0.7^{+2.0}_{-0.7}$ $\pm 2.0$ & $25.6 \pm 3.2$ $\pm 5.5$ & $ 0.1$ $\pm$ $0.0 ^{+0.3}_{-0.1}$  &  $ {<}0.2$  & $ {<}0.1$  & 30 \\\hline
$\mathrm{0j}-{20}$ & $1.4 \pm 0.7$ $\pm 1.0$ & $ {<}0.1$  & $0.5 \pm 0.3$ $\pm 0.5$ & $2.8 \pm 0.8$ $\pm 1.3$ & $0.2 \pm 0.1$ $\pm 0.2$ & $ {<}0.1$  & $4.9 \pm 1.1$ $\pm 1.7$ & $ 0.1$ $\pm$ $0.0 ^{+0.3}_{-0.1}$  &  $ {<}0.1$  & $ {<}0.1$  & 5 \\\hline
$\mathrm{0j}-{21}$ & $ 0.4$ $\pm$ $0.4 ^{+0.6}_{-0.4}$  & $ {<}0.1$  & $ {<}0.4$  & $3.5 \pm 0.8$ $\pm 1.4$ & $0.2 \pm 0.1$ $\pm 0.1$ & $1.6^{+1.9}_{-1.6}$ $\pm 2.1$ & $5.6 \pm 2.1$ $\pm 2.6$ & $ 0.3$ $\pm$ $0.0 ^{+0.6}_{-0.3}$  &  $ {<}0.1$  & $ {<}0.1$  & 4 \\\hline
$\mathrm{0j}-{22}$ & $2.4 \pm 0.9$ $\pm 1.3$ & $ {<}0.1$  & $0.7 \pm 0.3$ $\pm 0.5$ & $0.9 \pm 0.4$ $\pm 0.5$ &  ${<}0.1$  & $ {<}0.1$  & $4.1 \pm 1.0$ $\pm 1.4$ & $ {<}0.1$ &  $ {<}0.1$  & $ {<}0.1$  & 2 \\\hline
$\mathrm{1j}-{1}$ & $ 1.0$ $\pm$ $0.6 ^{+1.1}_{-1.0}$  & $ {<}0.1$  & $ 0.2$ $\pm$ $0.2 ^{+0.2}_{-0.2}$  & $ 0.2$ $\pm$ $0.2 ^{+0.8}_{-0.2}$  & $ 1.6$ $\pm$ $1.4 ^{+2.8}_{-1.6}$  & $3.6 \pm 2.7$ $\pm 3.3$ & $6.5 \pm 3.2$ $\pm 4.5$ & $ {<}0.1$  & $ {<}0.1$  & $ {<}0.1$  & 2 \\\hline
$\mathrm{1j}-{2}$  & $20.2 \pm 2.7$ $\pm 7.6$ & $ {<}0.1$  & $6.3 \pm 1.0$ $\pm 2.4$ & $10.1 \pm 1.4$ $\pm 3.1$ & $1.8 \pm 0.5$ $\pm 1.0$ & $0.3^{+5.3}_{-0.3}$ $\pm 5.3$ & $38.6 \pm 6.2$ $\pm 10.1$ & $ {<}0.1$  &  $ {<}0.1$  & $ {<}0.1$  & 43 \\\hline
$\mathrm{1j}-{3}$ & $138.1 \pm 7.0$ $\pm 40.1$ & $50.5 \pm 6.2$ $\pm 10.4$ & $52.3 \pm 3.0$ $\pm 15.0$ & $114.1 \pm 4.8$ $\pm 29.6$ & $23.0 \pm 7.0$ $\pm 10.5$ & $ {<}0.1$  & $378.0 \pm 13.0$ $\pm 54.1$ & $ 0.0$ $\pm$ $0.0 ^{+0.2}_{-0.0}$  & $0.2 \pm 0.1$ $\pm 0.1$ & $ 0.2$ $\pm$ $0.0 ^{+1.1}_{-0.2}$  & 382 \\\hline
$\mathrm{1j}-{4}$ & $121.1 \pm 6.6$ $\pm 36.9$ & $1.2 \pm 0.7$ $\pm 0.8$ & $48.0 \pm 2.9$ $\pm 13.8$ & $59.4 \pm 3.5$ $\pm 16.6$ & $5.7 \pm 2.0$ $\pm 4.2$ & $ {<}0.1$  & $235.3 \pm 8.3$ $\pm 43.0$ & $ {<}0.1$  & $0.1 \pm 0.1$ $\pm 0.1$ & $ {<}0.1$  & 211 \\\hline
$\mathrm{1j}-{5}$  & $6.6 \pm 1.5$ $\pm 3.3$ & $ 0.5$ $\pm$ $0.5 ^{+0.6}_{-0.5}$  & $2.2 \pm 0.6$ $\pm 1.0$ & $5.3 \pm 1.0$ $\pm 2.4$ & $0.7 \pm 0.4$ $\pm 0.7$ & $6.6 \pm 4.2$ $\pm 5.3$ & $22.0 \pm 4.7$ $\pm 6.8$ & $ {<}0.1$  & $ {<}0.1$  & $ {<}0.1$  & 20 \\\hline
$\mathrm{1j}-{6}$ & $49.3 \pm 4.2$ $\pm 15.3$ & $3.2 \pm 1.8$ $\pm 1.9$ & $9.7 \pm 1.3$ $\pm 3.3$ & $15.9 \pm 1.8$ $\pm 4.8$ & $ 2.8$ $\pm$ $1.1 ^{+5.0}_{-2.8}$  & $ {<}0.1$  & $80.8 \pm 5.2$ $\pm 17.3$ & $ {<}0.1$  &  $ {<}0.1$  & $ 0.0$ $\pm$ $0.0 ^{+0.2}_{-0.0}$  & 54 \\\hline
$\mathrm{1j}-{7}$ & $266.9 \pm 9.8$ $\pm 79.3$ & $0.5 \pm 0.4$ $\pm 0.4$ & $86.1 \pm 3.8$ $\pm 23.4$ & $165.0 \pm 5.8$ $\pm 42.5$ & $14.2 \pm 4.5$ $\pm 6.5$ & $17.7 \pm 11.7$ $\pm 14.6$ & $550.3 \pm 17.3$ $\pm 94.3$ & $ 0.3$ $\pm$ $0.1 ^{+0.7}_{-0.3}$  &  $ {<}0.3$  & $ 0.1$ $\pm$ $0.0 ^{+0.7}_{-0.1}$  & 511 \\\hline
$\mathrm{1j}-{8}$ & $35.9 \pm 3.6$ $\pm 11.5$ & $ {<}0.1$  & $6.4 \pm 1.0$ $\pm 3.0$ & $9.4 \pm 1.4$ $\pm 3.0$ & $ {<}0.1$  & $ {<}0.1$  & $51.7 \pm 4.0$ $\pm 12.3$ & $ 0.1$ $\pm$ $0.0 ^{+0.2}_{-0.1}$  &  $ {<}0.1$  & $ {<}0.1$  & 62 \\\hline
$\mathrm{1j}-{9}$ & $31.5 \pm 3.3$ $\pm 10.5$ & $ {<}0.1$  & $7.1 \pm 1.1$ $\pm 2.9$ & $9.9 \pm 1.4$ $\pm 3.1$ & $0.6 \pm 0.5$ $\pm 0.6$ & $2.0^{+2.8}_{-2.0}$ $\pm 3.0$ & $51.1 \pm 4.8$ $\pm 11.8$ & $ 0.0$ $\pm$ $0.0 ^{+0.2}_{-0.0}$  &  $ {<}0.1$  & $ {<}0.1$  & 40 \\\hline
$\mathrm{1j}-{10}$ & $68.1 \pm 4.9$ $\pm 21.3$ & $0.4 \pm 0.4$ $\pm 0.4$ & $20.7 \pm 1.9$ $\pm 5.9$ & $14.1 \pm 1.7$ $\pm 4.1$ & $1.4^{+1.8}_{-1.4}$ $\pm 2.7$ & $ {<}0.1$  & $104.8 \pm 5.8$ $\pm 22.7$ & $ 0.2$ $\pm$ $0.0 ^{+0.4}_{-0.2}$  &  $ {<}0.1$  & $ {<}0.1$  & 88 \\\hline
$\mathrm{1j}-{11}$ & $93.2 \pm 5.8$ $\pm 30.3$ & $ {<}0.1$  & $28.8 \pm 2.2$ $\pm 9.0$ & $25.8 \pm 2.3$ $\pm 7.4$ & $1.7 \pm 0.7$ $\pm 1.7$ & $ {<}0.1$  & $149.5 \pm 6.6$ $\pm 32.5$ & $ 0.4$ $\pm$ $0.1 ^{+1.0}_{-0.4}$  & $0.4 \pm 0.1$ $\pm 0.1$ & $ {<}0.1$  & 122 \\\hline
$\mathrm{1j}-{12}$ & $ {<}0.4$  & $ {<}0.1$  & $  {<}0.1$  & $0.4 \pm 0.3$ $\pm 0.4$ &  ${<}0.1$  & $ {<}0.1$  & $0.5 \pm 0.3$ $\pm 0.6$ & $ {<}0.1$  &  $ {<}0.1$  & $ {<}0.1$  & 0 \\\hline
$\mathrm{1j}-{13}$  & $2.6 \pm 0.9$ $\pm 1.2$ & $ {<}0.1$  & $1.1 \pm 0.4$ $\pm 0.5$ & $1.5 \pm 0.6$ $\pm 0.8$ &  ${<}0.1$  & $ {<}0.1$  & $5.4 \pm 1.2$ $\pm 1.6$ & $ {<}0.1$  &  $ {<}0.1$  & $ {<}0.1$  & 1 \\\hline
$\mathrm{1j}-{14}$ & $23.7 \pm 2.8$ $\pm 7.0$ & $ {<}0.1$  & $6.1 \pm 1.0$ $\pm 2.0$ & $6.4 \pm 1.1$ $\pm 2.1$ & $0.3 \pm 0.1$ $\pm 0.3$ & $ 2.8$ $\pm$ $2.8 ^{+3.1}_{-2.8}$  & $39.4 \pm 4.2$ $\pm 8.2$ & $ 0.1$ $\pm$ $0.0 ^{+0.3}_{-0.1}$  &  $ {<}0.1$  & $ {<}0.1$  & 30 \\\hline
$\mathrm{1j}-{15}$  & $250.0 \pm 9.2$ $\pm 73.0$ & $6.4 \pm 1.8$ $\pm 3.1$ & $48.0 \pm 2.8$ $\pm 12.8$ & $81.0 \pm 4.0$ $\pm 21.2$ & $10.6 \pm 2.4$ $\pm 7.9$ & $3.9^{+7.1}_{-3.9}$ $\pm 7.3$ & $399.8 \pm 13.0$ $\pm 77.9$ & $ 0.9$ $\pm$ $0.1 ^{+2.3}_{-0.9}$  & $1.0 \pm 0.1$ $\pm 0.2$ & $ 0.5$ $\pm$ $0.0 ^{+3.1}_{-0.5}$  & 353 \\\hline
$\mathrm{1j}-{16}$ & $74.3 \pm 5.0$ $\pm 21.1$ & $ {<}0.1$  & $21.1 \pm 1.9$ $\pm 6.1$ & $15.9 \pm 1.8$ $\pm 4.6$ & $1.7 \pm 0.7$ $\pm 0.7$ & $2.8^{+4.1}_{-2.8}$ $\pm 4.4$ & $115.8 \pm 7.0$ $\pm 22.9$ & $ 0.3$ $\pm$ $0.0 ^{+0.6}_{-0.3}$  &  $ {<}0.4$  & $ {<}0.1$  & 93 \\\hline
$\mathrm{1j}-{17}$  & $124.7 \pm 6.6$ $\pm 35.9$ & $ {<}0.1$  & $27.0 \pm 2.1$ $\pm 7.5$ & $23.3 \pm 2.2$ $\pm 6.8$ & $1.2 \pm 0.3$ $\pm 0.4$ & $2.0^{+5.7}_{-2.0}$ $\pm 5.8$ & $178.2 \pm 9.2$ $\pm 37.7$ & $ 0.3$ $\pm$ $0.0 ^{+0.7}_{-0.3}$  & $0.3 \pm 0.1$ $\pm 0.1$ & $ {<}0.1$  & 158 \\\hline
$\mathrm{1j}-{18}$ & $60.7 \pm 4.6$ $\pm 17.8$ & $ {<}0.1$  & $9.1 \pm 1.2$ $\pm 2.8$ & $11.8 \pm 1.5$ $\pm 3.6$ & $1.1 \pm 0.4$ $\pm 0.5$ & $2.8^{+3.6}_{-2.8}$ $\pm 3.9$ & $85.5 \pm 6.2$ $\pm 18.8$ & $ 0.1$ $\pm$ $0.0 ^{+0.3}_{-0.1}$  & $0.2 \pm 0.1$ $\pm 0.1$ & $ {<}0.1$  & 70 \\\hline
$\mathrm{1j}-{19}$ & $39.4 \pm 3.6$ $\pm 12.2$ & $ {<}0.1$  & $8.8 \pm 1.2$ $\pm 3.0$ & $10.0 \pm 1.4$ $\pm 3.6$ & $ {<}3.6$  & $1.3^{+2.9}_{-1.3}$ $\pm 3.0$ & $59.7 \pm 5.0$ $\pm 13.9$ & $ 0.3$ $\pm$ $0.0 ^{+0.7}_{-0.3}$  &  $ {<}0.4$  & $ {<}0.1$  & 57 \\\hline
$\mathrm{1j}-{20}$ & $5.2 \pm 1.3$ $\pm 3.3$ & $ {<}0.1$  & $1.6 \pm 0.5$ $\pm 0.7$ & $2.6 \pm 0.7$ $\pm 1.2$ & $0.4 \pm 0.1$ $\pm 0.2$ & $ {<}0.1$  & $9.8 \pm 1.6$ $\pm 3.6$ & $ 0.3$ $\pm$ $0.1 ^{+0.8}_{-0.3}$  &  $ {<}0.2$  & $ {<}0.1$  & 5 \\\hline
$\mathrm{1j}-{21}$ & $ 0.7$ $\pm$ $0.5 ^{+0.8}_{-0.7}$  & $ {<}0.1$  & $0.2 \pm 0.2$ $\pm 0.2$ & $2.6 \pm 0.7$ $\pm 1.0$ & $ {<}0.1$  & $ {<}0.1$  & $3.6 \pm 0.9$ $\pm 1.3$ & $ 0.6$ $\pm$ $0.1 ^{+1.5}_{-0.6}$  & $0.3 \pm 0.1$ $\pm 0.1$ & $ {<}0.1$  & 5 \\\hline
$\mathrm{1j}-{22}$ & $ 0.4$ $\pm$ $0.4 ^{+0.7}_{-0.4}$  & $ {<}0.1$  & $  {<}0.1$  & $0.7 \pm 0.4$ $\pm 0.5$ & $0.3 \pm 0.1$ $\pm 0.2$ & $ 1.9$ $\pm$ $1.9 ^{+2.1}_{-1.9}$  & $3.3 \pm 2.0$ $\pm 2.3$ & $ 0.2$ $\pm$ $0.0 ^{+0.4}_{-0.2}$  &  $ {<}0.1$  & $ {<}0.1$  & 1 \\
\hline
\end{tabular}}
\end{sidewaystable}
\cleardoublepage \section{The CMS Collaboration \label{app:collab}}\begin{sloppypar}\hyphenpenalty=5000\widowpenalty=500\clubpenalty=5000\vskip\cmsinstskip
\textbf{Yerevan Physics Institute, Yerevan, Armenia}\\*[0pt]
A.M.~Sirunyan, A.~Tumasyan
\vskip\cmsinstskip
\textbf{Institut f\"{u}r Hochenergiephysik, Wien, Austria}\\*[0pt]
W.~Adam, F.~Ambrogi, E.~Asilar, T.~Bergauer, J.~Brandstetter, E.~Brondolin, M.~Dragicevic, J.~Er\"{o}, A.~Escalante~Del~Valle, M.~Flechl, R.~Fr\"{u}hwirth\cmsAuthorMark{1}, V.M.~Ghete, J.~Hrubec, M.~Jeitler\cmsAuthorMark{1}, N.~Krammer, I.~Kr\"{a}tschmer, D.~Liko, T.~Madlener, I.~Mikulec, N.~Rad, H.~Rohringer, J.~Schieck\cmsAuthorMark{1}, R.~Sch\"{o}fbeck, M.~Spanring, D.~Spitzbart, A.~Taurok, W.~Waltenberger, J.~Wittmann, C.-E.~Wulz\cmsAuthorMark{1}, M.~Zarucki
% ...
\vskip\cmsinstskip
% ...
1:  Also at Vienna University of Technology, Vienna, Austria\\
% ...
\end{sloppypar}
\end{document}